\documentclass[fleqn,usenatbib]{mnras}

\usepackage[T1]{fontenc}
\usepackage{ae,aecompl}

\usepackage{graphicx}	% Including figure files
\usepackage{amsmath}	% Advanced maths commands
\usepackage{amssymb}	% Extra maths symbols
\usepackage{times}
\usepackage{booktabs}
\usepackage{multirow, makecell}
\usepackage{mathtools}
\usepackage{blkarray}
\usepackage{fix-cm}
\usepackage{cellspace}	% Control table cell spacing
\usepackage[inline]{enumitem}
\usepackage{xcolor}

\makeatletter
\newlength{\abovecaptionskip}%
\makeatother
\usepackage{threeparttable}
\usepackage{tabularx}
\makeatletter
\g@addto@macro{\UrlBreaks}{\UrlOrds}
\makeatother

\usepackage{newtxtext,newtxmath}

\defcitealias{eBOSS2021}{eBOSS21}
\defcitealias{Beutler2011}{6dFGS}
\defcitealias{Kazin2014}{WiggleZ}
\defcitealias{Ross2015}{SDSS MGS}
\defcitealias{Alam2017}{BOSS LRG}
\defcitealias{Bautista2021}{eBOSS LRG}
\defcitealias{Raichoor2021}{eBOSS ELG (Conf.)}
\defcitealias{deMattia2021}{eBOSS ELG (Fourier)}
\defcitealias{Hou2021}{eBOSS QSO}
\defcitealias{Bourboux2020}{eBOSS Ly$\alpha$}
\defcitealias{Riess2019}{Distance Ladder}

\newcommand{\mpc}{h^{-1}\,{\rm Mpc}}

\title[Cosmology with SDSS multi-tracer BAO]{The completed SDSS-IV extended Baryon Oscillation Spectroscopic Survey: Cosmological implications from multi-tracer BAO analysis with galaxies and voids}

\author[C. Zhao et al.]{
\parbox{\textwidth}{
Cheng Zhao$^{1,}$\thanks{E-mail: \texttt{\href{mailto:cheng.zhao@epfl.ch}{cheng.zhao@epfl.ch}}},
Andrei Variu$^1$,
Mengfan He$^{2,3,4}$,
Daniel Forero-S\'anchez$^1$,
Am\'elie Tamone$^1$,
Chia-Hsun Chuang$^5$,
Francisco-Shu Kitaura$^{6,7}$,
Charling Tao$^{8,9}$,
Jiaxi Yu$^1$,
Jean-Paul Kneib$^{1,10}$,
Will J. Percival$^{11,12,13}$,
Huanyuan Shan$^{3}$,
Gong-Bo Zhao$^{2,4}$,
Etienne Burtin$^{14}$,
Kyle S. Dawson$^{15}$,
Graziano Rossi$^{16}$,
Donald P. Schneider$^{17,18}$, and
Axel de la Macorra$^{19}$
}
\\
\vspace*{4pt} \\
\small $^{1}$Institute of Physics, Laboratory of Astrophysics, \'Ecole Polytechnique F\'ed\'erale de Lausanne (EPFL), Observatoire de Sauverny, CH-1290 Versoix, Switzerland\vspace*{-2pt} \\
\small $^{2}$National Astronomy Observatories, Chinese Academy of Science, Beijing 100101, P.R. China\vspace*{-2pt} \\
\small $^{3}$Key Laboratory for Research in Galaxies and Cosmology, Shanghai Astronomical Observatory, Shanghai 200030, P.R. China\vspace*{-2pt} \\
\small $^{4}$School of Astronomy and Space Science, University of Chinese Academy of Sciences, Beijing 100049, P.R. China\vspace*{-2pt} \\
\small $^{5}$Kavli Institute for Particle Astrophysics and Cosmology, Stanford University, 452 Lomita Mall, Stanford, CA 94305, USA\vspace*{-2pt} \\
\small $^{6}$Instituto de Astrofísica de Canarias, s/n, E-38205, La Laguna, Tenerife, Spain\vspace*{-2pt} \\
\small $^{7}$Departamento de Astrof\'isica, Universidad de La Laguna, E-38206, La Laguna, Tenerife, Spain\vspace*{-2pt} \\
\small $^{8}$CPPM, Aix-Marseille Universit\'e, CNRS/IN2P3, CPPM UMR 7346, F13288 Marseille, France\vspace*{-2pt} \\
\small $^{9}$Tsinghua Center for Astrophysics, Department of Astronomy, Tsinghua University, Beijing 100084, P.R. China\vspace*{-2pt} \\
\small $^{10}$Aix Marseille Univ, CNRS, CNES, LAM, F13388 Marseille, France\vspace*{-2pt} \\
\small $^{11}$Waterloo Centre for Astrophysics, University of Waterloo, Waterloo, ON N2L 3G1, Canada\vspace*{-2pt} \\
\small $^{12}$Department of Physics and Astronomy, University of Waterloo, Waterloo, ON N2L 3G1, Canada\vspace*{-2pt} \\
\small $^{13}$Perimeter Institute for Theoretical Physics, 31 Caroline St. North, Waterloo, ON N2L 2Y5, Canada\vspace*{-2pt} \\
\small $^{14}$IRFU,CEA, Universit\'e Paris-Saclay, F-91191 Gif-sur-Yvette, France\vspace*{-2pt} \\
\small $^{15}$Department Physics and Astronomy, University of Utah, 115 S 1400 E, Salt Lake City, UT 84112, USA\vspace*{-2pt} \\
\small $^{16}$Department of Physics and Astronomy, Sejong University, Seoul 143-747, Korea\vspace*{-2pt} \\
\small $^{17}$Department of Astronomy and Astrophysics, The Pennsylvania State University, University Park, PA 16802, USA\vspace*{-2pt} \\
\small $^{18}$Institute for Gravitation and the Cosmos, The Pennsylvania State University, University Park, PA 16802, USA\vspace*{-2pt} \\
\small $^{19}$Instituto de F\'{i}sica, Universidad Nacional Aut\'{o}noma de M\'{e}xico, Apdo. Postal 20-364, M\'{e}xico\vspace*{-2pt} \\
}

\date{Accepted XXX. Received YYY; in original form ZZZ}

\pubyear{2021}

\begin{document}
\label{firstpage}
\pagerange{\pageref{firstpage}--\pageref{lastpage}}
\maketitle

\begin{abstract}
We construct cosmic void catalogues with the \textsc{dive} void finder upon SDSS BOSS DR12 and eBOSS DR16 galaxy samples with BAO reconstruction, and perform a joint BAO analysis using different types of galaxies and the corresponding voids. The BAO peak is evident for the galaxy--galaxy, galaxy--void, and void--void correlation functions of all datasets, including the ones cross-correlating LRG and ELG samples. Two multi-tracer BAO fitting schemes are tested, one combining the galaxy and void correlation functions with a weight applied to voids, and the other using a single BAO dilation parameter for all clustering measurements. Both methods produce consistent results with mock catalogues, and on average $\sim$10 per cent improvements of the BAO statistical uncertainties are observed for all samples, compared to the results from galaxies alone. By combining the clustering of galaxies and voids, the uncertainties of BAO measurements from the SDSS data are reduced by 5 to 15 per cent, yielding 0.9, 0.8, 1.1, 2.3, and 2.9 per cent constraints on the distance $D_{_{\rm V}}(z)$, at effective redshifts 0.38, 0.51, 0.70, 0.77, and 0.85, respectively.
When combined with BAO measurements from SDSS MGS, QSO, and Ly$\alpha$ samples, as well as the BBN results, we obtain $H_0 = 67.58 \pm 0.91\,{\rm km}\,{\rm s}^{-1}\,{\rm Mpc}^{-1}$, $\Omega_{\rm m} = 0.290 \pm 0.015$, and $\Omega_\Lambda h^2 = 0.3241 \pm 0.0079$ in the flat-$\Lambda$CDM framework, where the 1\,$\sigma$ uncertainties are around 6, 6, and 17 per cent smaller respectively, compared to constraints from the corresponding anisotropic BAO measurements without voids and LRG--ELG cross correlations.

%The cosmological parameter constraints from these results are generally comparable to the galaxy-only results with the anisotropic $D_{_{\rm M}}$ and $D_{_{\rm H}}$ measurements. When combined with BAO results BBN, the SDSS BAO measurements with multi-tracer analysis yields tighter constraints on some parameters, such as $H_0$ and $\Omega_{\rm m}$ in the flat-$\Lambda$CDM framework.
\end{abstract}


%%%%%%%%%%%%%%%%%%%%%%%%%%%%%%%%%%%%%%%%%%%%%%%%%%

%%%%%%%%%%%%%%%%% BODY OF PAPER %%%%%%%%%%%%%%%%%%

\section{Introduction}

The baryon acoustic oscillation (BAO) is a pattern in the clustering of matter, imprinted by sound waves in the primordial baryon--photo plasma of the Universe. It characterises the sound horizon, a particular scale that can be used as a `standard ruler' for measuring cosmological distances \citep[][]{Blake2003}.
In addition, it can be measured from both the cosmic microwave background (CMB) anisotropies \citep[e.g.][]{Bennett2003}, as well as the distributions of large-scale structures at lower redshifts \citep[][]{Seo2003}.
The evolution of this scale across cosmic time traces the expansion history of the Universe, it is thus useful for constraining the properties of different cosmological components, such as the nature of dark energy \citep[e.g.][]{Weinberg2013,Aubourg2015}.

Since the first clear detections of BAO from galaxy clustering achieved by the Two Degree Field Galaxy Redshift Survey \citep[2dFGRS;][]{Cole2005} and the Sloan Digital Sky Survey \citep[SDSS;][]{Eisenstein2005}, measurement of the BAO scale has been a key science goal for many large-scale spectroscopic surveys, 
such as the 6dF Galaxy Survey \citep[6dFGS;][]{Jones2004},
the WiggleZ Dark Energy Survey \citep[][]{Drinkwater2010},
the SDSS-\uppercase\expandafter{\romannumeral 3} Baryon Oscillation Spectroscopic Survey \citep[BOSS;][]{Dawson2013}
and SDSS-\uppercase\expandafter{\romannumeral 4} Extended Baryon Oscillation Spectroscopic Survey \citep[eBOSS;][]{Dawson2016},
the Dark Energy Spectroscopic Instrument \citep[DESI;][]{DESI2016},
as well as future surveys like
the 4-metre Multi-Object Spectroscopic Telescope \citep[4MOST;][]{deJong2019}
and Euclid \citep[][]{Laureijs2011}.

Nowadays, the precision of BAO measurements has achieved one per cent level with the distributions of luminous red galaxies \citep[LRGs;][]{Beutler2017,Ross2017},
Moreover, the BAO feature has been detected in the clustering of different types of tracers, including star-forming emission line galaxies \citep[ELGs;][]{Raichoor2021}, quasi-stellar objects \citep[QSOs;][]{Ata2018}, Lyman $\alpha$ (Ly$\alpha$) forests \citep[][]{Busca2013}, and cosmic voids \citep[][]{Kitaura2016void}.
With the fast expansion of the spectroscopic databases, it is even possible to perform BAO measurements in a multi-tracer manner, with the cross correlations between different tracers taken into account, including but not limited to LRGs and ELGs \citep[][]{Wang2020,ZhaoGB2021}, Ly$\alpha$ forests and QSOs \citep[][]{Bourboux2020}, as well as galaxies and voids \citep[][]{Nadathur2019void,Zhao2020}.
In the multi-tracer approach, it has been shown that statistical uncertainties of various cosmological constraints due to cosmic variance effects are significantly reduced \citep[see also][]{McDonald2009,Seljak2009,Blake2013,Favole2019}.

Multi-tracer analysis typically requires observations of different types of tracers in the same cosmic volume, which are not always possible due to the inhomogeneity of tracer abundances at different redshifts and limitations of instruments. However, as an indirect type of tracers, cosmic voids can in principle be identified from all existing matter tracer catalogues without carrying out additional observations.
Thus, it is almost always possible to perform multi-tracer cosmological measurements with voids.
Voids are large regions that are devoid of luminous objects, and trace underdensities of the matter density field \cite[see][for a review]{Weygaert2011}, though there are various different void definitions and identification algorithms \citep[e.g.][and references therein]{Colberg2008,Zhao2016,Cautun2018}.
Therefore, the distribution of voids encodes the information from troughs of the density field, which are complementary to the peaks traced by matter tracers, such as galaxies and QSOs.
This is true even for a linearized matter tracer distribution, which can be obtained by reconstruction algorithms \citep[e.g.][]{Padmanabhan2012}, as these objects always trace only density peaks.
As the result, the combination of voids and matter tracers is foreseen to provide a more detailed description of density field, and yield better cosmological constraints.

This has been confirmed by \citet[][]{Zhao2020} in terms of BAO measurements, which makes use of the void definition based on Delaunay triangulation \citep[][]{Delaunay1934} for capturing the BAO feature from underdensities, using the Delaunay trIangulation Void findEr\footnote{\url{https://github.com/cheng-zhao/DIVE}} \citep[\textsc{dive};][]{Zhao2016}. These voids are allowed to overlap with each other, resulting in a high void abundance, which turns out to be crucial for evidencing the BAO peak in void clustering \citep[][]{Kitaura2016void}. Actually, it is so far the only void definition that can be used for BAO constraints.
By combining the two-point correlation functions (2PCFs) of galaxies and voids, \citet[][]{Zhao2020} presents a 10 per cent improvement on the BAO measurement precision based on studies using simulations, which is also achieved with the BOSS LRG data at low redshifts.

In this study, we extend the work in \citet[][]{Zhao2020}, and aim at a joint BAO analysis with all BOSS and eBOSS galaxies, as well as the corresponding voids.
The measurements are validated using previous studies with the same datasets, including those for BOSS LRGs \citep[][]{Alam2017}, eBOSS LRGs \citep[][]{Gilmarin2020,Bautista2021}, and eBOSS ELGs \citep[][]{Tamone2020,deMattia2021,Raichoor2021}.
In addition, measurements from the SDSS main galaxy sample \citep[MGS;][]{Ross2015}, eBOSS QSOs \citep[][]{Neveux2020,Hou2021}, and Ly$\alpha$ forests \citep[][]{Bourboux2020} are also included for cosmological parameter constraints.
A crucial component of these studies is the construction of approximate mock catalogues \citep[][]{Farr2020,Lin2020,Zhao2021} and galaxy mocks based on $N$-body simulations for assessing theoretical systematic errors \citep[][]{Avila2020,Smith2020,Alam2021,Rossi2021}. 
Our results are then compared to the cosmological analysis with the same datasets, but not including voids \citep[][hereafter eBOSS21]{eBOSS2021}.
In practice, existing SDSS BAO measurements and the cosmological interpretations are taken from the SDSS website\footnote{The BAO and RSD measurements are available at \url{https://sdss.org/science/final-bao-and-rsd-measurements}, and the cosmological results can be found in \url{https://sdss.org/science/cosmology-results-from-eboss} and \url{https://svn.sdss.org/public/data/eboss/DR16cosmo/tags/v1_0_1/}.}.
 
The datasets used in this work are described in Section~\ref{sec:data}, followed by detailed explanations of the analysis methodology in Section~\ref{sec:method}, including the construction of void catalogues, clustering measurement algorithms, as well as BAO measurement and cosmological parameter constraint schemes. The method is then validated using mock catalogues in Section~\ref{sec:mock_test}, and analysis with the SDSS data are presented in Section~\ref{sec:results}. Finally, the results are concluded in Section~\ref{sec:conclusion}.

%%%%%%%%%%%%%%%%% NEW SECTION %%%%%%%%%%%%%%%%%%

\section{Data}
\label{sec:data}

We introduce in this section the data we use for BAO measurements, i.e. the LRG and star-forming ELG samples from BOSS and eBOSS observations -- which are the same samples as used in \citetalias[][]{eBOSS2021} -- as well as the corresponding mock catalogues for covariance matrix estimations and methodology validations, including both approximate mocks and catalogues from $N$-body simulations.

\subsection{SDSS-\texorpdfstring{\uppercase\expandafter{\romannumeral 3}}{III} BOSS DR12 LRGs}

The Baryon Oscillation Spectroscopic Survey \citep[BOSS;][]{Dawson2013} of SDSS-\uppercase\expandafter{\romannumeral 3} \citep[][]{Eisenstein2011} measured the spectra of over 1.3\,million LRGs between 2009 and 2014, using double-armed spectrographs \citep[][]{Smee2013} as well as the 2.5-metre Sloan Telescope \citep[][]{Gunn2006} at the Apache Point Observatory.
The complete dataset of BOSS LRGs is released as part of the SDSS Data Release 12\footnote{\url{https://data.sdss.org/sas/dr12/boss/lss/}} \citep[DR12;][]{Alam2015}. This LRG sample consists of two major populations, LOWZ and CMASS, that are targeted using different algorithms \citep[][]{Reid2016}, for galaxies with redshifts $\lesssim 0.4$, and from 0.4 to 0.7, respectively.
Moreover, LRGs with redshifts measured by SDSS-\uppercase\expandafter{\romannumeral 1}/\uppercase\expandafter{\romannumeral 2} \citep[][]{Abazajian2009} are also included in DR12, as the BOSS target selections are designed to extend that of previous SDSS LRG samples \citep[][]{Eisenstein2001}.

\begin{figure}
\centering
\includegraphics[width=.98\columnwidth]{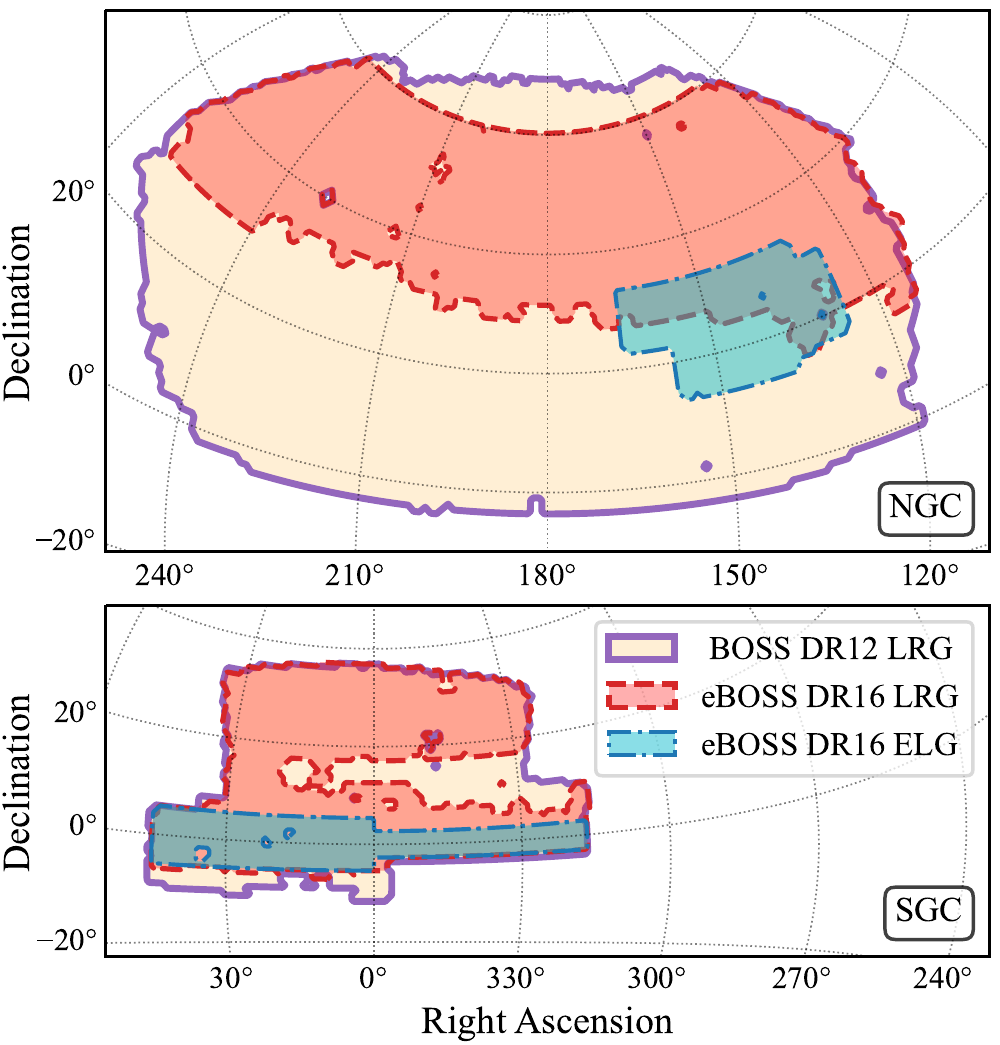}
\caption{Sky coverage of BOSS DR12 and eBOSS DR16 galaxies, for both northern ({\it upper}) and southern ({\it bottom}) galactic caps.}
\label{fig:footprint}
\end{figure}

\begin{figure}
\centering
\includegraphics[width=.98\columnwidth]{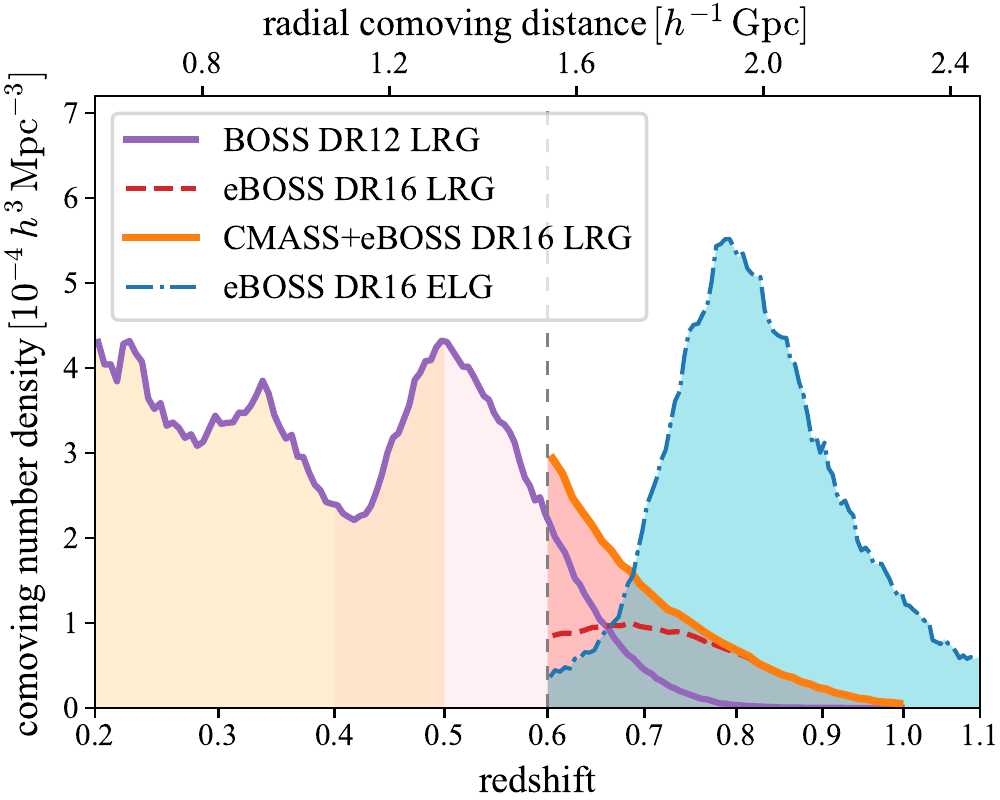}
\caption{Weighted comoving number densities of BOSS DR12 and eBOSS DR16 galaxies, with all the photometric and spectroscopic systematic weights included. Shadowed areas indicate redshift ranges of galaxy samples used in this work. The comoving distances and volumes are computed in the flat $\Lambda$CDM cosmology with $\Omega_{\rm m} = 0.31$.}
\label{fig:nbar}
\end{figure}

The final BOSS DR12 LRG catalogue for large-scale structure analysis is constructed by combining subsamples with different target selection rules, as the differences in clustering amplitudes are small \citep[][]{Reid2016, Alam2017}.
The footprints of this LRG sample for both northern and southern galactic caps (NGC and SGC, respectively) are shown in Figure~\ref{fig:footprint}, with a sky coverage of nearly 10000\,${\rm deg}^2$ in total.
Observational systematics, including photometric and spectroscopic effects, are corrected by various weights, and the total systematic weight is given by \citep[][]{Reid2016,Ross2017}
\begin{equation}
w_{\rm tot} = w_{\rm sys} ( w_{\rm cp} + w_{\rm noz} - 1 ),
\label{eq:boss_weight}
\end{equation}
where $w_{\rm sys}$ indicates the total angular photometric weights, while $w_{\rm cp}$ and $w_{\rm noz}$ are for correcting the fibre collision effect and redshift failures, respectively.
The overall weight for clustering measurements is then
\begin{equation}
w_{\rm all} = w_{\rm tot} w_{_{\rm FKP}} ,
\label{eq:weight_all}
\end{equation}
with $w_{_{\rm FKP}}$ being the FKP weight for reducing clustering variances \citep[][]{Feldman1994}, i.e.
\begin{equation}
w_{_{\rm FKP}} = \frac{1}{1 + n(z) P_0} .
\label{eq:weight_fkp}
\end{equation}
Here, $n(z)$ is the radial comoving number densities of galaxies (weighted by $w_{\rm tot}$, also known as the radial selection function, see Figure~\ref{fig:nbar}), and a $P_0$ value of $10000\,h^{-3}\,{\rm Mpc}^3$ is used for BOSS LRGs \citep[][]{Reid2016}.

Following \citet[][]{Alam2017} and 
\citetalias[][]{eBOSS2021}, we divide the LRG catalogue into two redshift bins. The redshift ranges are $0.2 < z < 0.5$ and $0.4 < z < 0.6$, and the samples are labelled `LRG (a)' and `LRG (b)' respectively.
The effective redshifts of the samples are defined as the weighted mean redshift of all galaxy pairs with separations in the range $(s_{\rm low}, s_{\rm high})$, as is done in eBOSS analysis \citep[e.g.][]{Bautista2021}:
\begin{equation}
z_{\rm eff} = \enspace \left. \sum_{\mathclap{\substack{i,j \\ s_{\rm low} < s_{ij} < s_{\rm high}}}}^{N_{\rm gal}} w_{{\rm all}, i} \, w_{{\rm all}, j}\, \frac{z_i + z_j}{2} \enspace \middle/ \quad \enspace \sum_{\mathclap{\substack{i,j \\ s_{\rm low} < s_{ij} < s_{\rm high}}}}^{N_{\rm gal}} w_{{\rm all}, i} \, w_{{\rm all}, j} \right. \, ,
\label{eq:zeff}
\end{equation}
where $N_{\rm gal}$ denotes the total number of galaxies in the given redshift ranges, and $s_{i j}$ indicates the separation between every galaxy pair.
We choose a separation range of $[50, 160]\,\mpc{}$ for effective redshift evaluations of all our samples, as it is close to the range used for our BAO measurements (see Section~\ref{sec:fit_range}).
The numbers of tracers and effective redshifts of our LRG samples are listed in Table~\ref{tab:sdss_ngal}. Note that when keeping two significant figures, the effective redshifts turn out to be identical to the ones used in BOSS DR12 analysis \citep[e.g.][]{Alam2017}, though their definitions are different.
Furthermore, BOSS LRGs with redshifts above 0.6 are not treated as an individual sample, but combined with eBOSS LRGs, as is detailed in the next subsection.

\begin{table}
\centering
\begin{threeparttable}
\caption{Redshift ranges and effective redshifts of the SDSS galaxy samples used in this paper, as well as the numbers of galaxies for each galactic cap.}
\begin{tabular}{ccccc}
\toprule
Sample & \makecell{Redshift\\range} & $z_{\rm eff}$ & \makecell{Galactic\\cap} & $N_{\rm gal}$ \\
\midrule
\multirowcell{3}{LRG (a)\\{\it(BOSS DR12)}} & \multirowcell{3}{$(0.2, 0.5)$} & \multirowcell{3}{0.38} & NGC & 429182 \\
& & & SGC & 174819 \\
& & & \textbf{Total} & \textbf{604001} \\
\midrule
\multirowcell{3}{LRG (b)\\{\it(BOSS DR12)}} & \multirowcell{3}{$(0.4, 0.6)$} & \multirowcell{3}{0.51} & NGC & 500872 \\
& & & SGC & 185498 \\
& & & \textbf{Total} & \textbf{686370} \\
\midrule
\multirowcell{3}{LRG (c)\\\textit{(CMASS+eBOSS DR16)}} & \multirowcell{3}{$(0.6, 1.0)$} & \multirowcell{3}{0.70} & NGC & 255741 \\
& & & SGC & 121717 \\
& & & \textbf{Total} & \textbf{377458} \\
\midrule
\multirowcell{3}{ELG\\{\it(eBOSS DR16)}} & \multirowcell{3}{$(0.6, 1.1)$} & \multirowcell{3}{0.85} & NGC & 83769 \\
& & & SGC & 89967 \\
& & & \textbf{Total} & \textbf{173736} \\
\bottomrule
\end{tabular}
\label{tab:sdss_ngal}
\end{threeparttable}
\end{table}

\subsection{SDSS-\texorpdfstring{\uppercase\expandafter{\romannumeral 4}}{IV} eBOSS DR16 LRGs and ELGs}

As the extension of BOSS, the Extended Baryon Oscillation Spectroscopic Survey \citep[eBOSS;][]{Dawson2016} of SDSS-\uppercase\expandafter{\romannumeral 4} \citep[][]{Blanton2017} probes galaxy distributions at higher redshifts.
It measured the spectra of over 0.2\,million LRGs and slightly more star-forming ELGs from 2014 to 2019, using the same instruments as BOSS.
The eBOSS LRG and ELG targets are selected to overlap with each other, as well as with the BOSS CMASS LRGs, in both angular and radial directions \citep[][]{Prakash2016,Raichoor2017}, thus permitting cross correlation measurements.
Moreover, the target selection algorithm for LRGs is designed to avoid BOSS CMASS targets \citep[][]{Prakash2016}.

The final eBOSS large-scale structure catalogues are released as DR16\footnote{\url{https://data.sdss.org/sas/dr16/eboss/lss/catalogs/DR16/}} \citep[][]{Ahumada2020}. The redshift ranges of LRGs and ELGs are $0.6 < z < 1.0$ and $0.6 < z < 1.1$, with the sky coverages of around 4200 and 720\,${\rm deg}^2$ respectively. The footprints are shown in Figure~\ref{fig:footprint}.
Similar to the case of BOSS data, observational systematics of eBOSS galaxies are corrected by weights accounting for angular photometric effects, fibre collisions, and redshift failures \citep[][]{Ross2020,Raichoor2021}. However, the combination of spectroscopic weights  are different from that of Eq.~\eqref{eq:boss_weight}, and the total weight for eBOSS galaxies is defined as
\begin{equation}
w_{\rm tot} = w_{\rm sys} w_{\rm cp} w_{\rm noz}.
\label{eq:eboss_weight}
\end{equation}
As stated before, eBOSS LRGs are combined with BOSS CMASS galaxies with $z > 0.6$ for cosmological analysis \citep[][]{Ross2020}. We denote this combined LRG sample by `LRG (c)' hereafter.
The weighted radial distributions of eBOSS DR16 galaxies are shown in Figure~\ref{fig:nbar}.
With $P_0$ for FKP weights (see Eq.~\eqref{eq:weight_fkp}) being 10000 and $4000\,h^{-3}\,{\rm Mpc}^3$ for the `LRG (c)' and eBOSS ELG samples (labelled `ELG') respectively \citep[][]{Ross2020,Raichoor2021}, the effective redshifts given by Eqs~\eqref{eq:weight_all}--\eqref{eq:eboss_weight} are listed in Table~\ref{tab:sdss_ngal}.
In addition, the effective redshift computed in the same way, for the cross correlation between these two samples, is $z_{\rm eff}^\times = 0.77$, which is consistent with the value reported in \citet[][]{Wang2020}.

The eBOSS DR16 data includes also 0.34 million QSOs \citep[][]{Ross2020}, which can in principle be used for the multi-tracer BAO measurements with voids as well. However, since the comoving number density of the QSOs is relatively low ($\sim 2 \times 10^{-5}\,h^3\,{\rm Mpc}^{-3}$), the radii of voids identified from the QSO sample are peaked at around $35\,\mpc$. The exclusion effect of such large voids may contaminate the BAO signature \citep[see][and discussions in Section~\ref{sec:radius_sel}]{Liang2016}, which requires careful inspections to avoid biases of BAO measurements. We leave a thorough multi-tracer BAO study with QSO voids to a future work.

\subsection{DR12 MultiDark-Patchy mock catalogues}
\label{sec:dr12_md_patchy}

We rely on the DR12 MultiDark-Patchy (MD-Patchy) mock catalogues\footnote{\url{https://data.sdss.org/sas/dr12/boss/lss/dr12_multidark_patchy_mocks/}} \citep[][]{Kitaura2016}, for estimating the covariance matrices of clustering measurements from the BOSS DR12 data, and validating our BAO fitting method (see Section~\ref{sec:bao_fit}).
These mocks are created based on halo catalogues generated at ten different redshifts in the range of $0.2 < z < 0.75$, that account for the evolution of clustering. The halo catalogues are constructed with the cosmological parameters listed in Table~\ref{tab:cosmology}, using the PerturbAtion Theory Catalogue generator of Halo and galaxY distributions \citep[\textsc{patchy};][]{Kitaura2014}, followed by halo mass assignment using the Halo mAss Distribution ReconstructiON method \citep[\textsc{hadron};][]{Zhao2015}. Both steps require calibrations with the BigMultiDark simulation \citep[BigMD;][see also Section~\ref{sec:data_sim}]{Klypin2016}.
Then, galaxy catalogues are constructed using the SUrvey GenerAtoR code \citep[\textsc{sugar};][]{Rodriguez2016}, which applies survey geometry to the halo samples, with observational effects such as stellar mass incompleteness and fibre collisions taken into account.

\begin{table}
\centering
\begin{threeparttable}
\caption{Flat-$\Lambda$CDM cosmology models used in this paper. The `fiducial` cosmology is for coordinate conversions and BAO fits of the SDSS data, MD-Patchy, and EZmock catalogues. The rest of the cosmology models are the ones for the construction of the corresponding datasets. Moreover, the BAO fits for the BigMD and OuterRim simulations are performed with their own cosmologies. Here, $r_{\rm d}$ denotes the comoving sound horizon at the drag epoch, computed using \textsc{camb} (see also Section~\ref{sec:bao_model}).}
\begin{tabular}{ccccc}
\toprule
Parameter & Fiducial & \makecell{BigMD \&\\MD-Patchy} & EZmock & OuterRim \\
\midrule
$h$ & 0.676 & 0.6777 & 0.6777 & 0.71 \\
$\Omega_{\rm m}$ & 0.31 & 0.307115 & 0.307115 & 0.26479 \\
$\Omega_{\rm b} h^2$ & 0.022 & 0.02214 & 0.02214 & 0.02258 \\
$\sigma_8$ & 0.8 & 0.8288 & 0.8225 & 0.8 \\
$n_{\rm s}$ & 0.97 & 0.9611 & 0.9611 & 0.963 \\
$\Sigma m_\nu$\,(eV) & 0.06 & 0 & 0 & 0 \\
\midrule
$r_{\rm d}\,({\rm Mpc})$ & 147.78 & 147.66 & 147.66 & 149.35 \\
\bottomrule
\end{tabular}
\label{tab:cosmology}
\end{threeparttable}
\end{table}

The \textsc{patchy} code implements accurate models of structure formation based on the Augmented Lagrangian Perturbation Theory \citep[ALPT;][]{Kitaura2013}, as well as galaxy biases that encodes nonlinear, stochastic, and nonlocal effects \citep[][]{Kitaura2014}.
As the result, MD-Patchy mocks reproduce the clustering of the BOSS DR12 data down to the scale of a few $\mpc$, including three-point statistics \citep[][]{Kitaura2016}.
There are 2048 independent realizations of DR12 MD-Patchy mocks in total, for each galactic cap. We use only 1000 of them (the indices are from 0001 to 1000) in this work, for both `LRG (a)' and `LRG (b)' samples.

\subsection{DR16 EZmock catalogues}

To assess the clustering covariances of the eBOSS DR16 data, including those of the cross correlations between LRGs and ELGs, we use the multi-tracer EZmock catalogues\footnote{\url{https://data.sdss.org/sas/dr17/eboss/lss/EZmocks/}} \citep[][]{Zhao2021} generated using the effective Zel’dovich approximation mock generator \citep[\textsc{EZmock};][]{Chuang2015EZ}. These mocks build upon the Zel'dovich approximation \citep[][]{Zeldovich1970} and effective descriptions of tracer biases that need to be calibrated with simulations or observations, which is similar to that of \textsc{patchy}. \textsc{EZmock} is highly efficient, and still achieves comparable precision as \textsc{patchy} on the same scales, for both two- and three-point clustering statistics \citep[][]{Chuang2015NIFTY}.

The DR16 EZmock catalogues are calibrated directly with the auto correlations of CMASS (for the combined LRG sample) and eBOSS data in different redshift bins.
In addition, the same white noises of the density fields are used for different tracers, to account for their cross correlations and covariances, as they reside in the same cosmic volume.
Mocks generated at different redshifts -- for reproducing the evolution of clustering -- are then combined and applied survey geometry using the \textsc{make\_survey}\footnote{\url{https://github.com/mockFactory/make\_survey}} \citep[][]{White2014} and \textsc{brickmask}\footnote{\url{https://github.com/cheng-zhao/brickmask}} \citep[][]{Zhao2021} codes.

Furthermore, various photometric and spectroscopic systematic effects are applied to the mocks, with corrections given by weights defined in the same way as those of the eBOSS data (see Eq.~\eqref{eq:eboss_weight}).
The clustering measurements of the final EZmock catalogues are in good agreement with those of the corresponding data on scales down to a few $\mpc$, including cross correlations in configuration space, and the discrepancies are generally less than 1\,$\sigma$ \citep[][]{Zhao2021}. We use all the 1000 realizations of DR16 EZmock catalogues in this paper, for the `LRG (c)' and `ELG' samples, as well as their cross correlations.

\subsection{\texorpdfstring{$\boldsymbol{N}$}{N}-body simulations}
\label{sec:data_sim}

Since the nonlinear damping of the BAO peak \citep[e.g.][]{Crocce2008,Prada2016} may not have been fully captured by the approximate mocks, which is a known issue for the DR16 EZmock catalogues \citep[][]{Bautista2021,Raichoor2021}, we use accurate mocks from high-resolution $N$-body simulations to model the BAO damping and validate our fitting method (see Section~\ref{sec:bao_model}).
To this end, we rely on galaxy catalogues constructed using $N$-body simulations at similar redshifts as our data samples, that reproduce the clustering statistics of the data.

For the LRG samples, we use galaxy catalogues from the BigMD simulation. They are actually the reference catalogues for the calibrations of DR12 MD-Patchy mocks (see Section~\ref{sec:dr12_md_patchy}), and are fitted to the BOSS DR12 data using the subhalo abundance matching \citep[SHAM; e.g.][]{Nuza2013} algorithm \citep[][]{Rodriguez2016}.
The BigMD simulation contains $3840^3$ particles with the mass of $2.36 \times 10^{10}\,h^{-1}\,{\rm M}_\odot$, in a box with the side length of $2.5\,h^{-1}\,{\rm Gpc}$ \citep[][]{Klypin2016}. It is run with the TreePM $N$-body code \textsc{gadget}-2 \citep[][]{Springel2005}, using the cosmological parameters listed in Table~\ref{tab:cosmology}. The halo catalogues used for SHAM are generated by the \textsc{rockstar} algorithm \citep[][]{Behroozi2013}.
We choose three of the ten BigMD galaxy catalogues for MD-Patchy calibrations, as is shown in Table~\ref{tab:nbody_mocks}, whose redshifts are the closest to the effective redshifts of our SDSS LRG samples (see Table~\ref{tab:sdss_ngal}).

\begin{table}
\centering
\begin{threeparttable}
\caption{$N\!$-body simulation galaxy catalogue used in this work, as well as the SDSS data that they are compared to (see Table~\ref{tab:sdss_ngal}). Properties of the simulation catalogues are also listed, including the side lengths of the periodic boxes, and redshifts and comoving number densities of the catalogues.}
\begin{tabular}{ccccc}
\toprule
\makecell{Simulation\\sample} & \makecell{SDSS\\sample} & \makecell{Boxsize\\($h^{-1}\,{\rm Gpc}$)} & Redshift & \makecell{Number density\\($h^3\,{\rm Mpc}^{-3}$)} \\
\midrule
BigMD (a) & LRG (a) & 2.5 & 0.392 & $2.66 \times 10^{-4}$ \\
BigMD (b) & LRG (b) & 2.5 & 0.505 & $4.28 \times 10^{-4}$ \\
BigMD (c) & LRG (c) & 2.5 & 0.638 & $1.98 \times 10^{-4}$ \\
OuterRim & ELG & 3 & 0.865 & $1.74 \times 10^{-3}$ \\
\bottomrule
\end{tabular}
\label{tab:nbody_mocks}
\end{threeparttable}
\end{table}

We use a galaxy catalogue generated using the halo occupation distribution \citep[HOD; e.g.][]{Berlind2003,Zheng2005} model \citep[`HOD-1' in][]{Avila2020}, based on the OuterRim simulation \citep[][]{Heitmann2019}, for our ELG sample.
The OuterRim simulation is performed using the Hardware/Hybrid Accelerated Cosmology Code \citep[HACC;][]{Habib2016}, with $10240^3$ particles in the volume of $(3\,h^{-1}\,{\rm Gpc})^3$. The particle mass is $1.89 \times 10^9\,h^{-1}\,{\rm M}_\odot$, and the cosmological parameters are shown in Table~\ref{tab:cosmology}.
Halo catalogues are constructed using the Friends-of-Friends \citep[FoF;][]{Davis1985} algorithm, and the one used in this work is at $z=0.865$.
The HOD catalogue is fitted to the eBOSS DR16 ELG data, with an enhanced number density (see Table~\ref{tab:nbody_mocks}) for reducing variances \citep[][]{Avila2020}.

%%%%%%%%%%%%%%%%% NEW SECTION %%%%%%%%%%%%%%%%%%

\section{Methodology}
\label{sec:method}

In this section we describe the recipe of measuring BAO peak positions from the two-point correlation functions of LRGs, ELGs, and the corresponding void catalogues, as well as the way for joint constraints using all tracers available. The methods are based on our previous analysis \citep[][]{Zhao2020}, but with various updates. Besides, we briefly introduce the method of extracting cosmological parameters from the BAO measurements.

\subsection{BAO reconstruction}

Since BAO reconstruction improves the significance of BAO peak in general, for both galaxies and voids \citep[see e.g.][]{Alam2017,Zhao2020}, we use this technique throughout this work.
It is a method introduced by \citet[][]{Eisenstein2007}, aiming at sharpening the BAO peak by mitigating nonlinear degradation effects, which are dominated by the bulk flows of large-scale structures and formations of superclusters.
This method is then extended \citep[][]{Noh2009,Padmanabhan2009} and validated using both simulations and observational data \citep[e.g.][]{Seo2010,Padmanabhan2012}.
In brief, the BAO reconstruction algorithm requires an estimation of the dark matter density field based on the distribution of biased tracers, usually in redshift space. This density field is smoothed with a Gaussian kernel, and converted to a displacement field using the Zel'dovich approximation. Finally, displacements are applied to galaxies and random samples to reverse the bulk flow motions. In particular, the shifted random catalogue is used for estimating the galaxy density field after reconstruction \citep[][]{Padmanabhan2012}.

The robustness of the BAO reconstruction technique and the improvement of BAO measurement precision have been verified by numerous studies \citep[e.g.][]{Xu2012,Burden2014,Vargas2015,White2015,Seo2016}.
Besides, a number of different reconstruction algorithms are proposed, that aim at better estimations of the displacement fields \citep[e.g.][]{Burden2015,Hada2018,Sarpa2019,Kitaura2021,Liu2021,Seo2021}.
BAO reconstruction is now a standard method for measuring BAO from spectroscopic galaxy data \citep[e.g.][]{Alam2017,Bautista2021,Raichoor2021}.

The BOSS DR12 galaxies and the corresponding MD-Patchy mocks, are reconstructed using the algorithm described in \citet[][]{Padmanabhan2012}. In particular, the reconstructions are run with the full redshift range of $0.2 < z < 0.75$, hence the `LRG (a)' and `LRG (b)' samples are reconstructed simultaneously \citep[][]{Alam2017}.
Meanwhile, the iterative reconstruction method\footnote{\url{https://github.com/julianbautista/eboss_clustering}} \citep[][]{Burden2015,Bautista2018} is used for the `LRG (c)' and `ELG' samples, as well as the corresponding EZmock catalogues \citep[][]{Bautista2021,Raichoor2021}.
Note that the CMASS and eBOSS DR16 LRGs are reconstructed as a whole sample, for both the observational data and each mock realization.
But the `LRG (c)' and `ELG' samples are reconstructed independently, even though they share the same survey volume to a large extent.
For the $N$-body simulation galaxy catalogues, we rely on the \textsc{Revolver}\footnote{\url{https://github.com/seshnadathur/Revolver}} \citep[][]{Nadathur2019} implementation of the iterative algorithm for BAO reconstruction.

The input parameters used for the reconstructions of all our samples are detailed in Table~\ref{tab:recon_param}.
In particular, the NGC and SGC samples are reconstructed separately -- as they are too far away from each other for cross correlations to matter -- but using the same set of parameters.
It is worth noting that small variances of the cosmological parameters and smoothing length do not significantly affect BAO measurements \citep[][]{Vargas2015,Sherwin2019,Carter2020}.
Moreover, though the reconstruction algorithms used for different samples in this work are not identical, there are no obvious biases on BAO measurements \citep[see][and Section~\ref{sec:mock_test}]{Vargas2018,Bautista2021,Raichoor2021}.
Actually, our reconstructed galaxy samples are the same as those used in \citetalias[][]{eBOSS2021}.

\begin{table}
\centering
\begin{threeparttable}
\caption{BAO reconstruction parameters for different samples. Here, $f$ denotes the growth factor for correcting redshift space distortions, $b$ denotes the galaxy bias parameter for estimating the matter density field, $N_{\rm grid}$ indicates the grid size for sampling the density field, and $\Sigma_{\rm r}$ is the Gaussian smoothing length.}
\begin{tabular}{ccccc}
\toprule
Sample & $f$ & $b$ & $N_{\rm grid}$ & $\Sigma_{\rm r}$ ($\mpc$) \\
\midrule
LRG (a) + (b) & 0.757 & 1.85 & $512^3$ & 15 \\
LRG (c) & 0.815 & 2.3 & $512^3$ & 15 \\
ELG & 0.820 & 1.4 & $512^3$ & 15 \\
\midrule
MD-Patchy & 0.757 & 2.2 & $512^3$ & 15 \\
EZmock for LRG (c) & 0.815 & 2.3 & $512^3$ & 15 \\
EZmock for ELG & 0.820 & 1.4 & $512^3$ & 15 \\
\midrule
BigMD (a) & 0.716 & 2.4 & $512^3$ & 10 \\
BigMD (b) & 0.757 & 2.5 & $512^3$ & 10 \\
BigMD (c) & 0.797 & 2.7 & $512^3$ & 10 \\
OuterRim & 0.823 & 1.8 & $512^3$ & 10 \\
\bottomrule
\end{tabular}
\label{tab:recon_param}
\end{threeparttable}
\end{table}

\subsection{Void catalogue creation}
\label{sec:void_cat}

We construct void catalogues upon the reconstructed galaxy samples for both observational data and approximate mocks. They are further selected by the void radii, to maximize the BAO significances.
We do not generate void catalogues using the $N$-body simulations, as the void clustering depends on the density and completeness of a galaxy sample, especially for the radial selection function \citep[][]{Forero2021}.
In this case, to reproduce the void 2PCFs of the observational data using catalogues from $N$-body simulations, one has to apply realistic geometric and radial selection effects to the simulated galaxy catalogues.
Consequently, the statistical variances of the catalogues generally become too large for precise BAO damping measurements.
Thus, we do not constrain the nonlinear damping of void BAO using $N$-body simulations in this work, but leave it to a future study.

\subsubsection{Void finding}

We rely on \textsc{dive} for identifying voids from a galaxy sample. It is a fast and parameter-free void finder, based on the \textsc{cgal}\footnote{\url{https://www.cgal.org}} \citep[][]{cgal} implementation of the Delaunay triangulation technique, which connects galaxies in comoving space to construct tetrahedra, such that their circumscribed spheres do not contain any galaxy inside. These circumspheres are used as our void tracers, which are essentially the largest spheres in the survey volume that are empty of galaxies.
It is worth noting that a high overlapping fraction is found for this type of voids \citep[][]{Zhao2016}, though the associated tetrahedra do not overlap, which turns out to be vital for detecting BAO from voids, as the BAO peak is not prominent for the clustering of disjoint voids \citep[][]{Kitaura2016void}.

Similar to the case of BAO reconstruction, the void finder is run on the full BOSS volume, so voids for the `LRG (a)' and `LRG (b)' samples are generated at once.
Besides, the `LRG (c)' void samples are constructed without distinguishing between CMASS and eBOSS LRGs, but independently of the `ELG' voids.
Void catalogues generated by \textsc{dive} are then trimmed to fit the survey volume: spheres are kept only if their centres are inside the footprints and redshift ranges, but outside veto masks, of the corresponding samples.
For the `LRG (c)' sample, we apply the union of the CMASS and eBOSS LRG footprints. There are however some ($< 3000$) very large spheres with radii larger than $100\,\mpc$, that reside inside the CMASS-only regions, and mostly at $z \gtrsim 0.85$, where there are only few galaxies. These objects are removed later during the void radius selection process (see Section~\ref{sec:radius_sel}), though their impacts on the clustering measurements are negligible on the BAO scale.

\begin{figure*}
\centering
\includegraphics[width=.95\textwidth]{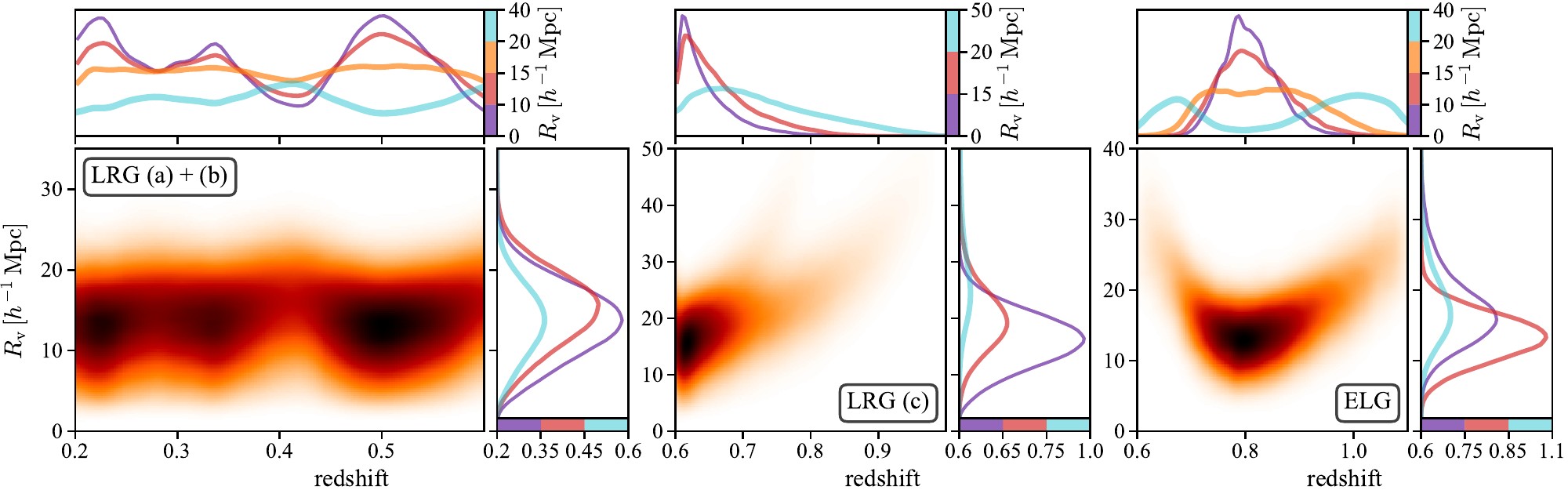}
\caption{Redshift and radius distributions of void samples generated by \textsc{dive}, for each SDSS sample. Colours in the heat maps indicate the comoving number density of voids, and the darker the colour, the higher the void density. Subplots to the right of the heat maps show the mean number density of voids with different radii, in three redshift bins indicated by the colours. The upper histograms show the radial void number density distributions, in radius bins encoded by different colours. Values of number densities are not shown, as the maps and curves are normalized arbitrarily.}
\label{fig:DTVoid_z_rad}
\end{figure*}

The redshift and radius distributions of the resulting void catalogue for each galaxy sample, are shown in Figure~\ref{fig:DTVoid_z_rad}.
It can be seen that the majority of the void radii are between 10 and 20\,$\mpc$, which is close the to mean separation of galaxies, but smaller than that of many other void definitions \citep[e.g.][]{Mao2017,Aubert2020,Hawken2020}. The is because we define voids purely geometrically, without removing any galaxies, and no merging process is performed even if voids overlap to a large extent.
The redshift dependence of void radius distributions is mostly due to the inhomogeneity of the radial number density of galaxies \citep[][]{Forero2021}. For the `LRG (c)' sample, since the number densities of CMASS and eBOSS LRGs differ significantly at high redshifts ($z \gtrsim 0.7$, see Figure~\ref{fig:nbar}), we observe two branches in the heat map shown in the middle panel of Figure~\ref{fig:DTVoid_z_rad}. These two void populations are found to be located in the overlapping regions between CMASS and eBOSS LRGs, and the CMASS only area (see Figure~\ref{fig:footprint}), respectively.

Moreover, our spheres with small radii show similar radial distributions as the corresponding galaxies (see Figure~\ref{fig:nbar}), while the distributions of large voids are in general less redshift-dependent, or anti-correlate with that of galaxies. These effects are consistent with the galaxy density dependences of voids in different sizes, studied by \citet[][]{Forero2021} using simulations. The behaviours can be explained by two distinct void populations -- voids-in-clouds and voids-in-voids -- that reside in regions with overdensities and underdensities respectively \citep[e.g.][]{Sheth2004,Zhao2016}. A void selection algorithm is then necessary, for mitigating contamination of the BAO signature from voids-in-clouds, and maximizing the statistical power of void BAO \citep[][]{Liang2016}. To this end, we have to construct random catalogues for our void samples first, to perform clustering measurements.

\subsubsection{Random catalogue construction}

The clustering signal is defined as the excess spatial correlation of tracers compared to that of a random distribution. Thus a random sample is required for clustering measurements, which should account for effects that may introduce additional clustering pattern, such as the survey geometry and sample completeness in both angular and radial directions.
The angular inhomogeneity of galaxies can be assessed by correlating galaxy densities with potential observational systematics, including but not limited to stellar density, seeing, airmass, and Galactic extinction \citep[e.g.][]{Ross2012,Ross2020,Raichoor2021}.
For the radial selection function, one can either spline interpolate the observed redshift distribution of data to sample redshifts for the random catalogue, or, randomly shuffle redshifts from the data and assign them to the randoms. Both methods show similar clustering measurements based on studies of the BOSS DR9 mocks, and results from the `shuffled' method are slightly closer to the expectations \citep[][]{Ross2012}. Recent studies have shown that identical redshifts in the data and random catalogues may bias the clustering measurements, especially when the survey area is small, which can be corrected using an additional integral constraint term \citep[see][]{deMattia2019,Tamone2020,Zhao2021}.

However, sample variations across the survey volume are not trivial for voids, as the number densities of these indirect tracers do not response straightforwardly to observational effects. Besides, the distributions of voids with different radii have to be taken into consideration. In general, the relationship between the void and galaxy number densities depends on the clustering of galaxies, and it is difficult to translate directly the completeness of galaxies to that of voids \citep[][]{Forero2021}. Thus, we follow the random generation scheme described in \citet[][]{Liang2016}, which makes use of the `shuffled' method.
In this manner, multiple mock realizations are stacked, and the coordinates are shuffled in both angular and radial directions. 
Moreover, we shuffle void radii along with their redshifts, in different radius and redshift bins,
since the void size distributions are sensitive to redshifts, especially for galaxy samples with dramatic variations of radial selection functions (see Figure~\ref{fig:DTVoid_z_rad}).

In practice, for each data sample, we stack 100 realizations of mocks catalogues, and divide the combined catalogue into subsamples with different radius and redshift bins. The bin sizes for radii and redshifts are $1\,\mpc$ and 0.05 respectively, and adjacent radius bins are merged if there are not enough objects ($< 1000$). Then, we shuffle the (radius, redshift) pairs together, for objects inside each bin. Finally, the shuffled subsamples are combined again, and 20 per cent of the objects are randomly chosen to construct the random catalogue for clustering measurements.
In this case, the number of random objects are roughly 20 times the number of voids. This is a typical random-to-data ratio for 2PCF estimators \citep[e.g.][]{Vargas2018}.
Moreover, the chance of having identical redshifts between the data and random catalogues is largely reduced, and given the fact that the biases caused by pairs with identical redshifts are small for the 2PCF monopole \citep[][]{Zhao2021}, we do not apply integral constraint corrections in this work.

Due to the large angular variations of galaxy densities for the `LRG (c)' sample, depending on whether the CMASS and eBOSS LRG samples overlap, we apply the `shuffled' random generation procedure to voids in the CMASS-only, eBOSS-only, and the overlapping regions individually, and combine the three random catalogues afterwards. Similarly, random samples for ELG chunks -- angular regions in which the plate and fibre assignments are performed independently \citep[][]{Raichoor2021} -- are generated separately as well.

\subsubsection{Optimal radius selection}
\label{sec:radius_sel}

With the void catalogues and the corresponding randoms, we are able to measure BAO from the 2PCFs of voids with different radii, using the estimators described in Section~\ref{sec:cf_estimator}.
In order to maximize the cosmological gain from void BAO, we have to avoid contaminations from voids-in-clouds that actually reside in overdensity regions, and keep as many voids-in-voids as possible to reduce the statistical uncertainties. To this end, we perform a grid search of the optimal void radius selection criteria, based on the signal-to-noise ratio (SNR) of the void BAO peak from mocks.

Following \citet[][]{Liang2016}, we use a model-independent definition of the BAO peak signal $S_{\rm BAO}$, which is essentially the height of the BAO peak with respect to the left and right dips:
\begin{equation}
S_{\rm BAO} = \left|\, \xi (s_{\rm peak}) - \frac{ \xi (s_{\rm left}) + \xi (s_{\rm right}) }{ 2} \,\right| ,
\label{eq:bao_snr}
\end{equation}
where $\xi (s)$ denotes the 2PCF monopole at separation $s$, and $s_{\rm left}$, $s_{\rm peak}$, and $s_{\rm right}$ indicate the bins at the left dip, the peak, and the right dip near the BAO scale ($\sim 100\,\mpc$), respectively.
The absolute value is necessary since the BAO `peak' of the galaxy--void cross correlation function is typically negative.
We compute 2PCFs with the separation range of $s \in (0, 200)\,\mpc$ in this work, with the bin size of $5\,\mpc$. The separation ranges we choose for $s_{\rm left}$, $s_{\rm peak}$, and $s_{\rm right}$ are then $(80,90)$, $(100,105)$, and $(115,125)\,\mpc$, respectively.
Note that there are two bins for each dip, and we take the average 2PCF value of these bins for $\xi (s_{\rm left})$ and $\xi (s_{\rm right})$, as is done in \citet[][]{Liang2016}. The noise of the BAO signature is then measured as the standard deviation of $S_{\rm BAO}$, evaluated from 100 mock relizations.

It has been shown by \citet[][]{Forero2021} that voids selected by constant radius cuts provide near-optimal BAO SNR for BAO reconstructed samples. Moreover, they are not sensitive to moderate observational systematics that introduces $< 20$ per cent incompleteness.
Thus, we explore only constant minimum radii $R_{\rm v, min}$ for void selections in this work. In particular, we examine 26 $R_{\rm v, min}$ values, from $12.5$ to $25\,\mpc$, with the step size of $0.5\,\mpc$.
The void exclusion effect introduces strong anti-correlations on scales below twice the void radii, for both void--void ($\xi_{\rm vv}$) and galaxy--void ($\xi_{\rm gv}$) 2PCFs \citep[][]{Liang2016}. To prevent its contamination on the BAO peak, which is at around $100\,\mpc$ given our current understanding of the Universe \citepalias[e.g.][]{eBOSS2021}, we set a maximum radius of $R_{\rm v, max} = 40\,\mpc$ for the selections of all our void samples.

\begin{figure}
\centering
\includegraphics[width=.8\columnwidth]{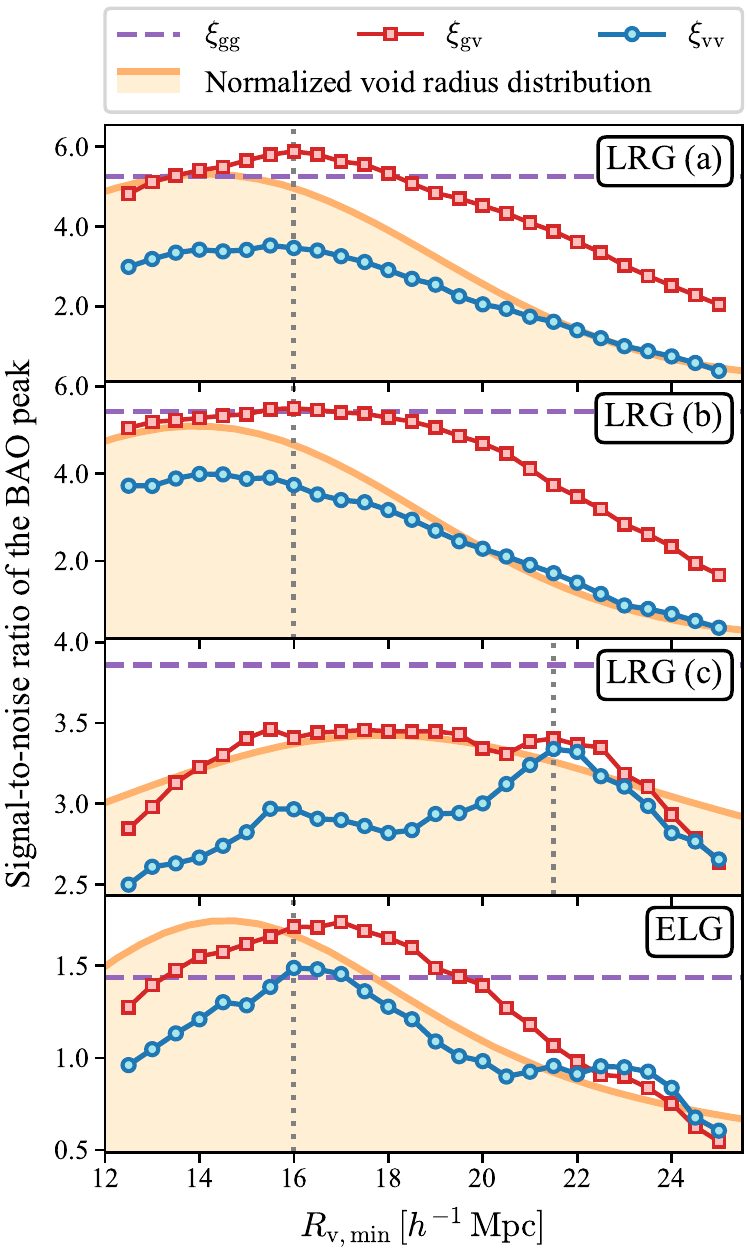}
\caption{Signal-to-noise ratio of the BAO peak, measured from 100 realizations of mock catalogues, for void--void ($\xi_{\rm vv}$), galaxy--void ($\xi_{\rm gv}$), and galaxy--galaxy ($\xi_{\rm gg}$) correlation functions. In particular, $\xi_{\rm vv}$ and $\xi_{\rm gv}$ are measured with different minimum void radii ($R_{\rm v, min}$), and the same maximum radius of $R_{\rm v, max} = 40\,\mpc$. The vertical dotted lines show the void radius cuts used in this work. Shadowed areas show normalized void radius distributions.}
\label{fig:DTVoid_SNR}
\end{figure}

The BAO SNR of $\xi_{\rm vv}$ and $\xi_{\rm gv}$ for voids with radii between $R_{\rm v, min}$ and $R_{\rm v, max}$, measured from 100 approximate mock realizations, are shown in Figure~\ref{fig:DTVoid_SNR}, together with the void radius distributions for different samples.
The optimal radius cuts for $\xi_{\rm vv}$ and $\xi_{\rm gv}$ are in general consistent, and are slightly larger than the most probable void radii, which are at the peaks of the void radius distributions.
Hence, we choose a single radius cut for each sample, which yields near-optimal BAO SNR for both $\xi_{\rm vv}$ and $\xi_{\rm gv}$.
It is worth noting that we use directly the radius cuts obtained from the `LRG (c)' and `ELG' samples separately, for their cross correlations (see Section~\ref{sec:cf_estimator}).
These radius cuts, as well as the corresponding void sample sizes, are listed in Table~\ref{tab:sdss_nvoid}.
It can also be seen that with the minimum radius cut, we remove 60--70 per cent voids, and the number of remaining voids is about twice the number of the corresponding galaxies (see Table~\ref{tab:sdss_ngal}) for each sample.

\begin{table}
\centering
\begin{threeparttable}
\caption{Void radius selection ranges $(R_{\rm v, min}, R_{\rm v, max})$, and the number of voids with different selection criteria, for each galaxy sample. $N_{\rm void}^{\rm max}$ indicates the number of voids with radii in the range of $(0, R_{\rm v, max})$, while $N_{\rm void}$ indicates the number of voids we use for the BAO analysis in this paper, i.e., with radii $R_{\rm v} \in (R_{\rm v, min}, R_{\rm v, max})$.}
\begin{tabular}{ccccc}
\toprule
Sample & \makecell{$(R_{\rm v, min}, R_{\rm v, max})$\\$(\mpc)$} & \makecell{Galactic\\cap} & $N_{\rm void}^{\rm raw}$ & $N_{\rm void}$ \\
\midrule
\multirowcell{3}{LRG (a)} & \multirowcell{3}{$(16, 40)$} & NGC & 2639522 & 902637 \\
& & SGC & 1047500 & 317737 \\
& & \textbf{Total} & \textbf{3687022} & \textbf{1220374} \\
\midrule
\multirowcell{3}{LRG (b)} & \multirowcell{3}{$(16, 40)$} & NGC & 3107491 & 995763 \\
& & SGC & 1130310 & 359202 \\
& & \textbf{Total} & \textbf{4237801} & \textbf{1354965} \\
\midrule
\multirowcell{3}{LRG (c)} & \multirowcell{3}{$(21.5, 40)$} & NGC & 1485650 & 551265 \\
& & SGC & 697680 & 243530 \\
& & \textbf{Total} & \textbf{2183330} & \textbf{794795} \\
\midrule
\multirowcell{3}{ELG} & \multirowcell{3}{$(16, 40)$} & NGC & 370356 & 172208 \\
& & SGC & 362127 & 149988 \\
& & \textbf{Total} & \textbf{732483} & \textbf{322196} \\
\bottomrule
\end{tabular}
\label{tab:sdss_nvoid}
\end{threeparttable}
\end{table}

Furthermore, the optimal BAO SNR of $\xi_{\rm gv}$ can be higher than that of the galaxy auto 2PCFs ($\xi_{\rm gg}$), which are shown as horizontal dashed lines in Figure~\ref{fig:DTVoid_SNR}.
This is consistent with the finding in \citet[][]{Forero2021}, and suggests that $\xi_{\rm gv}$ may yield better BAO constraints than $\xi_{\rm gg}$.
Thus, the galaxy--void cross correlation is a promising probe of cosmological parameters, considering the fact that it is also less sensitive to observational systematics than galaxy auto correlations \citep[][]{Forero2021}.

Note that the radius selection is the only degree of freedom for the construction of our void sample.
Even if the SNR estimated using Eq.\eqref{eq:bao_snr} does not represent faithfully the uncertainty of BAO scale determination, there should be no bias for the BAO measurements from voids, though the statistical errors may not be minimized.
In fact, we have checked the BAO measurements from the mean 2PCF of mocks with different $R_{\rm v, min}$ values that are close to our fiducial choice, and no significant bias is found.
It is also important to stress that apart from the redshift to distance conversion process, which is unavoidable for 3D clustering analysis, the whole void sample construction pipeline is model independent.

\subsection{Correlation function estimator}
\label{sec:cf_estimator}

Two-point correlation function quantifies the spatial distribution of objects on different scales. In practice, they are typically estimated through pair counts, which are histograms of tracer pairs binned by pair separations. Here, the tracers can be galaxies, voids, and randoms. In this work, all the pair counts are evaluated in three-dimensional comoving space, using the Fast Correlation Function Calculator\footnote{\url{https://github.com/cheng-zhao/FCFC}} (\textsc{fcfc}; Zhao in preparation). Besides, we consider only the monopole 2PCFs, i.e. 2PCFs in isotropic separation bins.
%, as reconstruction should have removed anisotropies of clustering on BAO scales.

Let $\widehat{\rm XY} (s)$ be the weighted number of tracer pairs with separation bin $s$, where members of the pairs are from catalogues $X$ and $Y$ respectively, the normalized pair count is given by
\begin{equation}
{\rm XY}(s) = \widehat{\rm XY}(s) / ( n_{_{\rm X}} n_{_{\rm Y}} ) .
\label{eq:pair_norm}
\end{equation}
Here, $n_{_{\rm X}}$ and $n_{_{\rm Y}}$ indicate the total weighted number of objects in $X$ and $Y$, respectively. Note that the denominator of this equation should be replaced by $n_{_{\rm X}} (n_{_{\rm X}} - 1)$ for auto pair counts, or the case of $X = Y$. But when $n_{_{\rm X}} \gg 1$, Eq.~\eqref{eq:pair_norm} holds for auto pair counts as well. We shall use this approximation throughout this work, as our numbers of tracers are sufficiently large (see Tables~\ref{tab:sdss_ngal} and \ref{tab:sdss_nvoid}).
In particular, we apply the weight expressed by Eq.~\eqref{eq:weight_all} to galaxies, but no weight (or a weight of 1) to voids, for the pair counting and normalization processes.

The 2PCF of a sample given the data ($D$) and random ($R$) catalogues can then be computed using the Landy--Szalay (LS) estimator \citep[][]{Landy1993}:
\begin{equation}
\xi = \frac{ {\rm DD} - 2 {\rm DR} + {\rm RR}}{{\rm RR}} .
\label{eq:xi_vv}
\end{equation}
Here, we have dropped the separation $s$ for simplicity.
This estimator applies to our void auto correlation functions. But for a reconstructed galaxy sample, due to the existence of the shifted random catalogue ($S$), which is generated by moving random objects with the displacement field for reversing galaxy motions, the 2PCF estimator is slightly different \citep[][]{Padmanabhan2012}:
\begin{equation}
\xi_{\rm gg} = \frac{{\rm DD} - 2 {\rm DS} + {\rm SS}}{{\rm RR}} ,
\label{eq:xi_gg}
\end{equation}
where $D$ denotes the reconstructed galaxy catalogue, and $R$ indicates the unshifted random sample.
%Actually, Eq.~\eqref{eq:xi_vv} can be seen as a special form of Eq.~\eqref{eq:xi_gg}, under the condition of $S = R$.
This equation is further generalised following the spirit of \citet[][]{Szapudi1997}, for the estimator of the galaxy--void cross correlation:
\begin{equation}
\xi_{\rm gv} = \frac{{\rm D_g D_v - D_g R_v - S_g D_v + S_g R_v}}{{\rm R_g R_v}} .
\label{eq:xi_gv}
\end{equation}
Here, the subscripts `g' and `v' indicate catalogues for galaxies and voids respectively. Furthermore, cross correlations between `LRG (c)' and `ELG' tracers are estimated using similar formulae.

Usually, we need to combine the 2PCFs of different datasets, such as those for NGC and SGC. This can be done by weighting 2PCFs with the volumes of the galactic caps (or equivalently areas as the redshift ranges are identical), as they are generally too far away from each other to have cross correlations on scales of interest.
However, we combine different samples at the pair count level in this work, as the cross correlations can be considered for tracers residing in the same volume, with optionally additional weights applied to the samples.
This is crucial for the combination of galaxy and void correlations, as is done in \citet[][]{Zhao2020}.

To combine two datasets -- indicated by subscripts `1' and `2' respectively, and both contain a data and a random catalogue -- with constant weights, we need only adding weights to one of them, as one can always renormalize weights applied to the two datasets simultaneously, without altering the clustering. Assume that a constant weight $w$ is to be applied to the second dataset, given Eq.~\eqref{eq:pair_norm}, the combined normalized data--data pair count can be expressed as
\begin{equation}
{\rm DD_{comb}} = \frac{ n_1^2 \cdot {\rm D_1 D_1} + w n_1 n_2 \cdot {\rm D_1 D_2} + w^2 n_2^2 \cdot {\rm D_2 D_2} }{ (n_1 + w n_2)^2 } ,
\end{equation}
where $n_1$ and $n_2$ are the number of tracers in the data catalogues of the two datasets, respectively. Note however that $w$ is generally not the correct weight for the random catalogue, since the ratios of the number of randoms to that of the data can be different for the two datasets. Thus, we introduce an additional weight for randoms of the second dataset, which corrects for the random-to-data ratio:
\begin{equation}
w_\alpha = \frac{ n_{\rm r1} / n_1 }{ n_{\rm r2} / n_2 } ,
\end{equation}
where $n_{\rm r1}$ and $n_{\rm r2}$ denote that number of random objects of the two datasets. Consequently, the total weight to be applied to the random catalogue of the second dataset is
\begin{equation}
w_{\rm r} = w \cdot w_\alpha
= \frac{ n_2 n_{\rm r1} }{ n_1 n_{\rm r2} } w ,
\end{equation}
and the combined random-random and data-random pair counts are then
\begin{equation}
{\rm RR_{comb}} = \frac{ n_{\rm r1}^2 \cdot {\rm R_1 R_1} + w_{\rm r} n_{\rm r1} n_{\rm r2} \cdot {\rm R_1 R_2} + w_{\rm r}^2 n_{\rm r2}^2 \cdot {\rm R_2 R_2} }{ (n_{\rm r1} + w_{\rm r} n_{\rm r2})^2 } ,
\end{equation}
\begin{equation}
\begin{aligned}
{\rm DR_{comb}} &= \frac{ n_1 n_{\rm r1} \cdot {\rm D_1 R_1} + w w_{\rm r} n_2 n_{\rm r2} \cdot {\rm D_2 R_2} }{ (n_1 + w n_2) (n_{\rm r1} + w_{\rm r} n_{\rm r2})} \\
&\; + \frac{ w_{\rm r} n_1 n_{\rm r2} \cdot {\rm D_1 R_2} + w n_{\rm r1} n_2 \cdot {\rm R_1 D_2} }{(n_1 + w n_2) (n_{\rm r1} + w_{\rm r} n_{\rm r2})} .
\end{aligned}
\end{equation}

The formulae of the combined pair counts are general, hence one can simply replace ${\rm D_1}$, ${\rm R_1}$, ${\rm D_2}$, and ${\rm R_2}$ by the data and random catalogues for galaxies and voids, including the shifted randoms, to estimate the combined pair counts for terms in Eqs~\eqref{eq:xi_vv}--\eqref{eq:xi_gv}.
In particular, the NGC and SGC pair counts can be combined with $w = 1$, in which case the cross pair counts between the two datasets are eliminated.
Note that correlation functions we use throughout this work are always with NGC and SGC combined, apart from those for consistency examinations in Appendix~\ref{sec:bao_caps}, .

\begin{figure*}
\centering
\includegraphics[width=.95\textwidth]{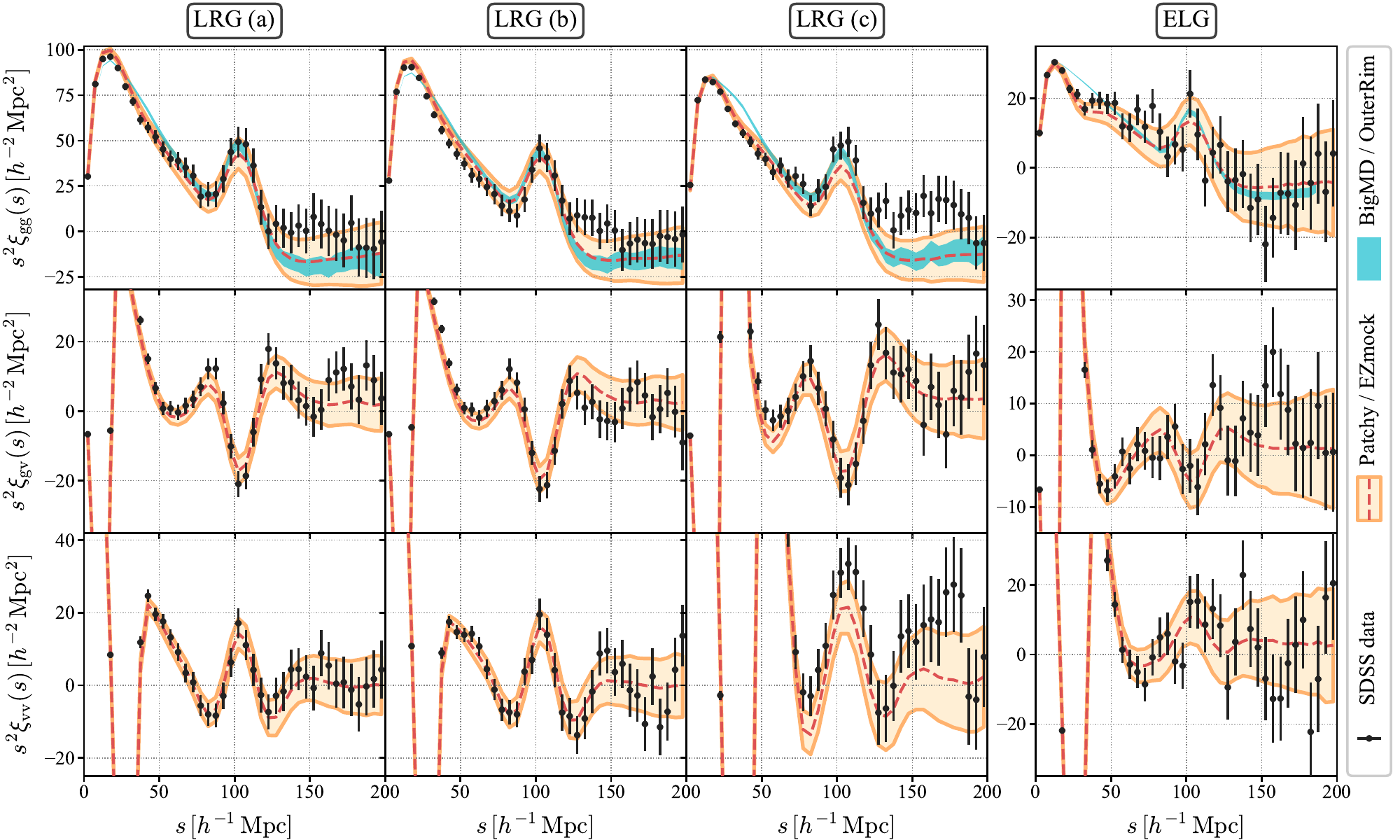}
\caption{galaxy--galaxy, galaxy--void, and void--void two-point correlation functions for different samples, with northern and southern galactic caps combined. Dots indicate measurements from the SDSS data, with error bars being the standard deviation of measurements from 1000 realizations of the corresponding approximate mocks (Patchy or EZmock). Red dashed lines and orange envelopes show the mean and 1\,$\sigma$ dispersions of 2PCFs from these mocks. Cyan regions denote jackknife error estimations, for $N$-body simulation galaxy catalogues, including BigMD and OuterRim, which are calibrated with the corresponding data. In particular, the 2PCF of the OuterRim simulation is shifted by $\alpha = 0.942$, to account for the difference of cosmology models (see Table~\ref{tab:cosmology}).}
\label{fig:xi_main}
\end{figure*}

\begin{figure}
\centering
\includegraphics[width=.95\columnwidth]{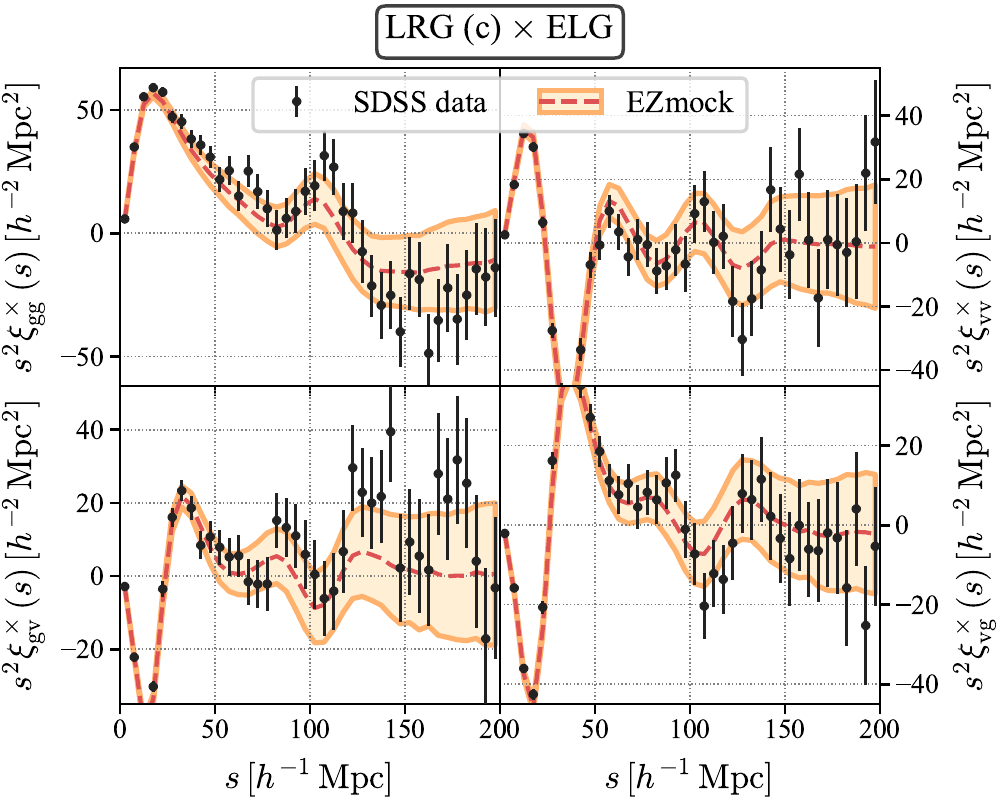}
\caption{Two-point cross correlation functions between tracers of the `LRG (c)' and `ELG' samples, including the ones for LRGs and ELGs ($\xi_{\rm gg}^\times$), LRG voids and ELG voids ($\xi_{\rm vv}^\times$), LRGs and ELG voids ($\xi_{\rm gv}^\times$), as well as LRG voids and ELGs ($\xi_{\rm vg}^\times$), with northern and southern galactics combined. Red dashed lines and orange envelopes show the mean and standard deviations measured from 1000 realizations of EZmock catalogues, and dots indicate measurements from the SDSS data, with error bars being the dispersions of mocks.}
\label{fig:xi_cross}
\end{figure}

The resulting galaxy--galaxy, galaxy--void, and void--void correlation functions are shown in Figure~\ref{fig:xi_main}.
Measurements from the SDSS data and the corresponding approximate mocks are generally consistent on BAO scales, with deviations smaller than 1\,$\sigma$ for most cases, though the BAO peak of the approximate mocks are typically not as sharp as that of the $N$-body simulations. Here, errors of the $N$-body simulation catalogues are estimated using the jackknife resampling method \citep[e.g.][]{Norberg2009}.
The discrepancies between data and mocks at $s \sim 150\,\mpc$ are possibly due to uncorrected observational systematics, but they do not bias BAO measurements provided models accounting for the broad-band amplitudes \citep[][]{Ross2017}.
The cross correlations between tracers of the `LRG (c)' and `ELG' samples are shown in Figure~\ref{fig:xi_cross}.
Again, the agreements between data and mocks are reasonably well on the BAO scale, and the BAO peaks of cross correlations between all different tracers are prominent.

\begin{figure}
\centering
\includegraphics[width=.95\columnwidth]{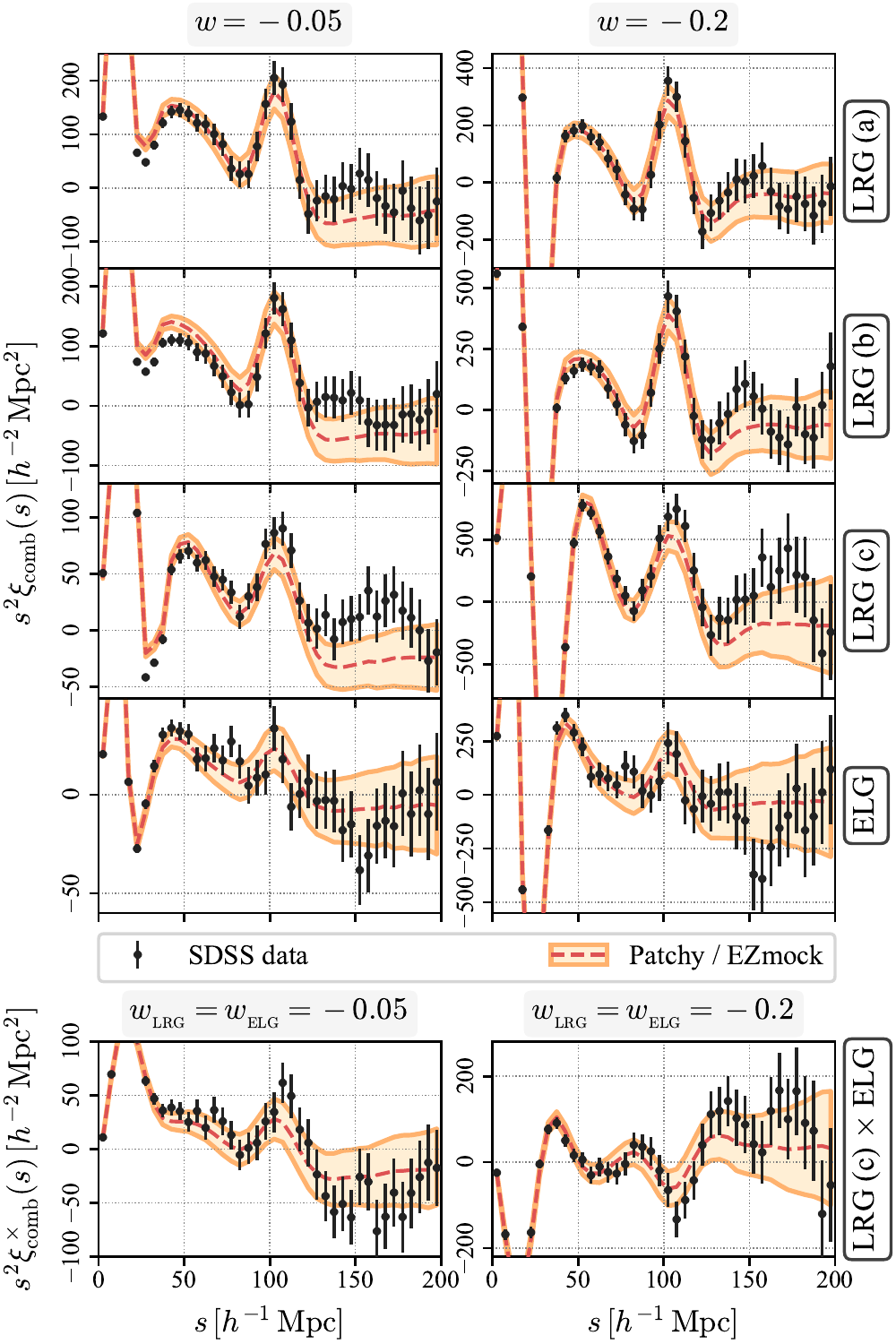}
\caption{Correlation functions evaluated by combining galaxy--galaxy, galaxy--void, and void--void pair counts, with constant weight applied to voids. They are equivalent to correlation functions of a joint galaxy and void sample, with the same weight for all voids. The left and right columns show results with the void weight of $-0.05$ and $-0.2$, respectively. Dots denotes measurements from the SDSS data, and redshift lines indicate mean results from 1000 realizations of EZmock catalogues. The standard deviations of these mocks, are shown as orange envelopes, as well as error bars on the data points.}
\label{fig:xi_weight}
\end{figure}

We then apply different negative weights to the void samples, and combine the galaxy--galaxy, galaxy--void, and void--void pair counts. This is because a negative weight enhances the BAO significance of the combined galaxy and void sample \citep[][]{Zhao2020}. The results for the weight values of $-0.05$ and $-0.2$ are shown in Figure~\ref{fig:xi_weight}.
Note that with a void weight of 0, the combined correlation function is essentially the galaxy--galaxy 2PCF.
The contribution of voids to the combined correlation function becomes larger when the absolute value of the weight increase.
The combined correlation functions we mention hereafter, always refer to the results for the combined galaxy and void samples, unless otherwise stated.

\subsection{BAO fitting}
\label{sec:bao_fit}

In order to measure the BAO peak positions from 2PCFs, we rely on the template fitting method introduced by \citet[][]{Xu2012}, and adapted for voids by \citet[][]{Zhao2020} and \citet[][]{Variu2021}.
Since the clustering measurements of galaxies and voids are strongly correlated (see Appendix~\ref{sec:cov_xi}), it is crucial to use a multi-tracer BAO fitting scheme that takes into account the cross covariances.
To this end, we introduce two multi-tracer approaches:
\begin{enumerate*}[label=(\alph*), itemjoin={{, }}, itemjoin*={{, and }}, labelwidth=0pt]
\item
fit the combined 2PCFs with weights applied to different tracers, as is done in \citet[][]{Zhao2020}
\item
fit the stacked 2PCFs of multiple tracers, with all their cross covariances included.
\end{enumerate*}

\subsubsection{BAO models}
\label{sec:bao_model}

The theoretical model we use for BAO fitting is based on a template correlation function $\xi_{\rm t} (s)$, and \citep[][]{Xu2012}
\begin{equation}
\xi_{\rm model} (s) = B^2 \xi_{\rm t} (\alpha s) + A(s),
\label{eq:xi_model}
\end{equation}
where $B$ is a normalization factor that controls the overall amplitude of the model, and $A(s)$ indicates a polynomial that accounts for the broad-band shape, which consists of three nuisance parameters $a_0$, $a_1$, and $a_2$:
\begin{equation}
A(s) = a_0 \, s^{-2} + a_1 \, s^{-1} + a_2 .
\label{eq:bao_poly}
\end{equation}
It has been shown that this polynomial term yields unbiased BAO measurements \citep[e.g.][]{Xu2012,Vargas2014}.
Lastly, $\alpha$ is the BAO dilation parameter, which quantifies the horizontal shift of the model curve, and is essentially the measurement of the BAO peak position.
Since $\alpha$ represents the relative difference between the model and template, it can be converted to a ratio of distance scales \citep[][]{Xu2012}, i.e.,
\begin{equation}
\alpha = \frac{ D_{_{\rm V}} (z) / r_{\rm d} }{ D_{\rm _V, fid} / r_{\rm d, fid}} ,
\label{eq:bao_alpha}
\end{equation}
where $r_{\rm d}$ is the sound horizon at the drag epoch, and $D_{_{\rm V}}$ indicates the volume-averaged angular diameter distance \citep[][]{Eisenstein2005}:
\begin{equation}
D_{_{\rm V}} (z) = \left[z D_{_{\rm M}}^2 (z) D_{_{\rm H}}(z) \right]^{1/3} = \left[ D_{_{\rm M}}^2 (z) \frac{ {\rm c} z}{H(z)} \right]^{1/3},
\label{eq:bao_dv}
\end{equation}
with $D_{_{\rm M}} (z)$ and $D_{_{\rm H}}(z)$ being the angular diameter distance and the Hubble distance respectively. Besides, $H(z)$ is the Hubble parameter.
The subscript `fid' in Eq.~\eqref{eq:bao_alpha} indicates parameters of the cosmology model used for generating the template $\xi_{\rm t} (s)$.

This template correlation function is actually the Hankel transform of a template power spectrum $P_{\rm t} (k)$:
\begin{equation}
\xi_{\rm t} (s) = \int \frac{k^2\,{\rm d} k}{2 {\uppi}^2} \, P_{\rm t} (k) j_0 (k s) \, {\rm e}^{-k^2 a^2} ,
\label{eq:xi_integral}
\end{equation}
where $j_0$ is the 0-order spherical Bessel function of the first kind, and $a$ is a parameter for high-$k$ damping, which reduces numerical instability of the integration. The value of $a$ used in this work is $2\,\mpc$, with which the BAO measurements are unbiased, and more robust against noises of the template power spectrum, compared to the results with smaller $a$ values \citep[][]{Variu2021}.

The template power spectrum for galaxy auto correlation is generated using the `de-wiggled' model \citep[][]{Xu2012}, i.e.
\begin{equation}
P_{\rm t, dw} (k) = \left[ P_{\rm lin} (k) - P_{\rm lin, nw} (k) \right] \, {\rm e}^{-k^2 \Sigma_{\rm nl}^2 / 2} + P_{\rm lin, nw} (k) ,
\label{eq:bao_model_dewiggle}
\end{equation}
where $P_{\rm lin} (k)$ is the linear matter power spectrum, $P_{\rm lin, nw}$ indicates its `non-wiggle' counterpart that is free of BAO wiggles, and $\Sigma_{\rm nl}$ denotes the BAO damping parameter. The linear matter power spectra we use for the fits, are generated using the \textsc{camb}\footnote{\url{https://camb.info/}} \citep[][]{Lewis2000} software. We use our fiducial cosmology to generate the templates for the SDSS data and approximate mocks, while for the $N$-body simulation catalogues, their corresponding cosmology models are used (see Table~\ref{tab:cosmology}). Then, $P_{\rm lin, nw} (k)$ is constructed by smoothing BAO wiggles of the corresponding $P_{\rm lin}$, using spline fits.

The `de-wiggled' BAO model, however, does not work well for galaxy--void, void--void, and combined correlation functions \citep[][]{Zhao2020, Variu2021}. This is because the void exclusion effect not only introduces high-$k$ oscillatory patterns in the void--void and galaxy--void power spectra, but also alter their broad-band shapes \citep[][]{Chan2014,Zhao2016}.
The polynomial term in Eq.~\eqref{eq:bao_poly} is not able to describe precisely the complicated shapes.
In order to account for these effects, we generalise Eq.~\eqref{eq:bao_model_dewiggle} by introducing a term $P_{\rm t, nw} (k)$, that models the broad-band power spectrum shape \citep[see][for details]{Zhao2020}:
\begin{equation}
P_{\rm t} (k) = P_{\rm t, dw} (k) \cdot \frac{ P_{\rm t, nw} (k) }{ P_{\rm lin, nw} (k) } .
\label{eq:bao_model_temp}
\end{equation}
This additional `non-wiggle' term encodes both the broad-band clustering signal, and geometric effects that are closely related to the void definition through Delaunay triangulations. Thus it is difficult to develop an accurate analytical model for this term.
To circumvent this problem, we rely on different models for the two multi-tracer analysis methods:
\begin{enumerate*}[label=(\alph*), itemjoin={{, }}, itemjoin*={{, and }}, labelwidth=0pt]
\item
fit the combined 2PCFs using a parabolic model with an additional free parameter, as is done in \citet[][]{Zhao2020}
\item
fit the stacked 2PCFs of multiple tracers using a template-based numerical model introduced by \citet[][]{Variu2021}.
\end{enumerate*}

\paragraph{Parabolic model for the combined 2PCFs}

\begin{figure}
\centering
\includegraphics[width=.95\columnwidth]{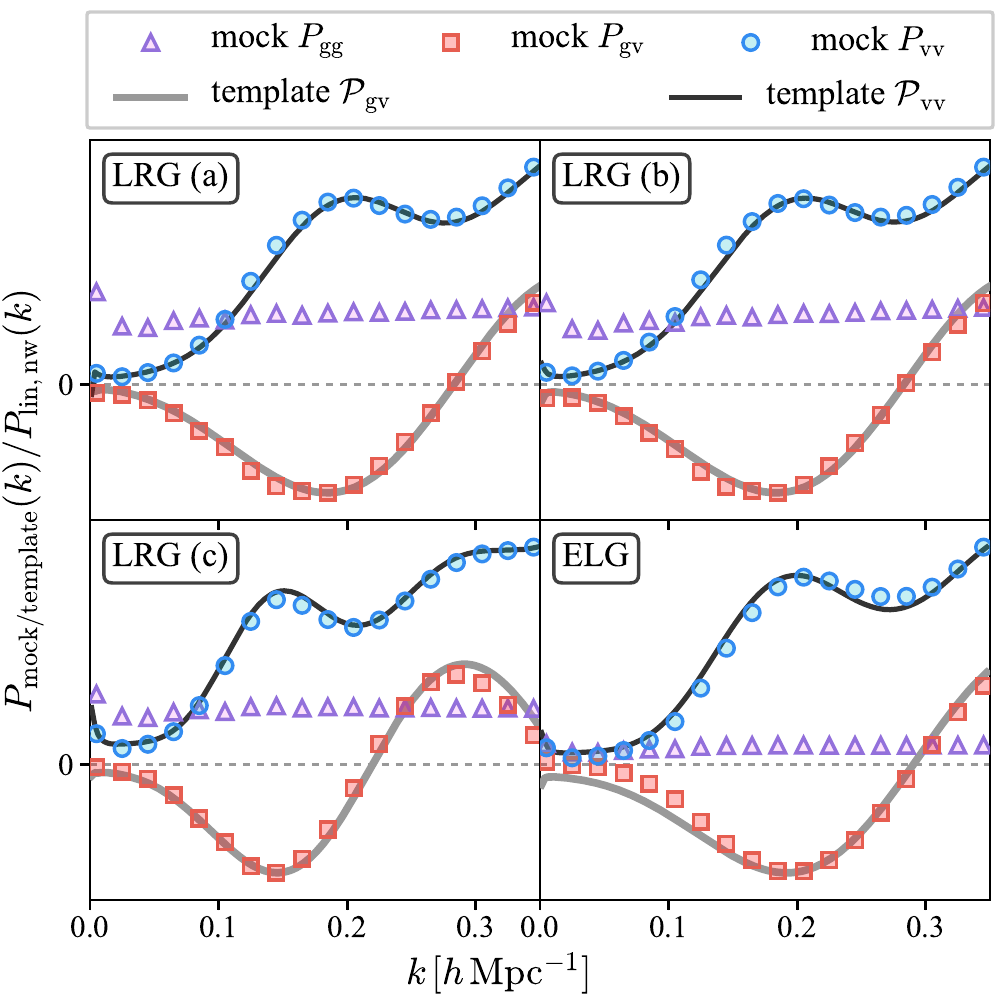}
\caption{Ratio of the mock or template power spectra, to the linear non-wiggle matter power spectrum. Open symbols indicate the mean results measured from 100 realizations of Patchy or EZmock catalogues, and black lines are obtained using the numerical power spectrum templates without BAO wiggles, which are used for the BAO fitting model in Eq.~\eqref{eq:bao_model_temp}. Values of the ratios are not shown, as the curves are normalized arbitrarily.}
\label{fig:bao_template}
\end{figure}

It has been shown by \citet[][]{Zhao2020} that the generalised BAO model in Eq.~\eqref{eq:bao_model_temp} works for the combined 2PCFs of galaxies and voids.
Since the $P_{\rm t, nw} (k)$ term depends on the weight applied to voids, in principle we have to find a formula that is able to describe well the `non-wiggle' term $P_{\rm t, nw} / P_{\rm lin, nw}$ for all the three correlations: galaxy--galaxy, galaxy--void, and void--void.
Figure~\ref{fig:bao_template} shows the measurements of $P_{\rm t} / P_{\rm lin, nw}$ from mocks for different correlations. Though the BAO wiggles are included, the broad-band shapes are the same as those of the corresponding `non-wiggle' terms \citep[see also][]{Zhao2020}.
It can be seen that the low-$k$ parts ($k \lesssim 0.15\,h\,{\rm Mpc}^{-1}$) of these measurements are consistent with parabolas.
Thus, we use the parabolic approximation of the $P_{\rm t, nw} / P_{\rm lin, nw}$ term introduced by \citet[][]{Zhao2020}:
\begin{equation}
P_{\rm t, para} (k) \approx P_{\rm t, dw} (k) \cdot ( 1 + c k^2 ).
\label{eq:bao_model_para}
\end{equation}
Here, $c$ is parameter that accounts for the shape of $P_{\rm t, nw} / P_{\rm lin, nw}$ at low $k$. Even though it does not model well the high-$k$ behaviour of measurements shown in Figure~\ref{fig:bao_template}, the model works reasonably well as the Gaussian damping term in Eq.~\eqref{eq:xi_integral} suppresses the high-$k$ discrepancies to a large extent \citep[see][and Section~\ref{sec:mock_test} for the performance of this model]{Zhao2020}.

\paragraph{Template-based numerical model for individual 2PCFs}
\label{sec:bao_model_temp}

Another way of modelling $P_{\rm t, nw} / P_{\rm lin, nw}$ is to use numerical templates that can be constructed using mocks that are free of BAO wiggles.
In particular, it is preferable to avoid model dependence during the generation of these templates, other than that of the matter power spectrum.
To this end, we construct mock galaxy catalogues upon Gaussian random fields that are generated with the `non-wiggle' power spectrum $P_{\rm lin, nw} (k)$, followed by survey geometry application, including in particular radial selection functions, for all our samples to be fitted. This is performed with the Cosmological GAussian Mock gEnerator\footnote{\url{https://github.com/cheng-zhao/CosmoGAME}} \citep[\textsc{CosmoGAME};][]{Variu2021}, which implements a simplified version of the galaxy bias model of \textsc{EZmock} \citep[][]{Chuang2015EZ,Zhao2021}.
We then construct the corresponding void catalogues following Section~\ref{sec:void_cat}, and measure both the void--void and galaxy--void power spectra. These power spectra are calibrated to those measured from the Patchy and EZmock catalogues, by tuning the galaxy bias parameters of \textsc{CosmoGAME}.
Finally, we generate 3000 realizations of Gaussian galaxy and void catalogues for each of our data sample, thanks to the high efficiency of \textsc{CosmoGAME}. The numerical templates of the $P_{\rm t, nw} (k)$ term in Eq.\eqref{eq:bao_model_temp}, denoted by $\mathcal{P}_{\rm gv}$ and $\mathcal{P}_{\rm vv}$ for galaxy--void and void--void power spectra respectively, are then generated by averaging the measurements of all the Gaussian mock realizations.
Here, the power spectra are all measured using the \textsc{powspec} code\footnote{\url{https://github.com/cheng-zhao/powspec}}.
We do not generate templates for $P_{\rm gg, nw}$, as the de-wiggled model works already well for galaxy auto correlations.

The resulting $\mathcal{P}_{\rm gv}$ and $\mathcal{P}_{\rm vv}$ templates for different samples, normalized by $P_{\rm lin, nw}$, are shown in Figure~\ref{fig:bao_template}, in comparisons with the corresponding power spectra measurements from Patchy and EZmock catalogues.
In general, these templates capture the main features of the measurements from the approximate mocks.
There is however an obvious discrepancy for the ELG void auto power spectrum at small $k$. This is not a critical problem, as small deviations of the template non-wiggle power spectra do not bias BAO measurements \citep[][]{Variu2021}. After all, residuals in simple shapes can be accounted for by the polynomial term in Eq.~\eqref{eq:bao_poly}.
Note that we do not construct templates specifically for the cross correlations between the `LRG (c)' and `ELG' samples, as the templates for the individual samples are found to be performing well (see Figure~\ref{fig:fit_range_cross}).

In principle, it is possible to measure the template power spectrum for the combined sample of galaxies and voids as well, by combining the Gaussian mocks of galaxies and voids with different void weights.
We will investigate the possibility of this approach in a future work.

\subsubsection{BAO fitting parameter inference}
\label{sec:param_infer}

To obtain constraints on the BAO fitting parameters $\boldsymbol{p}$, especially for the $\alpha$ parameter, we assume a multivariate Gaussian likelihood:
\begin{equation}
\mathcal{L} \approx {\rm e}^{ - \chi^2 (\boldsymbol{p}) / 2 } ,
\end{equation}
with the chi-squared function given by
\begin{equation}
\chi^2 (\boldsymbol{p}) = \boldsymbol{d}^{\sf T} {\bf C}^{-1} \boldsymbol{d} .
\label{eq:baofit_chi2}
\end{equation}
Here, ${\bf C}$ indicates the covariance matrix (see Appendix~\ref{sec:cov_xi}), and $\boldsymbol{d}$ denotes the difference between the data and model vectors:
\begin{equation}
\boldsymbol{d} = \boldsymbol{\xi}_{\rm data} - \boldsymbol{\xi}_{\rm model} (\boldsymbol{p}) .
\end{equation}
Furthermore, we use the reduced chi-squared, i.e., the best-fitting chi-squared value normalized by the degree of freedom (d.o.f.), for examining the goodness of the fits.

For the SDSS data and approximate mocks, the covariance matrices are estimated using the sample variances of mocks, i.e.
\begin{equation}
{\bf C}_{\rm s} = \frac{1}{N_{\rm m} - 1} {\bf M}^{\sf T} {\bf M} ,
\end{equation}
where $N_{\rm m}$ denotes the number of mock realizations, and ${\bf M}$ indicate differences between the 2PCFs of individual mocks and their mean. Elements of ${\bf M}$ are given by
\begin{equation}
{\bf M}_{i j} = \xi_i (s_j) - \bar{\xi} (s_j) ,  \quad i = 1, 2, \dots , N_{\rm m}, \quad s_j \in ( s_{\rm min}, s_{\rm max} ) ,
\label{eq:mock_matrix}
\end{equation}
where $\xi_i$ indicates the 2PCF of the $i$-th mock, $\bar{\xi}$ is the mean 2PCF of all mocks, and $( s_{\rm min}, s_{\rm max} )$ denotes the separation range used for the fit.
For $N$-body simulation catalogues, the covariance matrices are estimated using the jackknife resampling method. For $N_{\rm sub}$ subvolumes, with one being removed from the sample at each time, the covariance matrix can be estimated by \citep[][]{Norberg2009}
\begin{equation}
{\bf C}_{\rm jk} = \frac{ N_{\rm sub} - 1 }{ N_{\rm sub} } {{\bf M}'}^{\sf T} {\bf M}' ,
\end{equation}
with $M'$ computed from the 2PCFs of all $N_{\rm sub}$ subsamples, following Eq.~\eqref{eq:mock_matrix}.

Denoting the number of separation bins as $N_{\rm bin}$, the unbiased inverse covariance matrix in Eq.~\eqref{eq:baofit_chi2} is \citep[][]{Hartlap2007}
\begin{equation}
{\bf C}^{-1} = (1 - D) \, {\bf C}_{\rm s}^{-1},
\label{eq:hartlap}
\end{equation}
with the correction factor
\begin{equation}
D = \frac{N_{\rm bin} + 1}{ N_{\rm m} - 1 } .
\end{equation}
In addition, to assess the uncertainty propagated from the covariance matrix, the fitted parameter covariances are rescaled by \citep[][]{Percival2014}
\begin{equation}
m_1 = \frac{ 1 + B_0 ( N_{\rm bin} - N_{\rm par} ) }{ 1 + A_0 + B_0 (N_{\rm par} + 1)} ,
\label{eq:m1}
\end{equation}
and variances of the best-fitting parameters obtained from mocks used for ${\bf C}_{\rm s}$ estimation should be rescaled by
\begin{equation}
m_2 = ( 1 - D )^{-1} m_1 .
\label{eq:m2}
\end{equation}
Here, $N_{\rm par}$ is the number of free parameters for the fits, and the factors are
\begin{align}
A_0 &= \frac{2}{ ( N_{\rm m} - N_{\rm bin} - 1 )( N_{\rm m} - N_{\rm bin} - 4 ) } ,\\
B_0 &= \frac{N_{\rm m} - N_{\rm bin} - 2}{ ( N_{\rm m} - N_{\rm bin} - 1 )( N_{\rm m} - N_{\rm bin} - 4 ) } .
\end{align}

In practice, we use the BAO Fitter for muLtI-Tracers\footnote{\url{https://github.com/cheng-zhao/BAOflit}} (\textsc{BAOflit}) to constrain BAO fitting parameters.
It encodes all the BAO models described in Section~\ref{sec:bao_model}, and is able to fit a single $\alpha$ parameter for 2PCFs of multi-tracers, with their cross covariances taken into account.
The nuisance parameters in Eq.~\eqref{eq:bao_poly} are computed analytically using the least squares method; while for the rest of the parameters, \textsc{BAOflit} relies on the \textsc{MultiNest}\footnote{\url{https://github.com/farhanferoz/MultiNest}} tool \citep[][]{Feroz2008,Feroz2009,Feroz2019} for accurate and efficient Monte--Carlo (MC) Bayesian posterior sampling.
In fact, including the nuisance parameters in the MC sampler does not change the fitting results, but slows down the sampling process significantly due to the enlargement of the parameter space dimension \citep[see also][]{Zhao2020}.
The best-fitting (median) value and 1\,$\sigma$ confidence limits of the parameters -- defined as the 16th ($\sigma_{-}$) and 84th ($\sigma_{+}$) percentiles of the cumulative posterior distribution respectively -- are computed using the \textsc{PyMultiNest} package\footnote{\url{https://github.com/JohannesBuchner/PyMultiNest}} \citep[][]{Buchner2014}.

Moreover, \textsc{BAOflit} maximizes the numerical stability of matrix manipulations, and speeds up the chi-squared function evaluations, including the least squares calculations, using the QR decomposition technique (see Appendix~\ref{sec:chi2_optimize}).
This optimization can be used for all fitting applications that rely on covariance matrices from mocks or jackknife resampling, especially for those with nuisance parameters that can be calculated using the least squares method.

\subsection{Cosmological parameter constraints}
\label{sec:cosmo_par}

We consider in this work the standard flat-$\Lambda$CDM model, as well as its two one-parameter extensions, o$\Lambda$CDM and flat-$w$CDM, which add the $\Omega_k$ and $w$ parameters, respectively.
In particular, the Hubble parameter $H(z)$ in these models are given by
\begin{equation}
H^2 (z) = H_0^2 \left[ \Omega_{\rm m} (1+z)^3 + \Omega_{\rm k} (1+z)^2 + \Omega_{\rm de} (1+z)^{3(1+w)} \right] ,
\end{equation}
where $H_0 = H(z = 0) \equiv 100\, h$, and $\Omega_{\rm m} + \Omega_{\rm k} + \Omega_{\rm de} = 1$.
Here, $\Omega_k$ and $\Omega_{\rm de}$ denote the spatial curvature density and dark energy density at $z = 0$ respectively, and $w$ is the dark energy equation of state.
When $\Omega_k = 0$ and $w = -1$, the two extended models reduce to the flat-$\Lambda$CDM model.

We compare the parameter constraints with $D_{_{\rm V}} /r_{\rm d}$ measured in this work via the multi-tracer BAO analysis, and the corresponding BAO-only results in \citetalias[][]{eBOSS2021}.
Note that in the latter study, the anisotropic BAO measurements expressed by $D_{_{\rm M}} /r_{\rm d}$ and $D_{_{\rm H}} /r_{\rm d}$ are used whenever possible, to maximize the constraining power.
To break the degeneracies of cosmological parameters measured from BAO, we introduce also the primordial deuterium abundance \citep[][]{Cooke2018} for Big Bang Nucleosynthesis (BBN) calculations, the Planck CMB temperature and polarization data \citep[][]{Planck2020}, and the Pantheon type Ia supernovae (SNe Ia) sample \citep[][]{Scolnic2018}, and investigate different combinations of these probes.

In practice, we use a variant\footnote{\url{https://github.com/evamariam/CosmoMC_SDSS2020}} of the \textsc{CosmoMC} package\footnote{\url{http://cosmologist.info/cosmomc}} \citep[][]{Lewis2002} for cosmological fittings and model evaluations with different probes throughout this work.
For all cases we run four parallel Monte--Carlo Markov chains, with the first 30 per cent samples removed as the burn in period.
The Gelman and Rubin convergence statistics is chosen to be $R - 1 < 10^{-3}$, to have accurate confidence limits for comparing results from different BAO measurements.
The posterior distribution plots and marginalized statistics are generated using the \textsc{GetDist} tool\footnote{\url{https://github.com/cmbant/getdist}} \citep[][]{Lewis2019}.

\section{Robustness and systematic error analysis using mock catalogues}
\label{sec:mock_test}

In order to investigate the robustness of our analysis, and assess systematic uncertainties, we perform extensive tests of the multi-tracer BAO constraint methodology, using the large sets of approximate mocks.
As stated before, we apply two multi-tracer BAO fitting methods, including a simple fit to the combined correlation functions evaluated using the pair counts for different tracers, as well as a joint fit with the same $\alpha$ for all individual measurements.
We examine the performances of these two approaches using mocks, and choose the more reliable one for the SDSS data analysis.
To this end, we compare results with different fitting schemes, such as different fitting ranges and priors of parameters, to estimate potential biases of the measurements.

The fiducial BAO fitting schemes for different samples, as listed in Table~\ref{tab:fit_setting}, are chosen based on these studies.
When different clustering statistics are fitted simultaneously with the same $\alpha$ parameter, their own fitting methods are used for the corresponding segments of the data vector.
The covariance matrices we use for the studies, are all estimated from the same sets of mocks (see Appendix~\ref{sec:cov_xi}).
For simplicity, the fitted error on $\alpha$ from the mocks are often quoted as a single value: $\sigma_\alpha = (\sigma_{\alpha, +} - \sigma_{\alpha, -}) / 2$. This is because the posterior distributions of $\alpha$ from the mocks are typically highly symmetric. In this case, the median value of the posterior can be estimated as $\alpha_{\rm fit, med} \approx (\sigma_{\alpha, +} + \sigma_{\alpha, -}) / 2$.

\begin{table*}
\centering
\begin{threeparttable}
\setlength{\tabcolsep}{1.2\tabcolsep}
\caption{Fiducial BAO fitting scheme, including the theoretical model, fitting range, and parameter priors, for different samples in this work.
For the parameter priors, single numbers indicate fixed values for the fits, while ranges of numbers indicate the extents of flat priors.
$\Sigma_{\rm nl, prior}$ values in parenthesis denote best-fitting values from $N$-body simulations, while the rest of the figures are all from approximate mocks.
Optimal void weights used for the combined correlation functions are also shown, as the superscripts of $\xi_{\rm comb}$. In particular, for the cross correlations between `LRG (c)' and `ELG', the optimal weights for LRG voids and ELG voids are both $-0.1$.
Fitting schemes for the multi-tracer SDSS data analysis are highlighted in bold.}
\begin{tabular}{cScSlcccc}
\toprule
Sample & Clustering & Model & \makecell{Fitting range ($\mpc$)} & $B_{\rm prior}$ & \makecell{$\Sigma_{\rm nl, prior}$ ($\mpc$)} & \makecell{$c_{\rm prior}$ ($h^{-2}\,{\rm Mpc}^2$)} \\
\midrule
\multirowcell{4}[-2\aboverulesep]{LRG (a)} & $\xi_{\rm gg}$ & \eqref{eq:bao_model_dewiggle} & $[50, 160]$ & $[1.6, 1.8]$ & 4.9 (2.5) & -- \\
& $\xi_{\rm gv}$ & \eqref{eq:bao_model_temp} + $\mathcal{P}_{\rm gv}^{_{\rm LRG (a)}}$ & $[60, 160]$ & $[6.2,7.0]$ & 6.1 & -- \\
& $\xi_{\rm vv}$ & \eqref{eq:bao_model_temp} + $\mathcal{P}_{\rm vv}^{_{\rm LRG (a)}}$ & $[70, 160]$ & $[3.3,3.9]$ & 6.9 & -- \\
& $\boldsymbol{\xi}_{\bf comb}^{\boldsymbol{w}{\bf =-0.05}}$ & \eqref{eq:bao_model_para} & ${\bf [60, 150]}$ & ${\bf [2.6, 3.0]}$ & ${\bf [0,10]}$ & {\bf 160} \\
\midrule
\multirowcell{4}[-1.5\aboverulesep]{LRG (b)} & $\xi_{\rm gg}$ & \eqref{eq:bao_model_dewiggle} & $[50, 160]$ & $[1.6,1.8]$ & 4.7 (4.0) & -- \\
& $\xi_{\rm gv}$ & \eqref{eq:bao_model_temp} + $\mathcal{P}_{\rm gv}^{_{\rm LRG (b)}}$ & $[60, 160]$ & $[6.1,7.4]$ & 6.3 & -- \\
& $\xi_{\rm vv}$ & \eqref{eq:bao_model_temp} + $\mathcal{P}_{\rm vv}^{_{\rm LRG (b)}}$ & $[70, 160]$ & $[3.3,4.3]$ & 7.1 & -- \\
& $\boldsymbol{\xi}_{\bf comb}^{\boldsymbol{w}{\bf =-0.05}}$ & \eqref{eq:bao_model_para} & ${\bf [60, 150]}$ & ${\bf [2.5, 2.9]}$ & ${\bf [0,9.4]}$ & {\bf 150} \\
\midrule
\multirowcell{4}[-1.5\aboverulesep]{LRG (c)} & $\xi_{\rm gg}$ & \eqref{eq:bao_model_dewiggle} & $[50, 160]$ & $[1.5,1.7]$ & 4.8 (4.0) & -- \\
& $\xi_{\rm gv}$ & \eqref{eq:bao_model_temp} + $\mathcal{P}_{\rm gv}^{_{\rm LRG (c)}}$ & $[60, 160]$ & $[9.5,11]$ & 9.5 & -- \\
& $\xi_{\rm vv}$ & \eqref{eq:bao_model_temp} + $\mathcal{P}_{\rm vv}^{_{\rm LRG (c)}}$ & $[70, 160]$ & $[7.9,8.5]$ & 13 & -- \\
& $\boldsymbol{\xi}_{\bf comb}^{\boldsymbol{w}{\bf =-0.1}}$ & \eqref{eq:bao_model_para} & ${\bf [60, 150]}$ & ${\bf [4.1, 4.8]}$ & ${\bf [0,11]}$ & {\bf 13} \\
\midrule
\multirowcell{4}[-1.5\aboverulesep]{ELG} & $\xi_{\rm gg}$ & \eqref{eq:bao_model_dewiggle} & $[50, 160]$ & $[0.86,0.97]$ & 3.7 (1.5) & -- \\
& $\xi_{\rm gv}$ & \eqref{eq:bao_model_temp} + $\mathcal{P}_{\rm gv}^{_{\rm ELG}}$ & $[60, 160]$ & $[4.1,5.6]$ & 6.6 & -- \\
& $\xi_{\rm vv}$ & \eqref{eq:bao_model_temp} + $\mathcal{P}_{\rm vv}^{_{\rm ELG}}$ & $[70, 160]$ & $[3.2,3.9]$ & 10 & -- \\
& $\boldsymbol{\xi}_{\bf comb}^{\boldsymbol{w}{\bf =-0.1}}$ & \eqref{eq:bao_model_para} & ${\bf [60, 150]}$ & ${\bf [1.5, 2.0]}$ & ${\bf [0,13]}$ & {\bf 70} \\
\midrule
\multirowcell{5}[-1.5\aboverulesep]{LRG (c) $\times$ ELG} & $\xi_{\rm gg}^\times$ & \eqref{eq:bao_model_dewiggle} & $[50, 160]$ & $[1.1,1.4]$ & 6.0 & -- \\
& $\xi_{\rm gv}^\times$ & \eqref{eq:bao_model_temp} + $\mathcal{P}_{\rm gv}^{_{\rm ELG}}$ & $[60, 160]$ & $[5.5,7.0]$ & 7.9 & -- \\
& $\xi_{\rm vg}^\times$ & \eqref{eq:bao_model_temp} + $\mathcal{P}_{\rm gv}^{_{\rm LRG (c)}}$ & $[60, 160]$ & $[5.4,6.3]$ & 7.7 & -- \\
& $\xi_{\rm vv}^\times$ & \eqref{eq:bao_model_temp} + $\mathcal{P}_{\rm vv}^{_{\rm LRG (c)}}$ & $[70, 160]$ & $[3.0,3.6]$ & 7.9 & -- \\
& $\boldsymbol{\xi}_{\bf comb}^{\boldsymbol{\times}, \boldsymbol{w}{\bf =\{-0.1,-0.1\}}}$ & \eqref{eq:bao_model_para} & ${\bf [60, 150]}$ & ${\bf [1.6, 2.1]}$ & ${\bf [0,14]}$ & {\bf 320} \\
\bottomrule
\end{tabular}
\label{tab:fit_setting}
\end{threeparttable}
\end{table*}

\subsection{Optimal void weight}
\label{sec:optimal_weight}

The negative BAO `peak' of the galaxy--void cross correlation (see Figures~\ref{fig:xi_main} and \ref{fig:xi_cross}) suggests that galaxies and voids are actually anti-correlated on the BAO scale.
Therefore, the BAO signal will be diluted if combining galaxies and voids by merging the catalogues directly.
Instead, a negative weight can be applied to one of the tracers, to maximize the BAO significance of the combined sample \citep[][]{Zhao2020}.
Actually, weights are generally needed when combining multiple tracers with different biases, to optimize the clustering measurements \citep[see e.g.][]{Percival2004}.
To this end, we measure the combined correlation functions with different void weights for pair count combinations (see Section~\ref{sec:cf_estimator}), and explore the optimal weight that minimizes the fitted error of $\alpha$.
In particular, we focus on negative void weights, and examine weights from 0 down to $-0.2$, with a step size of $0.01$.
This range has already been shown to be sufficiently wide for the optimal $w$ value determination with BOSS LRGs \citep[cf.][for the BOSS results with larger $w$ ranges]{Zhao2020, Forero2021}.

\begin{figure}
\centering
\includegraphics[width=.98\columnwidth]{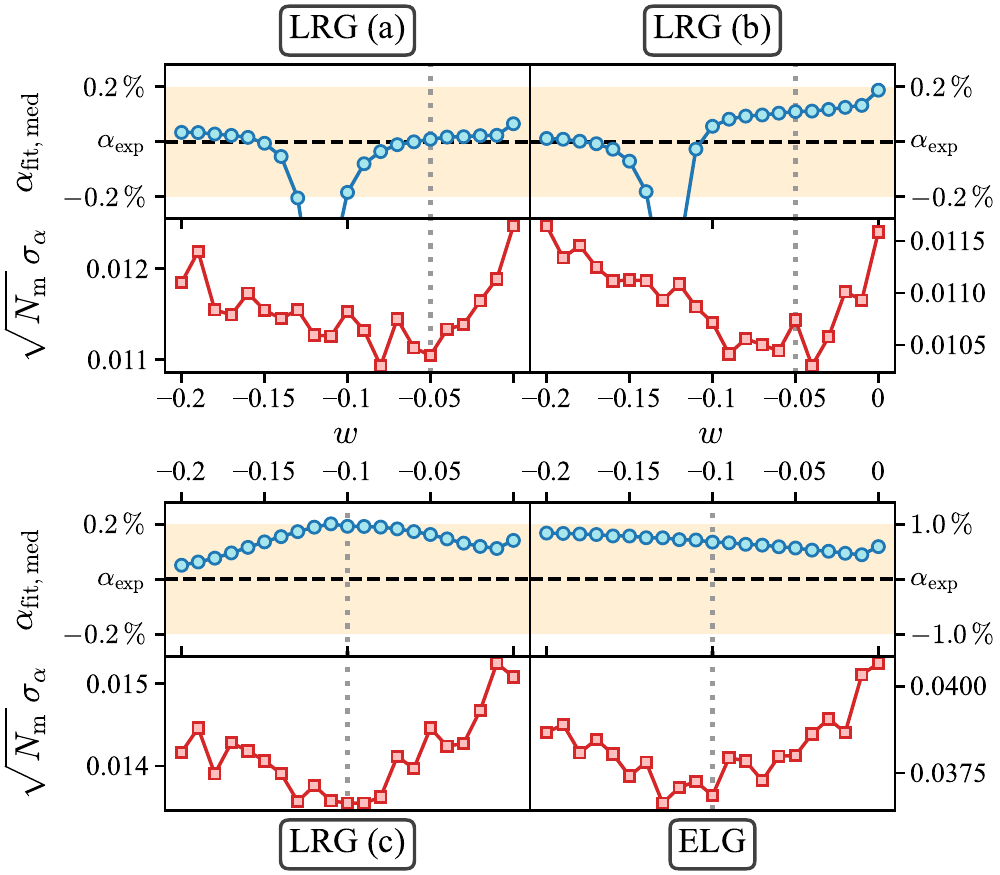}
\caption{The best-fitting (median) value and 1\,$\sigma$ error of $\alpha$, obtained from fits to the mean combined correlation functions of all mocks, evaluated with different void weights, for each sample. The best-fitting values are shown as relative differences to $\alpha_{\rm exp}$, which indicates the expected $\alpha$ value given the cosmology models of the mocks. $N_{\rm m} = 1000$ denotes the total number of mock realizations. Vertical dotted lines indicate the void weights for different samples used in this work.}
\label{fig:optimal_weight}
\end{figure}

We then fit the mean combined correlation functions of all 1000 mock realizations with the fitting range of $[60,150]\,\mpc$, which is chosen based on Appendix~\ref{sec:result_fit_range}.
Here, the covariances are rescaled by $1/1000$, to allow for tight constraints on all fitting parameters, which are used later for the prior investigations in Appendix~\ref{sec:result_fit_prior}.
The resulting median values of $\alpha$ from the posterior distributions, and the corresponding fitted errors, are shown in Figure~\ref{fig:optimal_weight}.
The optimal void weights we choose for the `LRG (a)' and `LRG (b)' samples are both $-0.05$, which is consistent with the results in \citet[][]{Zhao2020}, while for `LRG (c)' and `ELG' the chosen values are $-0.1$.
With these weights, the best-fitting $\alpha$ values of the combined correlation functions are consistent with those of the corresponding galaxy auto correlations (i.e., the combined 2PCF with $w = 0$), while the fitted errors are significantly reduced.
The difference between the measured $\alpha$ and the expected value is the largest for the `ELG' sample, possibly because of the observational systematics applied to the mocks \citep[][]{Raichoor2021, Zhao2021}.

\begin{figure}
\centering
\includegraphics[width=.98\columnwidth]{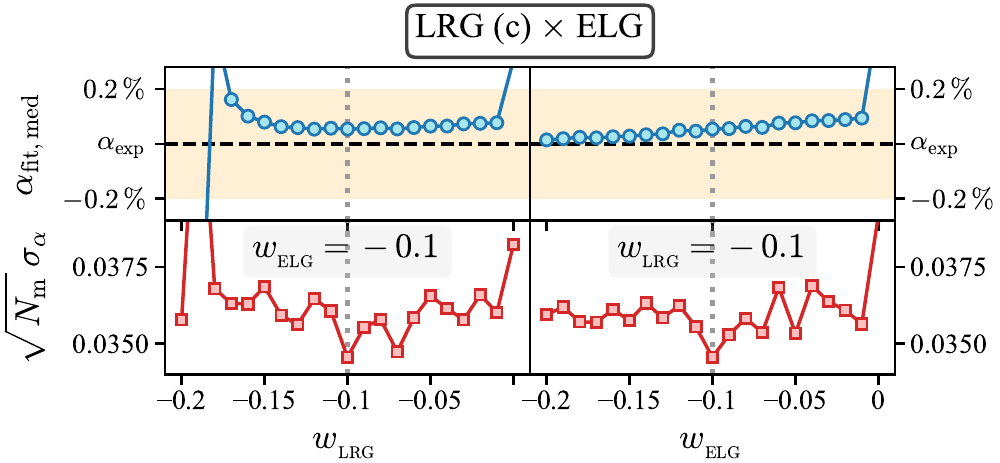}
\caption{Same as Figure~\ref{fig:optimal_weight}, but for the cross correlation between the `LRG (c)' and `ELG' samples. The left panel shows results with a fixed ELG void weight of $-0.1$, when varying the weight for LRG voids. In contrast, the right panel is for different ELG void weights, with a fixed LEG void weight of $-0.1$.}
\label{fig:optimal_cross_weight}
\end{figure}

For the cross correlations between the `LRG (c)' and `ELG' sample, in principle the optimal weights of the LRG voids ($w_{_{\rm LRG}}$) and ELG voids ($w_{_{\rm ELG}}$) can be different.
For simplicity, we vary in each time the weight for one void sample only, and fix the other void weight to $-0.1$, which is the optimal value for both individual samples.
The fitting results are shown in Figure~\ref{fig:optimal_cross_weight}, where the fitted error is minimized when both void weights are $-0.1$.
The bias of $\alpha$ is smaller than that of the `ELG' sample. This can be explained by the fact that the cross correlation suffers less observational systematic effects \citep[e.g.][]{Wang2020,Zhao2021}.
The combined correlation functions discussed hereafter, are always the ones with the optimal void weights shown in Table~\ref{tab:fit_setting}, unless otherwise stated.

\subsection{Fitting ranges}
\label{sec:fit_range}

The choice of the fitting range depends on scales that are useful for extracting the BAO peak position, and how well the correlation functions on different scales are reproduced by the templates and broad-band terms of our theoretical models.
To understand how they affect our BAO measurements, we run BAO fits with different fitting ranges, to the mean 2PCFs computed using 1000 realizations of mocks, for all the samples.
Again, the covariance matrices are divided by 1000.
The BAO measurements with different fitting ranges for individual clustering measurements, as well as the combined 2PCFs, are presented in Appendix~\ref{sec:result_fit_range}.
Actually, our fiducial fitting ranges are chosen based on these results.

We then fit multiple correlation functions simultaneously with the same $\alpha$, using the corresponding fiducial fitting ranges of individual clustering statistics.
In particular, since the fitting results for void--void correlations are sometimes sensitive to the fitting ranges (see Appendix~\ref{sec:result_fit_range}), including the void auto correlations ($\xi_{\rm vv}$), and the cross correlation between LRG voids and ELG voids ($\xi_{\rm vv}^\times$),
we present two sets of joint fits that differ in whether the void--void correlation is included.
The fitting results for different samples are shown in Table~\ref{tab:fit_range}, together with measurements from the combined correlation functions.
Here, the fitted errors on $\alpha$ rescaled by $\sqrt{1000}$, for approximating the precision of BAO measurements with the observational data.

\begin{table*}
\centering
\begin{threeparttable}
\setlength{\tabcolsep}{1.2\tabcolsep}
\caption{Fitting results for the mean 2PCFs of 1000 approximate mock realizations, with the covariances evaluated from the same set of mocks, but rescaled by $1/1000$. When a data vector consists of multiple clustering statistics, the 2PCFs share the same $\alpha$ parameter during the fit.
$\alpha_{\rm exp}$ denotes the expected $\alpha$ value for each sample, given the cosmology models for constructing the mocks, as well as the fiducial cosmology used for coordinate conversions and BAO fits (see Table~\ref{tab:cosmology}). $\alpha_{\rm fit}$ indicates results with the fiducial fitting ranges -- $[s_{\rm smin}, s_{\rm max}]$, listed in Table~\ref{tab:fit_setting} -- with the fitted errors rescaled by $\sqrt{1000}$.
$N_{\rm bin, fid}$ indicates the length of the data vector with the fiducial fitting range.
$\Delta_{s_{\rm min}, s_{\rm max}} \alpha_{\rm fit, med}$ denotes the differences of the fitted median values of $\alpha$, when varying $s_{\rm min}$ or $s_{\rm max}$ by $5\,\mpc$.}
\begin{tabular}{ccSlcSccc}
\toprule
Sample & $\alpha_{\rm exp}$ & Data vector & $N_{\rm bin, fid}$ & $\alpha_{\rm fit}$ & $\alpha_{\rm fit, med} - \alpha_{\rm exp}$ & $\max | \Delta_{s_{\rm min}, s_{\rm max}} \alpha_{\rm fit, med} | $ \\
\midrule
\multirowcell{5}[-3\aboverulesep]{MD-Patchy\\LRG (a)} & \multirowcell{5}[-3\aboverulesep]{$0.9993$} & $\{\boldsymbol{\xi}_{\rm gg}\}$ & 22 & $0.9999_{-0.0127}^{+0.0124}$ & $0.0005$ & $0.0004$ \\
& & $\{\boldsymbol{\xi}_{\rm gv}\}$ & 20 & $0.9995_{-0.0121}^{+0.0119}$ & $0.0002$ & $0.0002$ \\
& & $\{\boldsymbol{\xi}_{\rm gg},\boldsymbol{\xi}_{\rm gv}\}$ & 42 & $1.0003_{-0.0111}^{+0.0114}$ & $0.0010$ & $0.0003$ \\
& & $\{\boldsymbol{\xi}_{\rm gg},\boldsymbol{\xi}_{\rm gv},\boldsymbol{\xi}_{\rm vv}\}$ & 60 & $1.0007_{-0.0113}^{+0.0111}$ & $0.0014$ & $0.0007$ \\
& & $\{\boldsymbol{\xi}_{\rm comb}\}$ & 18 & $0.9994_{-0.0114}^{+0.0112}$ & $0.0001$ & $0.0001$ \\
\midrule
\multirowcell{5}[-3\aboverulesep]{MD-Patchy\\LRG (b)} & \multirowcell{5}[-3\aboverulesep]{$0.9996$} & $\{\boldsymbol{\xi}_{\rm gg}\}$ & 22 & $1.0017_{-0.0118}^{+0.0111}$ & $0.0020$ & $0.0002$ \\
& & $\{\boldsymbol{\xi}_{\rm gv}\}$ & 20 & $1.0006_{-0.0113}^{+0.0118}$ & $0.0009$ & $0.0000$ \\
& & $\{\boldsymbol{\xi}_{\rm gg},\boldsymbol{\xi}_{\rm gv}\}$ & 42 & $1.0017_{-0.0105}^{+0.0103}$ & $0.0021$ & $0.0002$ \\
& & $\{\boldsymbol{\xi}_{\rm gg},\boldsymbol{\xi}_{\rm gv},\boldsymbol{\xi}_{\rm vv}\}$ & 60 & $1.0014_{-0.0106}^{+0.0106}$ & $0.0018$ & $0.0005$ \\
& & $\{\boldsymbol{\xi}_{\rm comb}\}$ & 18 & $1.0007_{-0.0105}^{+0.0105}$ & $0.0011$ & $0.0001$ \\
\midrule
\multirowcell{5}[-3\aboverulesep]{EZmock\\LRG (c)} & \multirowcell{5}[-3\aboverulesep]{$1.0000$} & $\{\boldsymbol{\xi}_{\rm gg}\}$ & 22 & $1.0012_{-0.0151}^{+0.0154}$ & $0.0012$ & $0.0004$ \\
& & $\{\boldsymbol{\xi}_{\rm gv}\}$ & 20 & $0.9988_{-0.0169}^{+0.0168}$ & $-0.0012$ & $0.0004$ \\
& & $\{\boldsymbol{\xi}_{\rm gg},\boldsymbol{\xi}_{\rm gv}\}$ & 42 & $1.0011_{-0.0143}^{+0.0145}$ & $0.0011$ & $0.0005$ \\
& & $\{\boldsymbol{\xi}_{\rm gg},\boldsymbol{\xi}_{\rm gv},\boldsymbol{\xi}_{\rm vv}\}$ & 60 & $0.9993_{-0.0135}^{+0.0138}$ & $-0.0007$ & $0.0022$ \\
& & $\{\boldsymbol{\xi}_{\rm comb}\}$ & 18 & $1.0019_{-0.0145}^{+0.0137}$ & $0.0019$ & $0.0016$ \\
\midrule
\multirowcell{5}[-3\aboverulesep]{EZmock\\ELG} & \multirowcell{5}[-3\aboverulesep]{$1.0003$} & $\{\boldsymbol{\xi}_{\rm gg}\}$ & 22 & $1.0059_{-0.0409}^{+0.0395}$ & $0.0056$ & $0.0005$ \\
& & $\{\boldsymbol{\xi}_{\rm gv}\}$ & 20 & $1.0021_{-0.0428}^{+0.0419}$ & $0.0018$ & $0.0013$ \\
& & $\{\boldsymbol{\xi}_{\rm gg},\boldsymbol{\xi}_{\rm gv}\}$ & 42 & $1.0057_{-0.0377}^{+0.0374}$ & $0.0054$ & $0.0019$ \\
& & $\{\boldsymbol{\xi}_{\rm gg},\boldsymbol{\xi}_{\rm gv},\boldsymbol{\xi}_{\rm vv}\}$ & 60 & $1.0065_{-0.0374}^{+0.0384}$ & $0.0062$ & $0.0008$ \\
& & $\{\boldsymbol{\xi}_{\rm comb}\}$ & 18 & $1.0068_{-0.0383}^{+0.0349}$ & $0.0066$ & $0.0020$ \\
\midrule
\multirowcell{4}[-2.5\aboverulesep]{EZmock\\LRG (c) $\times$ ELG} & \multirowcell{4}[-2.5\aboverulesep]{$1.0001$} & $\{\boldsymbol{\xi}_{\rm gg}^\times\}$ & 22 & $1.0023_{-0.0375}^{+0.0385}$ & $0.0021$ & $0.0001$ \\
& & $\{\boldsymbol{\xi}_{\rm gg}^\times,\boldsymbol{\xi}_{\rm gv}^\times,\boldsymbol{\xi}_{\rm vg}^\times\}$ & 62 & $1.0022_{-0.0371}^{+0.0369}$ & $0.0021$ & $0.0013$ \\
& & $\{\boldsymbol{\xi}_{\rm gg}^\times,\boldsymbol{\xi}_{\rm gv}^\times,\boldsymbol{\xi}_{\rm vg}^\times,\boldsymbol{\xi}_{\rm vv}^\times\}$ & 80 & $1.0017_{-0.0389}^{+0.0405}$ & $0.0015$ & $0.0039$ \\
& & $\{\boldsymbol{\xi}_{\rm comb}^\times\}$ & 18 & $1.0006_{-0.0377}^{+0.0376}$ & $0.0005$ & $0.0005$ \\
\bottomrule
\end{tabular}
\label{tab:fit_range}
\end{threeparttable}
\end{table*}

In general, results from the galaxy--void cross correlations ($\xi_{\rm gv}$) and galaxy--galaxy auto correlations ($\xi_{\rm gg}$) are similar, in terms of both the median values and fitted errors.
Including galaxy--void cross correlations in the data vectors reduces the BAO measurement uncertainties, compared to the results from galaxies alone.
However, it does not necessarily help when further including the void--void correlation.
Results from the combined correlation functions are generally consistent with the ones from joint fits to $\{\boldsymbol{\xi}_{\rm gg},\boldsymbol{\xi}_{\rm gv},\boldsymbol{\xi}_{\rm vv}\}$ and $\{\boldsymbol{\xi}_{\rm gg}^\times,\boldsymbol{\xi}_{\rm gv}^\times,\boldsymbol{\xi}_{\rm vg}^\times,\boldsymbol{\xi}_{\rm vv}^\times\}$.
For all cases, no obvious biases of $\alpha$ are observed, when taking the differences between the fitted and expected $\alpha$ values of the mocks, given the uncertainties of $\alpha$ measurements.
It reveals that with the BAO fitting methods and reconstruction algorithms used in this work, we are able to recover the `true' cosmology of the mocks through both multi-tracer BAO fitting methods, despite of the observational systematics encoded in the catalogues.

In order to assess the potential biases of $\alpha$ due to the fitting range choices, we further vary either the lower or upper bound of the fitting ranges by one separation bin of the 2PCFs ($5\,\mpc$), resulting in four additional fitting range settings.
The systematic uncertainties due to the fitting ranges are then estimated as the maximum difference between the resulting $\alpha$ of these four cases, and that of the fiducial fitting range. The biases measured in this way are not significant as well, as is shown in Table~\ref{tab:fit_range}.

\subsection{Parameter priors}
\label{sec:fit_prior}

We introduce priors of the BAO fitting parameters based on the results from the mean 2PCFs of mocks. The priors are then verified by comparing the fitted errors and the dispersion of $\alpha$ fitted from individual mocks.
In particular, we measure the distribution of the pull quantity of $\alpha$ \citep[see e.g.][]{Xu2012,Bautista2021},
\begin{equation}
g(\alpha) = \frac{ \alpha - \bar{\alpha} }{\sigma_{\alpha}} ,
\label{eq:pull_alpha}
\end{equation}
and compare it to a standard normal distribution.
Here, $\alpha$ and $\sigma_\alpha$ denote the fitted median and 1\,$\sigma$ error of $\alpha$ respectively, for each mock realization. $\bar{\alpha}$ indicates the averaged best-fitting $\alpha$ over all realizations.

Throughout this work we use a flat prior of $\alpha$, in the range of $[0.8, 1.2]$. The priors of the rest of the parameters are listed in Table~\ref{tab:fit_setting}, based on investigations in Appendix~\ref{sec:result_fit_prior}.
In summary, we use always flat priors on the $B$ parameter in Eq.~\eqref{eq:xi_model}, with ranges chosen based on the pull quantities obtained from fits to individual mocks.
Since the BAO damping of approximate mocks are generally overestimated compared to the results from $N$-body simulations (see Table:~\ref{tab:fit_setting}), flat ${\Sigma_{\rm nl}}$ priors are used for the combined correlation functions, with a lower limit of 0, and upper limits estimated using approximate mocks.
Even though this type of ${\Sigma_{\rm nl}}$ prior works well for the individual clustering measurements, as well as the combined correlation function, the fitted errors are underestimated for joint fits to stacked 2PCFs (see Table~\ref{tab:fit_prior_multi}).
Therefore, we follow the traditional BAO fitting scheme \citep[e.g.][]{Xu2012, Alam2017, Raichoor2021}, and fix the $\Sigma_{\rm nl}$ parameter when fitting multiple 2PCFs simultaneously. 
%For fits to the mocks, the $\Sigma_{\rm nl}$ values are measured from fits to the mean 2PCFs of mocks.
%However, for the galaxy auto correlations of the observational data, we use the $\Sigma_{\rm nl}$ values from $N$-body simulations, as they are more accurate on the nonlinear BAO evolutions.
%These values are shown in Table~\ref{tab:fit_setting}, and the numbers from $N$-body simulations are indeed always smaller than those from the approximate mocks, indicating stronger BAO peaks with full gravity solvers.
Lastly, for the combined 2PCFs, the $c$ parameter in Eq.~\eqref{eq:bao_model_para} is always fixed, with the value measured from the mean 2PCFs of mocks.
Given these priors, the difference between the dispersion of best-fitting $\alpha$ and the median fitted error are mostly less than 5 per cent (see Tables~\ref{tab:fit_prior_weight} and \ref{tab:fit_prior_multi}).

\begin{figure}
\centering
\includegraphics[width=.98\columnwidth]{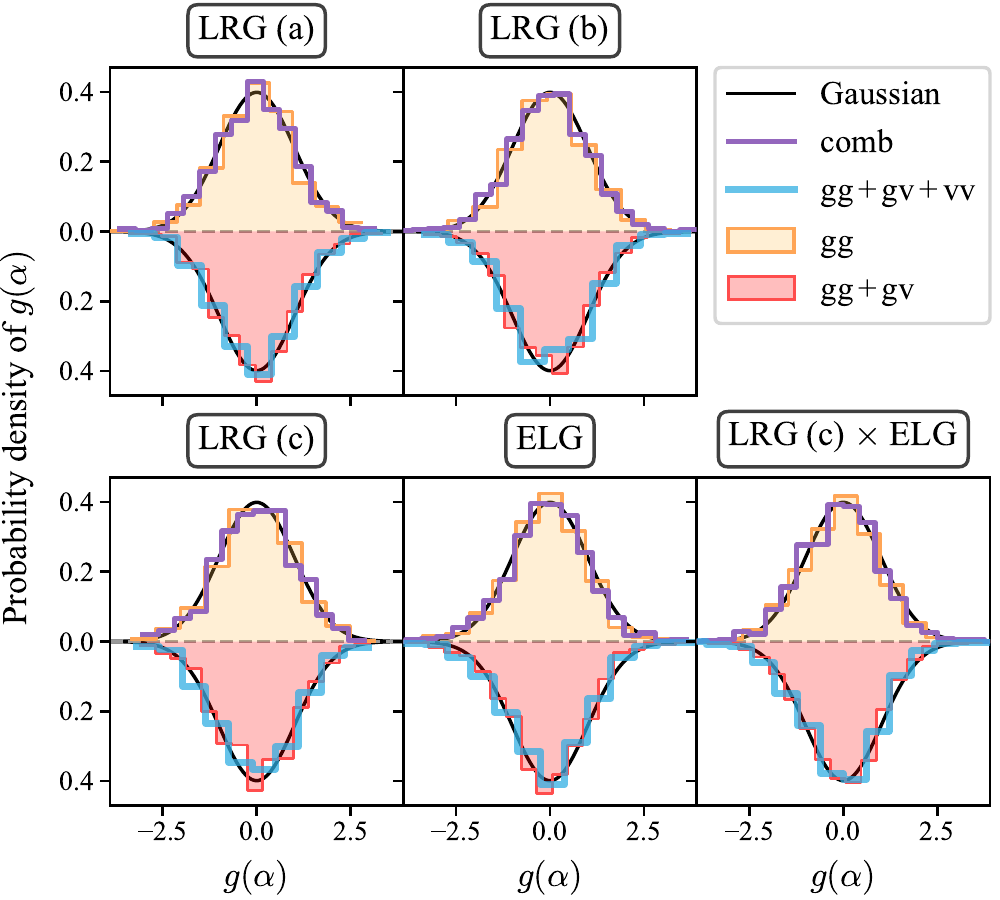}
\caption{Distributions of the pull statistic defined in Eq.~\eqref{eq:pull_alpha}, based on fits to individual mock realizations, with parameter priors given in Table~\ref{tab:fit_setting}. Black lines shown standard normal distributions, and histograms show results with different fitting data vectors. `gg', `gg\,+\,gv', and `gg\,+\,gv\,+\,vv' denote results from fits to $\{\boldsymbol{\xi}_{\rm gg}\}$, $\{\boldsymbol{\xi}_{\rm gg},\boldsymbol{\xi}_{\rm gv}\}$, and $\{\boldsymbol{\xi}_{\rm gg},\boldsymbol{\xi}_{\rm gv},\boldsymbol{\xi}_{\rm vv}\}$ respectively, for the individual samples. While for the cross correlations between `LRG (c)' and `ELG', they are for $\{\boldsymbol{\xi}_{\rm gg}^\times\}$, $\{\boldsymbol{\xi}_{\rm gg}^\times,\boldsymbol{\xi}_{\rm gv}^\times,\boldsymbol{\xi}_{\rm vg}^\times\}$, and $\{\boldsymbol{\xi}_{\rm gg}^\times,\boldsymbol{\xi}_{\rm gv}^\times,\boldsymbol{\xi}_{\rm vg}^\times,\boldsymbol{\xi}_{\rm vv}^\times\}$, respectively.
`comb' indicates results for the combined correlation functions, with optimal void weights.}
\label{fig:pull_dist}
\end{figure}

With the fiducial choices of priors, the distributions of $g(\alpha)$ for different samples are shown in Figure~\ref{fig:pull_dist}. They are generally in good agreements with a standard normal distribution.
Besides, the distributions of the fitted median values of $\alpha$ are shown in Figure~\ref{fig:alpha_dist}. The histograms are well consistent with the posterior distributions of $\alpha$ from fits to the mean 2PCFs of mocks, i.e., the ones for the results in Table~\ref{tab:fit_range}. Here, the widths of the posterior distributions are rescaled by $\sqrt{1000}$, with respect to the median values.
These results validate that our measurements with the fiducial priors are robust, including both best-fitting values and error estimates.

To assess potential systematic uncertainties due to the choices of parameter priors, we vary our fiducial priors and perform fits to individual mock realizations.
In practice, we exchange the prior types used for fitting multiple 2PCFs and the combined correlation function.
That is to say, we use flat priors of $\Sigma_{\rm nl}$ for joint fits to multiple 2PCFs, and fixed $\Sigma_{\rm nl}$ values for the combined correlation functions, with ranges and values taken from Tables~\ref{tab:fit_prior_weight} and \ref{tab:fit_prior_multi}.
In addition, to examine possible biases due to the $c$ parameter for fitting combined 2PCFs, we vary the $c$ parameter values in Table~\ref{tab:fit_setting}, by 10 per cent for the BOSS tracers and 30 per cent for the eBOSS tracers respectively, while keeping the fiducial priors of $B$ and $\Sigma_{\rm nl}$.
These fractions are roughly the relative errors on $c$, from fits to the mean 2PCFs of mocks.
The biases of the median value of best-fitting $\alpha$ from all mocks, compared to the results with the fiducial priors, are shown in Table~\ref{tab:prior_syst}.
The results suggest that our BAO measurements are not sensitive to the prior types and ranges, as the biases are small, especially for the multi-tracer approach with combined 2PCFs.

\begin{table}
\centering
\begin{threeparttable}
\setlength{\tabcolsep}{.5\tabcolsep}
\caption{Fitting results of individual mock realizations with different parameter priors. $\langle\alpha_{\rm fid}\rangle$ denotes the median value of best-fitting $\alpha$ for all mocks, with the fiducial priors in Table~\ref{tab:fit_setting}. $\Delta_{B,\Sigma_{\rm nl}} \langle\alpha\rangle$ indicates the bias of $\langle\alpha\rangle$ when using a different set of priors for $B$ and $\Sigma_{\rm nl}$, taken from Tables~\ref{tab:fit_prior} and \ref{tab:fit_prior_weight}. $\Delta_c \langle\alpha\rangle$ denotes the differences of $\langle\alpha\rangle$, when increasing or reducing the fiducial $c$ parameter value, by 10\,\% for the BOSS tracers and 30\,\% for the eBOSS tracers.}
\begin{tabular}{cSlccc}
\toprule
Sample & Data vector & $\langle\alpha_{\rm fid}\rangle$ & $|\Delta_{B,\Sigma_{\rm nl}} \langle\alpha\rangle|$ & $\max |\Delta_c \langle\alpha\rangle |$ \\
\midrule
\multirowcell{4}[-2\aboverulesep]{MD-Patchy\\LRG (a)} & $\{\boldsymbol{\xi}_{\rm gg}\}$ & $1.0000$ & $0.0006$ & -- \\
& $\{\boldsymbol{\xi}_{\rm gg},\boldsymbol{\xi}_{\rm gv}\}$ & $1.0010$ & $0.0010$ & -- \\
& $\{\boldsymbol{\xi}_{\rm gg},\boldsymbol{\xi}_{\rm gv},\boldsymbol{\xi}_{\rm vv}\}$ & $1.0008$ & $0.0007$ & -- \\
& $\{\boldsymbol{\xi}_{\rm comb}\}$ & $0.9992$ & $0.0004$ & $0.0002$ \\
\midrule
\multirowcell{4}[-2\aboverulesep]{MD-Patchy\\LRG (b)} & $\{\boldsymbol{\xi}_{\rm gg}\}$ & $1.0020$ & $0.0005$ & -- \\
& $\{\boldsymbol{\xi}_{\rm gg},\boldsymbol{\xi}_{\rm gv}\}$ & $1.0026$ & $0.0010$ & -- \\
& $\{\boldsymbol{\xi}_{\rm gg},\boldsymbol{\xi}_{\rm gv},\boldsymbol{\xi}_{\rm vv}\}$ & $1.0017$ & $0.0008$ & -- \\
& $\{\boldsymbol{\xi}_{\rm comb}\}$ & $1.0004$ & $0.0007$ & $0.0002$ \\
\midrule
\multirowcell{4}[-2\aboverulesep]{EZmock\\LRG (c)} & $\{\boldsymbol{\xi}_{\rm gg}\}$ & $1.0017$ & $0.0005$ & -- \\
& $\{\boldsymbol{\xi}_{\rm gg},\boldsymbol{\xi}_{\rm gv}\}$ & $1.0019$ & $0.0003$ & -- \\
& $\{\boldsymbol{\xi}_{\rm gg},\boldsymbol{\xi}_{\rm gv},\boldsymbol{\xi}_{\rm vv}\}$ & $1.0002$ & $0.0007$ & -- \\
& $\{\boldsymbol{\xi}_{\rm comb}\}$ & $1.0018$ & $0.0008$ & $0.0004$ \\
\midrule
\multirowcell{4}[-2\aboverulesep]{EZmock\\ELG} & $\{\boldsymbol{\xi}_{\rm gg}\}$ & $1.0090$ & $< 10^{-4}$ & -- \\
& $\{\boldsymbol{\xi}_{\rm gg},\boldsymbol{\xi}_{\rm gv}\}$ & $1.0087$ & $0.0015$ & -- \\
& $\{\boldsymbol{\xi}_{\rm gg},\boldsymbol{\xi}_{\rm gv},\boldsymbol{\xi}_{\rm vv}\}$ & $1.0105$ & $0.0016$ & -- \\
& $\{\boldsymbol{\xi}_{\rm comb}\}$ & $1.0103$ & $< 10^{-4}$ & $0.0003$ \\
\midrule
\multirowcell{4}[-2\aboverulesep]{EZmock\\LRG (c)\\$\times$\\ELG} & $\{\boldsymbol{\xi}_{\rm gg}^\times\}$ & $1.0084$ & $0.0004$ & -- \\
& $\{\boldsymbol{\xi}_{\rm gg}^\times,\boldsymbol{\xi}_{\rm gv}^\times,\boldsymbol{\xi}_{\rm vg}^\times\}$ & $1.0069$ & $0.0001$ & -- \\
& $\{\boldsymbol{\xi}_{\rm gg}^\times,\boldsymbol{\xi}_{\rm gv}^\times,\boldsymbol{\xi}_{\rm vg}^\times,\boldsymbol{\xi}_{\rm vv}^\times\}$ & $1.0067$ & $0.0018$ & -- \\
& $\{\boldsymbol{\xi}_{\rm comb}^\times\}$ & $1.0036$ & $0.0006$ & $0.0004$ \\
\bottomrule
\end{tabular}
\label{tab:prior_syst}
\end{threeparttable}
\end{table}

\subsection{BAO fitting results}
\label{sec:mock_result}

\begin{figure*}
\centering
\includegraphics[width=.98\textwidth]{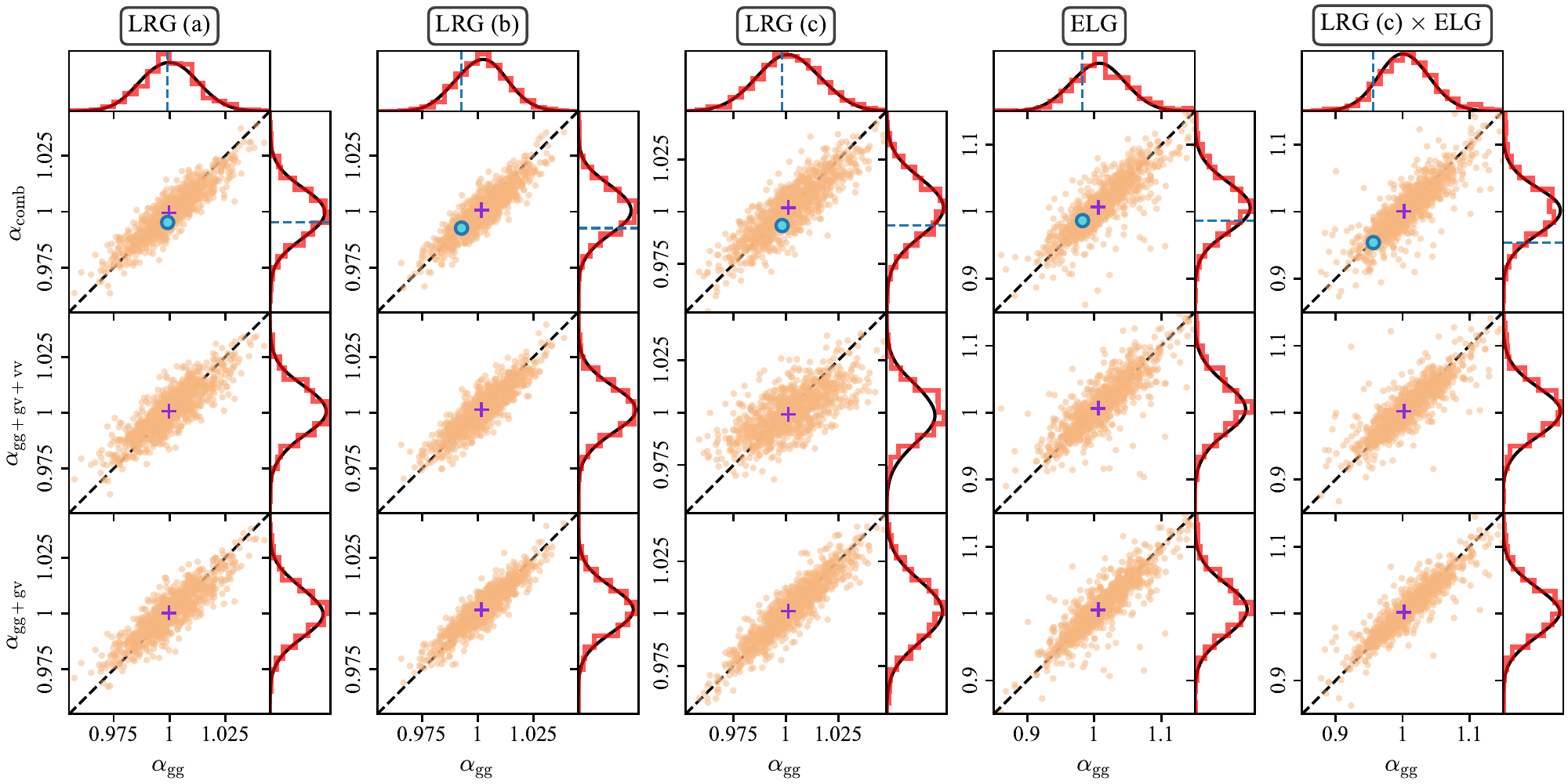}
\caption{Distributions of the best-fitting (median) values of $\alpha$. Scatter plots show comparisons of measurements from individual mocks, with different data vectors (see Figure~\ref{fig:pull_dist} for the interpretations of subscripts), and the corresponding histograms are shown in red. Black solid curves show posterior distributions of $\alpha$, obtained from fits to the mean 2PCFs of mocks, with the widths rescaled by $\sqrt{1000}$.
Purple plus symbols indicate median values drawn from the black curves. Circles and dashed lines in blue indicate results from SDSS data.}
\label{fig:alpha_dist}
\end{figure*}

\begin{figure*}
\centering
\includegraphics[width=.98\textwidth]{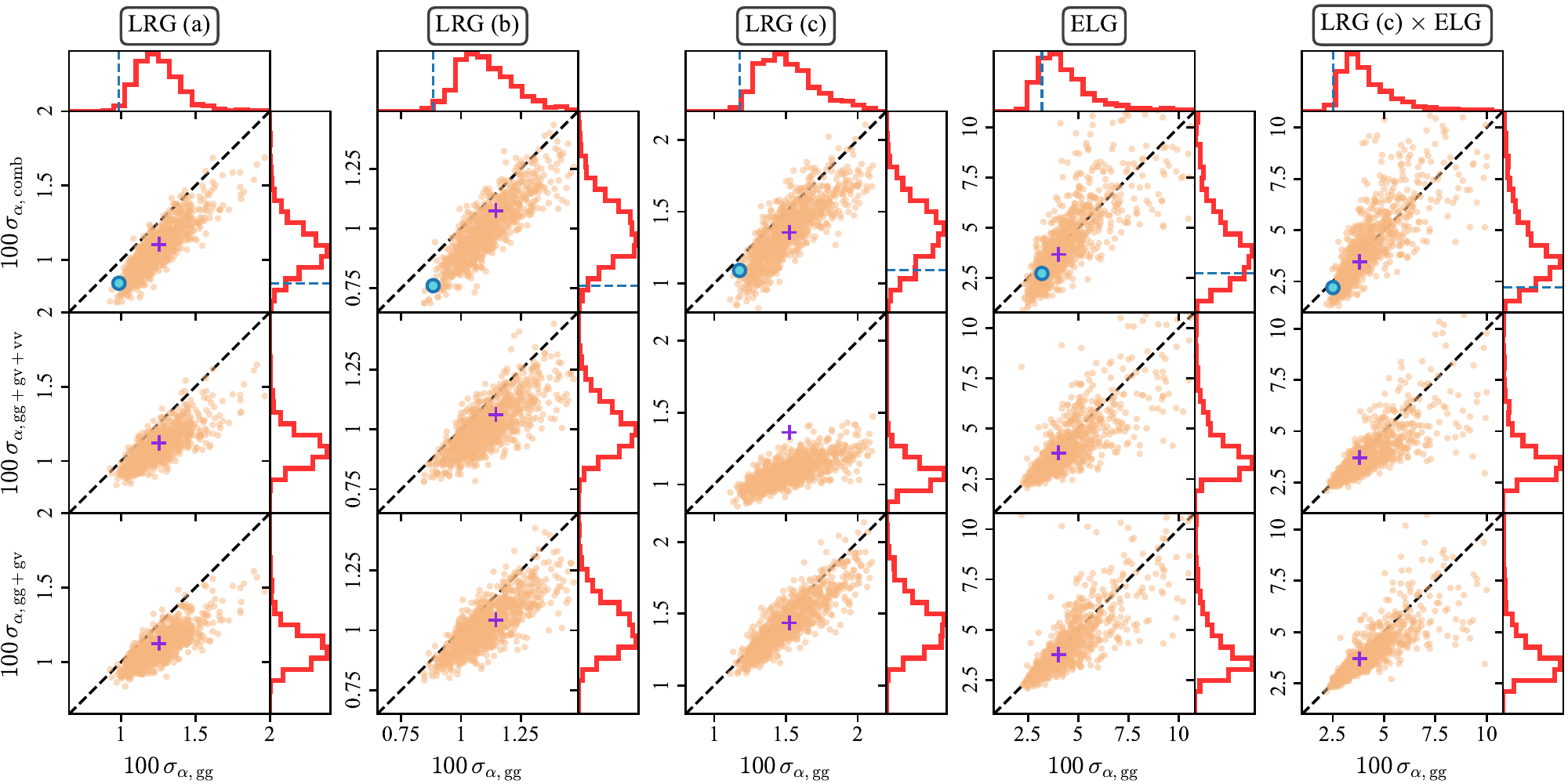}
\caption{Distributions of the fitted error on $\alpha$. Orange points indicate results from individual mocks, with different data vectors (see Figure~\ref{fig:pull_dist}), and red histograms show their distributions.
Purple plus denote measurements from fits to the mean 2PCFs of all mocks. Circles and dash lines in blue indicate results from SDSS data.}
\label{fig:sigma_dist}
\end{figure*}

With the optimal void weights and our fiducial fitting ranges and priors detailed in Table~\ref{tab:fit_setting}, the fitting results from individual mock realizations are summarised in Table~\ref{tab:mock_result}, for different data vectors.
To examine the improvements on BAO constraints with voids, we measure the relative difference of $\sigma_\alpha$ as
\begin{equation}
\delta \sigma_{\alpha, {\rm gg}}
= \frac{ \sigma_{\alpha, {\rm gg}} - \sigma_\alpha }{\sigma_{\alpha, {\rm gg}} } ,
\label{eq:alpha_improve}
\end{equation}
where $\sigma_{\alpha, {\rm gg}}$ indicates the fitted errors from galaxies alone.
In addition, the best-fitting $\alpha$ parameters of each individual mock realization, as well as the fitted errors, are shown in Figures~\ref{fig:alpha_dist} and \ref{fig:sigma_dist}, respectively.
In general, the dispersions of best-fitting $\alpha$ are consistent with the median fitted errors.
%Figure~\ref{fig:alpha_dist} shows also that the distributions of the best-fitting $\alpha$ from individual mocks, are consistent with the fitted posterior of $\alpha$ for the mean 2PCFs of all mocks, for most cases.
It suggests that the statistical errors estimated from the fitted posteriors are unbiased.

\begin{table*}
\centering
\begin{threeparttable}
\caption{A summary of BAO fitting results for the approximate mocks used in this work. $m_1$ and $m_2$ are the factors for correcting fitted parameter covariances, and the scatter of the best-fitting parameters of mocks, respectively (see Section~\ref{sec:param_infer}).
The symbol $\langle \cdot \rangle$ denotes the median value of results from all 1000 individual mocks.
$\Delta \alpha_{1\sigma -}$ and $\Delta \alpha_{1\sigma +}$ indicate the 16th and 84th percentile of the best-fitting $\alpha$ of individual mocks, and $\sigma_\alpha$ denotes the fitted error of $\alpha$.
$\sigma_{\alpha, {\rm syst}}$ is the systematic error on $\alpha$, obtained from the biases of $\alpha$ shown in Tables~\ref{tab:fit_range} and \ref{tab:prior_syst}, which are added in quadrature.
$\delta \sigma_{\alpha, {\rm gg}}$ indicates the relative difference of the fitted error, compared to the result from galaxy--galaxy 2PCF of each mock (see Eq.\eqref{eq:alpha_improve}), and $\eta_{\rm improve}$ denotes the proportion of mocks with $\delta \sigma_{\alpha, {\rm gg}} > 0$, i.e., with smaller $\sigma_\alpha$ when including contributions from voids.}
\begin{tabular}{cSlccScccccc}
\toprule
Sample & Data vector & $m_1$ & $m_2$ & $\langle\alpha\rangle_{-\Delta\alpha_{1\sigma -}}^{+{\Delta\alpha_{1\sigma +}}}$ & $\langle \sigma_\alpha \rangle$ & $\langle \chi^2 \rangle / {\rm d.o.f.}$ & $ \sigma_{\alpha, {\rm syst}} $ & $\langle \delta \sigma_{\alpha, {\rm gg}} \rangle$ & $ \eta_{\rm improve} $ \\
\midrule
%%%%% LRG (a) %%%%%
\multirowcell{4}[-2\aboverulesep]{MD-Patchy\\LRG (a)} & $\{\boldsymbol{\xi}_{\rm gg}\}$ & 1.0173 & 1.0413 & $1.0000_{-0.0118}^{+0.0117}$ & 0.0123 & $16.4 / (22 - 5) = 0.96$ & 0.0009 & -- & -- \\
& $\{\boldsymbol{\xi}_{\rm gg},\boldsymbol{\xi}_{\rm gv}\}$ & 1.0342 & 1.0807 & $1.0010_{-0.0110}^{+0.0100}$ & 0.0109 & $32.1 / (42 - 9) = 0.97$ & 0.0014 & 11.7\,\% & 97.6\,\% \\
& $\{\boldsymbol{\xi}_{\rm gg},\boldsymbol{\xi}_{\rm gv},\boldsymbol{\xi}_{\rm vv}\}$ & 1.0494 & 1.1177 & $1.0008_{-0.0105}^{+0.0110}$ & 0.0108 & $45.0 / (60 - 13) = 0.96$ & 0.0017 & 12.4\,\% & 97.3\,\% \\
& $\{\boldsymbol{\xi}_{\rm comb}\}$ & 1.0122 & 1.0318 & $0.9992_{-0.0106}^{+0.0107}$ & 0.0107 & $11.4 / (18 - 6) = 0.95$ & 0.0005 & 13.7\,\% & 98.2\,\% \\
\midrule
%%%%% LRG (b) %%%%%
\multirowcell{4}[-2\aboverulesep]{MD-Patchy\\LRG (b)} & $\{\boldsymbol{\xi}_{\rm gg}\}$ & 1.0173 & 1.0413 & $1.0020_{-0.0107}^{+0.0114}$ & 0.0109 & $16.3 / (22 - 5) = 0.96$ & 0.0021 & -- & -- \\
& $\{\boldsymbol{\xi}_{\rm gg},\boldsymbol{\xi}_{\rm gv}\}$ & 1.0342 & 1.0807 & $1.0026_{-0.0099}^{+0.0095}$ & 0.0101 & $31.9 / (42 - 9) = 0.97$ & 0.0023 & 7.7\,\% & 90.6\,\% \\
& $\{\boldsymbol{\xi}_{\rm gg},\boldsymbol{\xi}_{\rm gv},\boldsymbol{\xi}_{\rm vv}\}$ & 1.0494 & 1.1177 & $1.0017_{-0.0098}^{+0.0102}$ & 0.0100 & $45.0 / (60 - 13) = 0.96$ & 0.0020 & 8.0\,\% & 91.8\,\% \\
& $\{\boldsymbol{\xi}_{\rm comb}\}$ & 1.0122 & 1.0318 & $1.0004_{-0.0095}^{+0.0102}$ & 0.0099 & $11.6 / (18 - 6) = 0.97$ & 0.0013 & 10.1\,\% & 94.8\,\% \\
\midrule
%%%%% LRG (c) %%%%%
\multirowcell{4}[-2\aboverulesep]{EZmock\\LRG (c)} & $\{\boldsymbol{\xi}_{\rm gg}\}$ & 1.0173 & 1.0413 & $1.0017_{-0.0145}^{+0.0159}$ & 0.0149 & $16.4 / (22 - 5) = 0.96$ & 0.0013 & -- & -- \\
& $\{\boldsymbol{\xi}_{\rm gg},\boldsymbol{\xi}_{\rm gv}\}$ & 1.0342 & 1.0807 & $1.0019_{-0.0145}^{+0.0144}$ & 0.0142 & $32.6 / (42 - 9) = 0.99$ & 0.0013 & 4.8\,\% & 80.8\,\% \\
& $\{\boldsymbol{\xi}_{\rm gg},\boldsymbol{\xi}_{\rm gv},\boldsymbol{\xi}_{\rm vv}\}$ & 1.0494 & 1.1177 & $1.0002_{-0.0107}^{+0.0112}$ & 0.0107 & $46.3 / (60 - 13) = 0.99$ & 0.0024 & 27.7\,\% & 100\,\% \\
& $\{\boldsymbol{\xi}_{\rm comb}\}$ & 1.0122 & 1.0318 & $1.0018_{-0.0139}^{+0.0139}$ & 0.0136 & $11.8 / (18 - 6) = 0.98$ & 0.0027 & 10.2\,\% & 89.4\,\% \\
\midrule
%%%%% ELG %%%%%
\multirowcell{4}[-2\aboverulesep]{EZmock\\ELG} & $\{\boldsymbol{\xi}_{\rm gg}\}$ & 1.0173 & 1.0413 & $1.0090_{-0.0364}^{+0.0440}$ & 0.0409 & $16.7 / (22 - 5) = 0.98$ & 0.0057 & -- & -- \\
& $\{\boldsymbol{\xi}_{\rm gg},\boldsymbol{\xi}_{\rm gv}\}$ & 1.0342 & 1.0807 & $1.0087_{-0.0359}^{+0.0419}$ & 0.0376 & $31.8 / (42 - 9) = 0.96$ & 0.0059 & 6.1\,\% & 66.1\,\% \\
& $\{\boldsymbol{\xi}_{\rm gg},\boldsymbol{\xi}_{\rm gv},\boldsymbol{\xi}_{\rm vv}\}$ & 1.0494 & 1.1177 & $1.0105_{-0.0361}^{+0.0430}$ & 0.0369 & $45.3 / (60 - 13) = 0.96$ & 0.0065 & 8.1\,\% & 72.1\,\% \\
& $\{\boldsymbol{\xi}_{\rm comb}\}$ & 1.0122 & 1.0318 & $1.0103_{-0.0389}^{+0.0393}$ & 0.0393 & $11.9 / (18 - 6) = 0.99$ & 0.0069 & 7.8\,\% & 62.9\,\% \\
\midrule
%%%%% CROSS %%%%%
\multirowcell{4}[-2\aboverulesep]{EZmock\\LRG (c)\\$\times$\\ELG} & $\{\boldsymbol{\xi}_{\rm gg}^\times\}$ & 1.0173 & 1.0413 & $1.0084_{-0.0389}^{+0.0377}$ & 0.0387 & $16.2 / (22 - 5) = 0.95$ & 0.0022 & -- & -- \\
& $\{\boldsymbol{\xi}_{\rm gg}^\times,\boldsymbol{\xi}_{\rm gv}^\times,\boldsymbol{\xi}_{\rm vg}^\times\}$ & 1.0516 & 1.1224 & $1.0069_{-0.0359}^{+0.0376}$ & 0.0360 & $46.3 / (62 - 13) = 0.95$ & 0.0024 & 6.8\,\% & 74.3\,\% \\
& $\{\boldsymbol{\xi}_{\rm gg}^\times,\boldsymbol{\xi}_{\rm gv}^\times,\boldsymbol{\xi}_{\rm vg}^\times,\boldsymbol{\xi}_{\rm vv}^\times\}$ & 1.0674 & 1.1616 & $1.0067_{-0.0342}^{+0.0378}$ & 0.0351 & $59.0 / (80 - 17) = 0.94$ & 0.0046 & 8.7\,\% & 75.8\,\% \\
& $\{\boldsymbol{\xi}_{\rm comb}^\times\}$ & 1.0122 & 1.0318 & $1.0036_{-0.0361}^{+0.0386}$ & 0.0376 & $11.6 / (18 - 6) = 0.97$ & 0.0010 & 6.3\,\% & 61.1\,\% \\
\bottomrule
\end{tabular}
\label{tab:mock_result}
\end{threeparttable}
\end{table*}

We sum the biases of $\alpha$ from the fits with different fitting ranges and parameter priors in quadrature, to assess the systematic errors on the $\alpha$ measurements.
Compared to the magnitude of statistical errors, systematic errors on the BAO measurements are typically fairly small.
When including clustering measurements with voids, either by fitting simultaneously multiple 2PCFs, or by measuring BAO from the combined correlation function of galaxies and voids, the fitted error is smaller than that from galaxy--galaxy correlation alone, for the majority of the mock realizations. The typical improvement on the statistical uncertainty of BAO peak position is around 5 to 10 per cent.
This confirms that as tracers of underdensities, the clustering of voids bring additional information compared to that of galaxies.
Since our voids are defined by quadruples of galaxies, the information should be encoded in the galaxy distribution already, but exacted only from high-order statistics such as four-point correlations.
Thus, the relative improvement from voids can be larger without BAO reconstruction, which transfers BAO information from higher-point statistics into two-point clustering \citep[][]{Schmittfull2015}.
Nevertheless, the joint BAO constraint after reconstruction should still be better, as long as the galaxy density field is not fully Gaussianized after reconstruction. This has been confirmed by the pre- and post-reconstruction measurements in \citet[][]{Forero2021}.

Furthermore, the improvement is relatively larger and detected from more mock realizations, when the BAO fitting uncertainty is smaller.
This is consistent with the studies in \citet[][]{Zhao2020}, and suggests that the contribution of voids will be more promising for future surveys with larger volumes and smaller statistical errors.
By comparing results from joint fits to the stacked data vectors with different clustering measurements, we conclude that the improvement of the BAO constraint is dominated by the contribution of galaxy--void cross correlation. This is not surprising given the BAO signal-to-noise analysis in Section~\ref{sec:radius_sel}.
However, for the `LRG (c)' sample, an over 20 per cent improvement is observed when including voids.
This is possibly due to the small fitted errors of void auto correlations, which are actually smaller than those of both galaxy--galaxy and galaxy--void correlations on average (see Table~\ref{tab:fit_prior}).
We have examined different parameter priors for both the void 2PCFs of individual mocks, as well as the mean result from all mocks, but no obvious problem is found (see Appendix~\ref{sec:result_fit_prior}).
In particular, for the mean void 2PCF of mocks, the fitted error with our fiducial priors is smaller than that with sufficiently large prior ranges, and agrees with the distribution of results from individual mocks. It implies that a prior is needed even for the fit to the mean void 2PCFs of the `LRG (c)' mocks.

For the combined correlation function, however, the relative BAO precision improvement for `LRG (c)' is around 10 per cent. This is more consistent with the results of the other samples.
Actually, results from the combined 2PCFs are in good agreements with those from fits to multiple clustering measurements for all the other samples.
This suggests that results from the combined 2PCFs are more robust.
Moreover, these results benefit from the smaller lengths of data vector, compared to those of fits to the stacked 2PCFs.
In this case, the correction factors for the uncertainties of the covariance matrices estimated using mocks are smaller, and statistical errors from the fits are more reliable.
Therefore, we choose the multi-tracer fitting method based on the combined correlation functions with void weights, for BAO measurements and cosmological parameter constraints with the SDSS data, as is detailed in Section~\ref{sec:results}.

\subsection{Cosmological parameter measurements}

\begin{figure}
\centering
\includegraphics[width=.98\columnwidth]{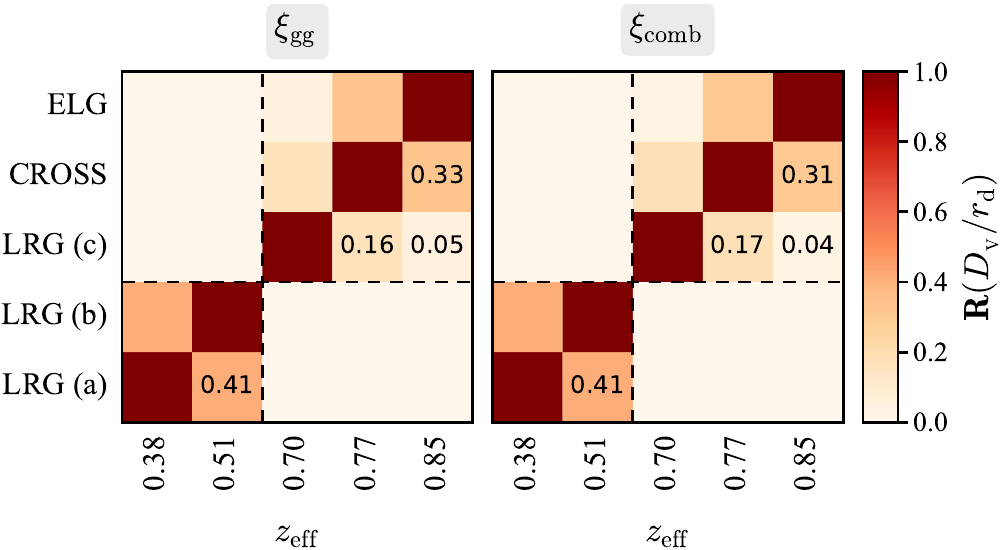}
\caption{Correlation coefficients of $D_{_{\rm V}} / r_{\rm d}$ for different samples, based on the $\alpha$ measurements from 1000 individual mock realizations. `CROSS' denotes results from the cross correlations between `LRG (c)' and `ELG'. $\xi_{\rm gg}$ and $\xi_{\rm comb}$ indicate results from galaxy--galaxy correlations and the combined 2PCFs respectively.}
\label{fig:corr_coef}
\end{figure}

In order to perform cosmological analysis with the BAO fitting results, we convert the best-fitting $\alpha$ of individual mocks to $D_{_{\rm V}} / r_{\rm d}$ measurements, and plot the correlation coefficients between different samples in Figure~\ref{fig:corr_coef}.
The results from galaxy--galaxy correlations and the combined 2PCFs are almost identical. This is consistent with the cross covariances of 2PCFs for different samples in Appendix~\ref{sec:cov_xi}, and can be explained by the fact that adding galaxy--void and void--void correlations does not bias the $\alpha$ measurement, but only reduces the fitted error.
The correlation coefficient matrices can be divided into two blocks, one for the BOSS LRG samples that are below redshift 0.6, and the other for galaxies with $z > 0.6$. We do not consider the cross correlations between these two populations, as they do not overlap in the 3D volume.
Since the `LRG (a)' and `LRG (b)' samples share the same galaxies in the redshift range $0.4 < z < 0.5$, their cross coefficient is relatively large.
In contrast, the coefficient between the `LRG (c)' and `ELG' samples is small, as these two samples do not contain the same objects.
Actually, these two samples are only correlated through their common volume, which is only a small proportion of the LRG volume, but involves the majority of the ELGs. This explains why the $D_{_{\rm V}} / r_{\rm d}$ measured from the cross correlations between `LRG (c)' and `ELG' is more correlated with the `ELG' sample.

\begin{figure}
\centering
\includegraphics[width=.98\columnwidth]{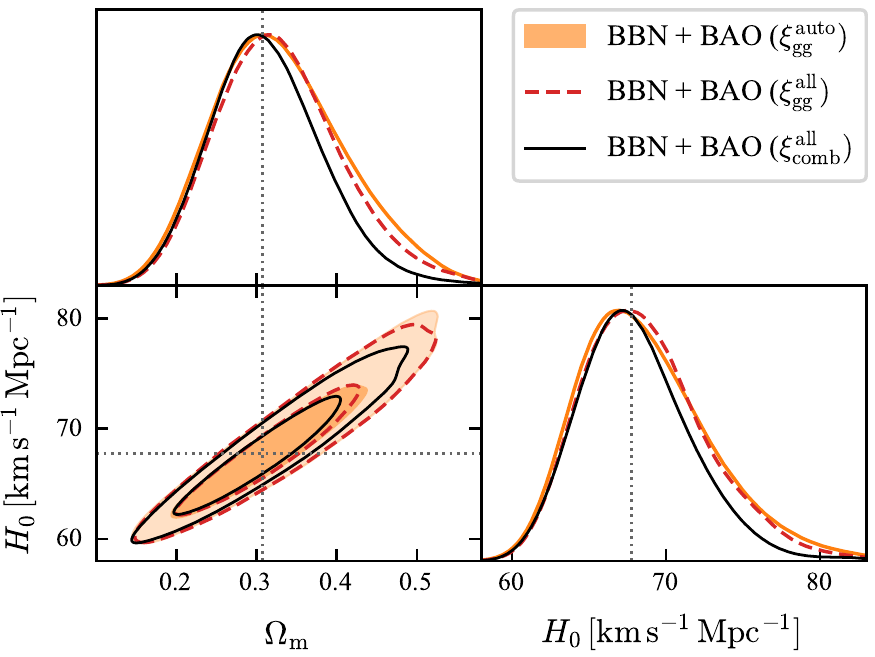}
\caption{Constraints on $H_0$ and $\Omega_{\rm m}$ in a flat-$\Lambda$CDM model, obtained from the combination of BBN and different BAO measurements from mocks. $\xi_{\rm gg}^{\rm auto}$ indicates results from galaxy auto correlation only, while $\xi_{\rm gg}^{\rm all}$ includes also the contribution of galaxy cross correlation between the `LRG (c)' and `ELG' samples. $\xi_{\rm comb}^{\rm all}$ denotes BAO measurements from all samples, with both galaxies and voids. Dotted lines indicate the cosmological parameters for constructing the mocks. The contours show the 68\,\% and 95\,\% confidence intervals.}
\label{fig:mock_cosmo}
\end{figure}

\begin{table}
\centering
\begin{threeparttable}
\setlength{\tabcolsep}{.95\tabcolsep}
\caption{Best-fitting cosmological parameters and 1\,$\sigma$ confidence intervals for the mocks, drawn from the posterior distributions in Figure~\ref{fig:mock_cosmo}. The `true' cosmological parameters denote the ones used for mock construction.}
\begin{tabular}{ccScScSc}
\toprule
Parameter & `True' & BAO\,($\xi_{\rm gg}^{\rm auto}$) & BAO\,($\xi_{\rm gg}^{\rm all}$) & BAO\,($\xi_{\rm comb}$) \\
\midrule
$H_0$ & 67.77 & $68.7_{-5.1}^{+3.1}$ & $68.7_{-4.7}^{+3.0}$ & $68.0_{-4.1}^{+2.9}$ \\
$\Omega_{\rm m}$ & 0.307115 & $0.330_{-0.095}^{+0.069}$ & $0.329_{-0.087}^{+0.065}$ & $0.313_{-0.076}^{+0.061}$ \\
\bottomrule
\end{tabular}
\label{tab:mock_cosmo}
\end{threeparttable}
\end{table}

We construct also the covariance matrices of $D_{_{\rm V}}/r_{\rm d}$ from individual mocks, and rescaled them by the factor $m_2$ given by Eq.~\eqref{eq:m2}. The covariances for different samples are then used together with the median distance measurements of all mocks, for constraining $H_0$ and $\Omega_{\rm m}$ in a flat-$\Lambda$CDM cosmological model.
Apart from comparing measurements with galaxies alone and a combination of galaxies and voids, we examine also the galaxy-only results with and without the cross correlations between the `LRG (c)' and `ELG' samples, to investigate improvements from multi-tracer analysis with different types of galaxies residing in the same cosmic volume.
The constraints from combinations of BBN and BAO measurements are shown in Figure~\ref{fig:mock_cosmo} and Table~\ref{tab:mock_cosmo}.
In all cases, the `true' cosmological parameters of the mocks are well recovered.
When including only the contribution from the cross correlation between LRGs and ELGs, the uncertainties of both parameters are reduced by around 6 per cent, and a further 10 per cent improvement is achieved after taking into account the clustering of voids.

The results confirm that the multi-tracer BAO analysis approach is promising for extracting more information from the cosmic density field, compared to the results with matter tracers alone.
In addition, the unbiased cosmological parameter constraints with the mocks validate the robustness of our methodology, which can then be applied to the observational data.

%%%%%%%%%%%%%%%%% NEW SECTION %%%%%%%%%%%%%%%%%%

\section{Results}
\label{sec:results}

As is discussed in the previous section, the multi-tracer BAO fitting approach based on the combined correlation functions is preferred over the joint fit to the stacked 2PCFs of multiple tracers.
We then perform BAO fits with the combined 2PCFs of galaxies and voids from the SDSS data, with the fiducial BAO fitting scheme shown in bold in Table~\ref{tab:fit_setting}, and compare the results to those from galaxy correlations alone.
In particular, when fitting to the galaxy--galaxy correlations, we use the $\Sigma_{\rm nl}$ values obtained from $N$-body simulations.
The fitting results are converted to distance measurements, which are finally used for cosmological parameter constraints.

\subsection{Distance measurements}

\begin{figure}
\centering
\includegraphics[width=.98\columnwidth]{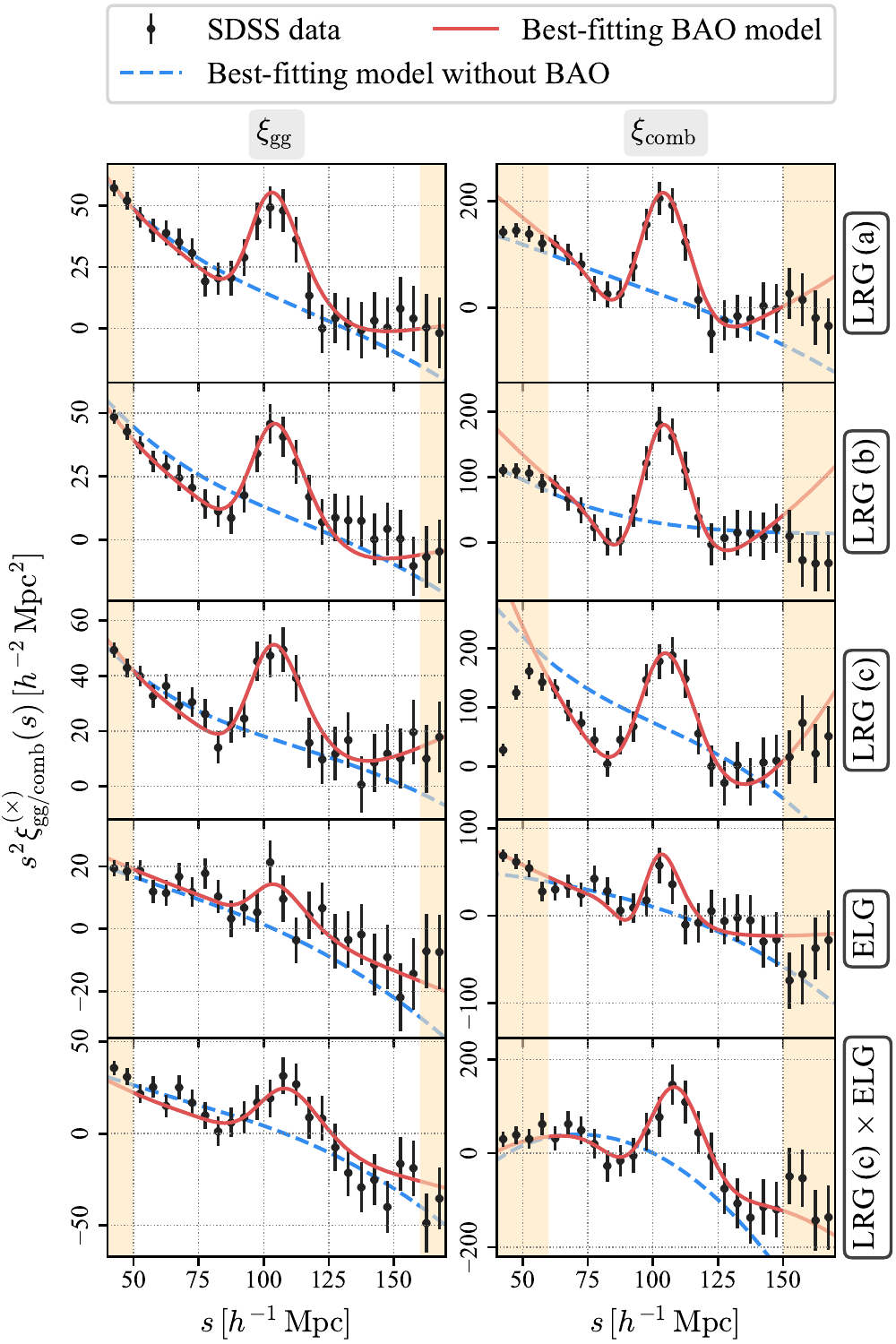}
\caption{Galaxy--galaxy correlations and combined 2PCFs measured from SDSS data, as well as the best-fitting model curves with and without BAO peaks. In particular, the combined correlations are computed with the optimal void weights measured from mocks. Shadowed areas indicate regions outside the fitting ranges.}
\label{fig:xi_best}
\end{figure}

\begin{figure}
\centering
\includegraphics[width=.95\columnwidth]{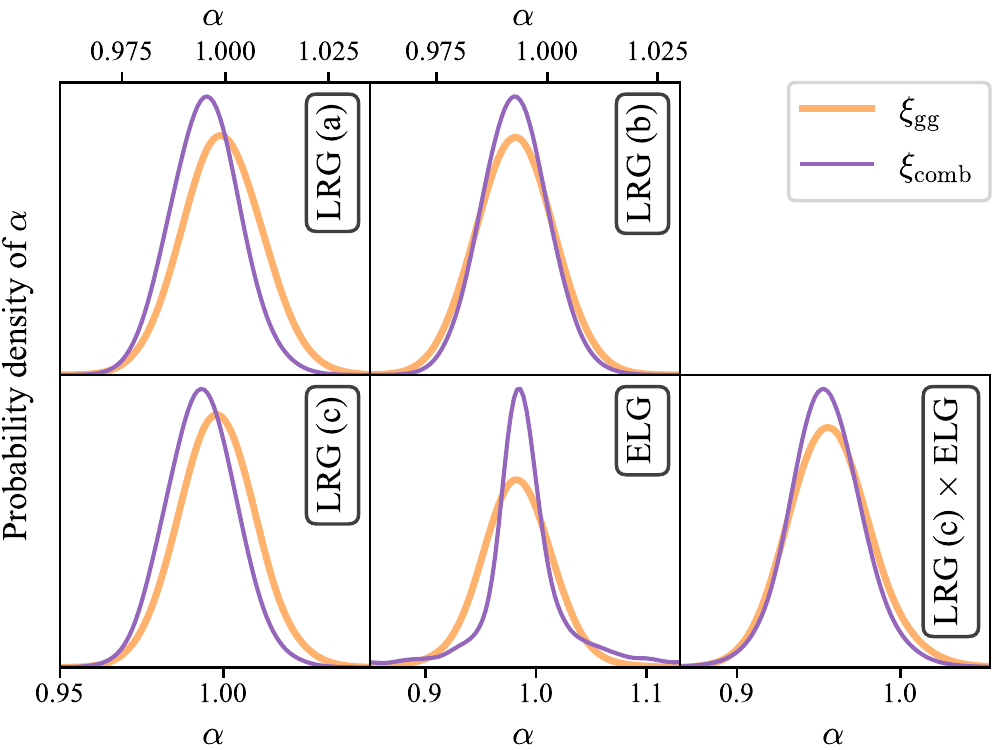}
\caption{Marginalized posterior distributions of $\alpha$ for different SDSS samples. Orange lines show results from galaxies only, while purple lines indicate measurements from the combined 2PCFs of galaxies and voids.}
\label{fig:alpha_data}
\end{figure}

With the fit settings listed in Table~\ref{tab:fit_setting}, we run BAO fits for both the galaxy--galaxy correlation ($\xi_{\rm gg}$) and combined correlation functions with galaxies and voids ($\xi_{\rm comb}$) for all the SDSS samples studied in this work, with covariance matrices estimated using the corresponding approximate mocks.
The best-fitting BAO model curves agree well with the observational data in the fitting ranges, as shown in Figure~\ref{fig:xi_best}, with the corresponding posterior distributions of the $\alpha$ parameters illustrated in Figure~\ref{fig:alpha_data}.
The best-fitting $\alpha$ parameters, their statistical uncertainties drawn from the posterior distributions, as well as the reduced chi-squared values, are listed in Table~\ref{tab:data_alpha}.
For all the samples, the $\alpha$ measurements from $\xi_{\rm gg}$ and $\xi_{\rm comb}$ agree within 0.5\,$\sigma$ statistical uncertainties.
Besides, the fitted errors for the combined correlation functions are always smaller than those from galaxies alone, with the relative differences ranging from 5 to 15 per cent.

The improvements on the BAO measurements are significant, compared to the systematic errors estimated using approximate mocks, as well as  uncertainties on the fitted errors, which are generally less than 5 per cent (see Appendix~\ref{sec:result_fit_prior}).
Figure~\ref{fig:xi_best} shows also the best-fitting models without BAO peaks, obtained by replacing $P_{\rm lin}(k)$ with $P_{\rm lin, nw} (k)$ in Eq.~\eqref{eq:bao_model_dewiggle}.
The smoothness of best-fitting `non-wiggle' curves confirm that our BAO models for both the galaxy--galaxy and combined 2PCFs are able to describe well the BAO peak, without introducing additional patterns on the same scale.
The differences of chi-squared values, for models without and with BAO peaks, are presented in Table~\ref{tab:data_alpha}. These results suggest that the BAO significances of the combined correlation functions are generally higher than the ones of $\xi_{\rm gg}$, despite the smaller number of $\xi_{\rm comb}$ bins.

\begin{table}
\centering
\begin{threeparttable}
\setlength{\tabcolsep}{\tabcolsep}
\caption{BAO fitting results for the SDSS data. The best-fitting value, lower and upper 1\,$\sigma$ confidence limites are evaluated from the 50th, 16th, and 84th percentiles of the marginalized posterior distribution of $\alpha$. The systematic error $\sigma_{\alpha, {\rm syst}}$ is estimated using mocks, as is shown in Table~\ref{tab:mock_result}.
$(\chi^2 / {\rm d.o.f.})$ indicates the reduced chi-squared obtained from the model with BAO peaks, and $\Delta \chi^2 = \chi_{\rm no\,BAO}^2 - \chi_{\rm BAO}^2$, where $\chi_{\rm no\,BAO}^2$ and $\chi_{\rm BAO}^2$ denote the best-fitting chi-squared values from the model without and with BAO peaks respectively.
}
\begin{tabular}{ccSccccc}
\toprule
Sample & \makecell{Data\\vector} & $\alpha_{\rm fit}$ & $\sigma_{\alpha, {\rm syst}}$ & $\dfrac{\chi^2 }{\rm d.o.f.}$ & $\Delta \chi^2$ \\
\midrule
%%%%% LRG (a) %%%%%
\multirowcell{2}[-\aboverulesep]{LRG (a)} & $\{\boldsymbol{\xi}_{\rm gg}\}$ & $0.9992_{-0.0097}^{+0.0100}$ & 0.0009 & 1.51 & 55.9 \\
& $\{\boldsymbol{\xi}_{\rm comb}\}$ & $0.9952_{-0.0086}^{+0.0083}$ & 0.0005 & 1.59 & 62.8 \\
\midrule
%%%%% LRG (b) %%%%%
\multirowcell{2}[-\aboverulesep]{LRG (b)} & $\{\boldsymbol{\xi}_{\rm gg}\}$ & $0.9928_{-0.0089}^{+0.0088}$ & 0.0021 & 1.18 & 57.2 \\
& $\{\boldsymbol{\xi}_{\rm comb}\}$ & $0.9927_{-0.0076}^{+0.0076}$ & 0.0013 & 0.46 & 70.7 \\
\midrule
%%%%% LRG (c) %%%%%
\multirowcell{2}[-\aboverulesep]{LRG (c)} & $\{\boldsymbol{\xi}_{\rm gg}\}$ & $0.9982_{-0.0119}^{+0.0116}$ & 0.0013 & 1.46 & 37.9 \\
& $\{\boldsymbol{\xi}_{\rm comb}\}$ & $0.9934_{-0.0109}^{+0.0110}$ & 0.0027 & 1.26 & 41.0 \\
\midrule
%%%%% ELG %%%%%
\multirowcell{2}[-\aboverulesep]{ELG} & $\{\boldsymbol{\xi}_{\rm gg}\}$ & $0.9825_{-0.0318}^{+0.0321}$ & 0.0057 & 1.34 & 2.2 \\
& $\{\boldsymbol{\xi}_{\rm comb}\}$ & $0.9867_{-0.0236}^{+0.0310}$ & 0.0069 & 0.98 & 5.7 \\
\midrule
%%%%% CROSS %%%%%
\multirowcell{2}[-\aboverulesep]{LRG (c)\\$\times$ ELG} & $\{\boldsymbol{\xi}_{\rm gg}^\times\}$ & $0.9565_{-0.0246}^{+0.0254}$ & 0.0022 & 1.10 & 14.3 \\
& $\{\boldsymbol{\xi}_{\rm comb}^\times\}$ & $0.9539_{-0.0212}^{+0.0228}$ & 0.0010 & 0.56 & 11.6 \\
\bottomrule
\end{tabular}
\label{tab:data_alpha}
\end{threeparttable}
\end{table}

The fitting results of the SDSS data are also shown in Figures~\ref{fig:alpha_dist} and \ref{fig:sigma_dist}, in comparison with results from the approximate mocks. The best-fitting $\alpha$ values from the data and mocks are generally in good agreements.
However, the fitted errors of the data, computed as the mean of the lower and upper 1\,$\sigma$ confidence limits, are always smaller than the median BAO errors of mocks. This is consistent with previous studies \citep[e.g.][]{Vargas2018,Zhao2020,Bautista2021}, and can be explained by the fact that the approximate mocks generally overestimate the nonlinear damping of the BAO peak, which magnifies the uncertainty of BAO position.
We have further checked results for different Galactic caps, and good consistencies are found in general (see Appendix~\ref{sec:bao_caps}).

\begin{table}
\centering
\begin{threeparttable}
\setlength{\tabcolsep}{.85\tabcolsep}
\caption{Distance measurements from the BAO fitting results of the SDSS data at different effective redshifts. Systematic errors estimated using mocks are also included, for both results from galaxies alone ($\xi_{\rm gg}$), and measurements from a joint analysis with galaxies and voids ($\xi_{\rm comb}$).}
\begin{tabular}{ccccc}
\toprule
\multirowcell{2}{Sample} & \multirowcell{2}{$z_{\rm eff}$} & $\xi_{\rm gg}$ & \multicolumn{2}{c}{$\xi_{\rm comb}$} \\
\cmidrule(lr){3-3}
\cmidrule(lr){4-5}
& & $D_{_{\rm V}} / r_{\rm d}$ & $D_{_{\rm V}} / r_{\rm d}$ & Precision\\
\midrule
LRG (a) & 0.38 & $9.98\pm0.10$ & $9.94\pm0.08$ & 0.9\,\% \\
LRG (b) & 0.51 & $12.67\pm0.12$ & $12.67\pm0.10$ & 0.8\,\% \\
LRG (c) & 0.70 & $16.25\pm0.19$ & $16.17\pm0.18$ & 1.1\,\% \\
LRG (c) $\times$ ELG & 0.77 & $16.67\pm0.44$ & $16.63\pm0.39$ & 2.3\,\% \\
ELG & 0.85 & $18.33\pm0.61$ & $18.41\pm0.53$ & 2.9\,\% \\
\bottomrule
\end{tabular}
\label{tab:data_dist}
\end{threeparttable}
\end{table}

Given the $D_{_{\rm V}} / r_{\rm d}$ value in our fiducial cosmology at different redshifts, we convert the fitting results of $\alpha$ from the SDSS data to cosmological distance measurements, and present the results in Table~\ref{tab:data_dist}.
Since the upper and lower 1\,$\sigma$ errors of $\alpha$ are similar for almost all cases, we quote only symmetric errors of $D_{_{\rm V}} / r_{\rm d}$.
Moreover, the statistical errors are rescaled by $\sqrt{m_1}$, following Eq.~\eqref{eq:m1}, and then combined with systematic errors by summing in quadrature.
The cross covariances between different samples are estimated using the correlation coefficients measured from 1000 mock realizations (see Figure~\ref{fig:corr_coef}), and are presented in Appendix~\ref{sec:dist_cov}.

\begin{figure}
\centering
\includegraphics[width=220pt]{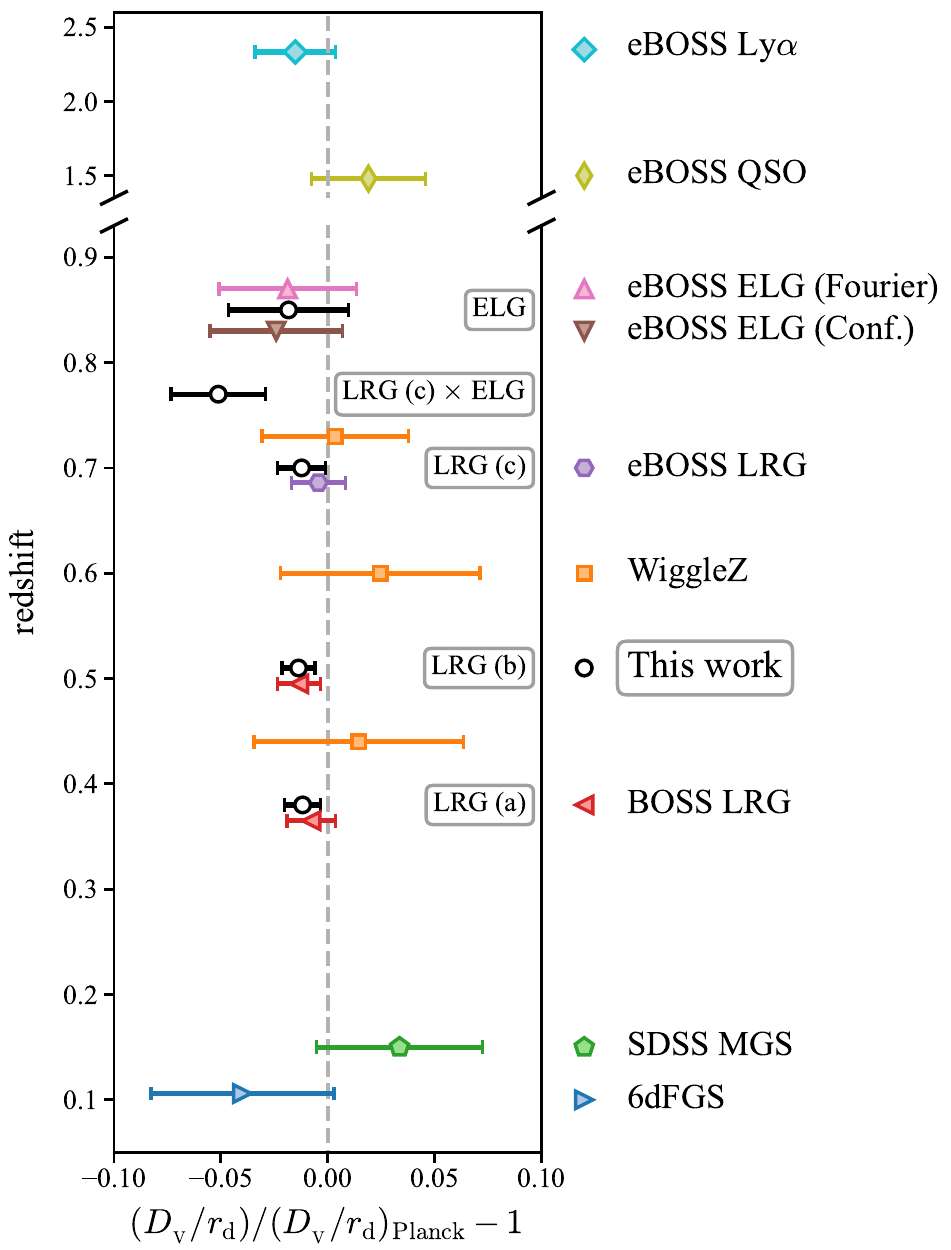}
\begin{picture}(0,0)
\put(-80,32){\colorbox{white}{\citetalias[][]{Beutler2011}}}
\put(-80,44.5){\colorbox{white}{\citetalias[][]{Ross2015}}}
\put(-80,100.5){\colorbox{white}{\citetalias[][]{Alam2017}}}
\put(-80,155){\colorbox{white}{\citetalias[][]{Kazin2014}}}
\put(-80,179){\colorbox{white}{\citetalias[][]{Bautista2021}}}
\put(-80,210.5){\colorbox{white}{\citetalias[][]{Raichoor2021}}}
\put(-80,220.5){\colorbox{white}{\citetalias[][]{deMattia2021}}}
\put(-80,246.5){\colorbox{white}{\citetalias[][]{Hou2021}}}
\put(-80,275.5){\colorbox{white}{\citetalias[][]{Bourboux2020}}}
\end{picture}
\caption{BAO distance measurements from different spectroscopic data, normalized by predictions of the \citet[][]{Planck2020} best-fitting base-$\Lambda$CDM parameters ($\Omega_{\rm m} = 0.3158$, $h = 0.6732$). The results are all from BAO-only analysis.}
\label{fig:dist_comp}
\end{figure}
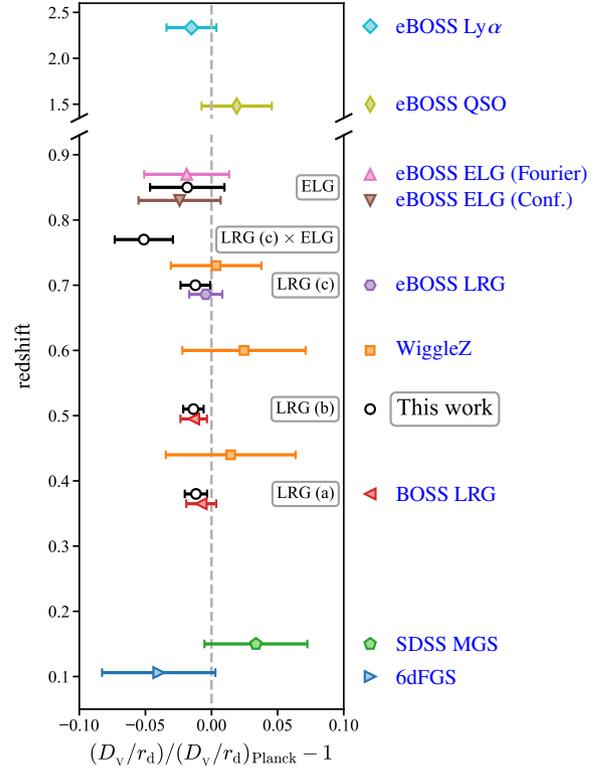

We then compare our distance measurements from the combined correlation functions with those from previous studies -- that are based on data from different spectroscopic surveys -- in Figure~\ref{fig:dist_comp}. In particular, the measurements are collected from BAO-only analysis of 6dFGS \citep[][]{Beutler2011}, WiggleZ \citep[][]{Kazin2014}, SDSS MGS \citep[][]{Ross2015}, BOSS LRGs \citep[][]{Alam2017}, eBOSS LRGs \citep[][]{Gilmarin2020,Bautista2021}, eBOSS ELGs in configuration space \citep[][]{Raichoor2021} and Fourier space \citep[][]{deMattia2021}, eBOSS QSOs \citep[][]{Neveux2020,Hou2021}, as well as eBOSS Ly$\alpha$ forests \citep[][]{Bourboux2020}.
Here, the consensus results based on configuration and Fourier space measurements are used whenever possible, and the Ly$\alpha$ forest measurement is based on the combination of Ly$\alpha$ auto correlations and Ly$\alpha$--QSO cross correlations.
Previous BOSS/eBOSS results that are given as $D_{_{\rm M}}/r_{\rm d}$ and $D_{_{\rm H}}/r_{\rm d}$ measurements, are combined into $D_{_{\rm V}}/r_{\rm d}$ following Eq.~\eqref{eq:bao_dv}, with their cross covariances taken into account.
The distance measurements are then normalized by predictions of the best-fitting base-$\Lambda$CDM parameters from \citet[][]{Planck2020}, for better comparisons of the sizes of error bars.
Besides, measurements at the same effective redshifts are added small vertical offsets for illustration purposes.

In general, the results in this work agree well with previous BOSS/eBOSS analyses, apart from the smaller uncertainties due to the contributions of voids.
One can also see that the central values of our measurements are always smaller than the predictions of Planck results, though the differences are not significant.
In particular, the result from the cross correlations between `LRG (c)' and `ELG' shows the largest tension, which is around 2$\,\sigma$.
We then study the implications on cosmological parameters in the following subsections.

\subsection{Hubble parameter and density parameters}

\begin{figure}
\centering
\includegraphics[width=235pt]{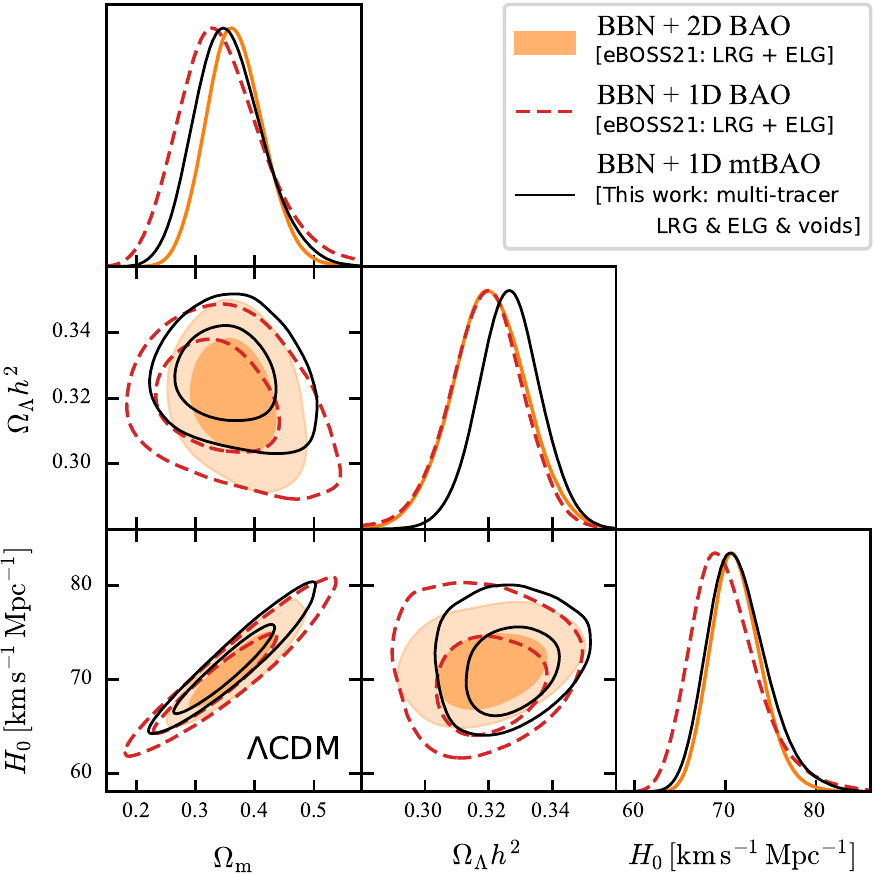}
\begin{picture}(0,0)
\setlength{\fboxsep}{0.5pt}
%\put(-71.2,136){\scalebox{0.8}{\colorbox{white}{\textsf{\citetalias[][]{eBOSS2021}}}}}
%\put(-71.2,156.2){\scalebox{0.8}{\colorbox{white}{\textsf{\citetalias[][]{eBOSS2021}}}}}
\put(-75,218.3){\scalebox{0.8}{\colorbox{white}{\textsf{\citetalias[][]{eBOSS2021}}}}}
\put(-75,199.5){\scalebox{0.8}{\colorbox{white}{\textsf{\citetalias[][]{eBOSS2021}}}}}
\end{picture}
\caption{Constraints on $H_0$, $\Omega_{\rm m}$, and $\Omega_\Lambda h^2$ in a flat-$\Lambda$CDM model, obtained from the combination of BBN and different BAO measurements.
Orange contours and lines show the results with $D_{_{\rm M}} / r_{\rm d}$ and $D_{_{\rm H}} / r_{\rm d}$ measured from LRGs and ELGs in \citetalias[][]{eBOSS2021}.
Red dashed lines indicate constraints with the corresponding $D_{_{\rm V}} / r_{\rm d}$, obtained by combining the 2D distance measurements, with their cross covariances taken into account.
Black solid lines indicate results with $D_{_{\rm V}} / r_{\rm d}$ measured in a multi-tracer manner, using the combined correlation functions of galaxies and voids, including the cross correlations between LRGs and ELGs.}
\label{fig:Dv_vs_DmDh}
\end{figure}
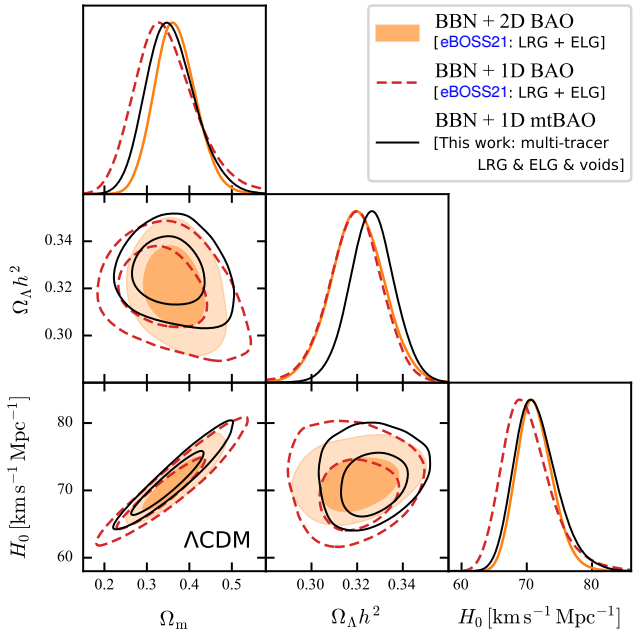

To highlight the contribution of the multi-tracer approach, we compare firstly the cosmological parameters constrained using the distances measured from the LRG and ELG samples in this work, to the corresponding measurements used in \citetalias[][]{eBOSS2021}.
In addition, we combine the $D_{_{\rm M}} / r_{\rm d}$ and $D_{_{\rm H}} / r_{\rm d}$ measured from BOSS/eBOSS LRGs into $D_{_{\rm V}} / r_{\rm d}$, to have 1D BAO results for fair comparisons.
The posterior distributions of $H_0$, $\Omega_{\rm m}$, and $\Omega_\Lambda h^2$ from joint BBN and BAO analyses in a flat-$\Lambda$CDM cosmology are shown in Figure~\ref{fig:Dv_vs_DmDh}. Note that throughout this work, the contours show the 68 and 95 per cent confidence intervals.
It can be seen that for the 1D BAO constraints based on $D_{_{\rm V}} / r_{\rm d}$, the multi-tracer results with cosmic voids improve significantly the precision of all parameters.
However, the uncertainties of parameters measured from multi-tracer 1D BAOs can be larger than the ones from 2D BAOs of galaxies alone.
For instance, the marginalized 1\,$\sigma$ error of the $\Omega_{\rm m}$ parameter from the multi-tracer 1D BAOs is around 18 per cent smaller than the measurements from the 1D BAOs of galaxies alone, but 19 per cent larger than the corresponding galaxy 2D BAO results.
This suggests that the decomposed $D_{_{\rm M}} / r_{\rm d}$ and $D_{_{\rm H}} / r_{\rm d}$ parameters encode more cosmological information than the spherical averaged $D_{_{\rm V}} / r_{\rm d}$.

Nevertheless, it is still possible that the multi-tracer 1D BAOs yields better constraints for certain parameters, compared to the results from 2D BAOs of galaxies alone.
As an example, the anisotropic BAO do not constrain $\Omega_{\Lambda} h^2$ better than the isotropic BAO in the flat-$\Lambda$CDM framework at low redshifts, since both $D_{_{\rm M}} / r_{\rm d}$ and $D_{_{\rm H}} / r_{\rm d}$ depend on this parameter in approximately the same way, in the dark-energy-dominated era.
Actually, we observe similar $\Omega_\Lambda h^2$ uncertainties for both 1D and 2D galaxy BAOs in Figure~\ref{fig:Dv_vs_DmDh}.
In contrast, a 16 per cent reduction of the 1\,$\sigma$ error is observed for the 1D BAO with voids.
This implies that by combining the multi-tracer 1D BAOs with different datasets, the cosmological constraints can be tighter than those from the galaxy 2D BAOs, due to different parameter degeneracies.
We start with combining the multi-tracer BAOs measured in this work, with the BAO-only results from the rest of the SDSS large-scale structure samples, including MGS \citep[][]{Ross2015}, QSOs \citep[][]{Neveux2020,Hou2021}, and Ly$\alpha$ forests \citep[][]{Bourboux2020}. The resulting BAO measurements are dubbed `mtBAO' henceforth.
It is worth noting that the same datasets are used for cosmological parameter constraints with BAO in \citetalias[][]{eBOSS2021}, which serves as our baseline for examining the contributions from cosmic voids.

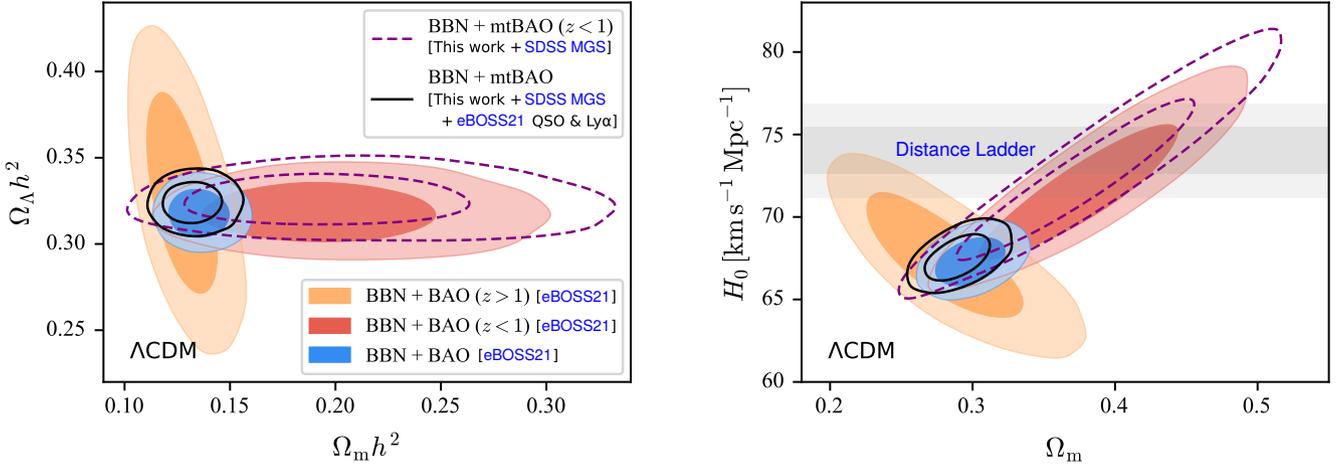
\begin{figure*}
\centering
\includegraphics[width=235pt]{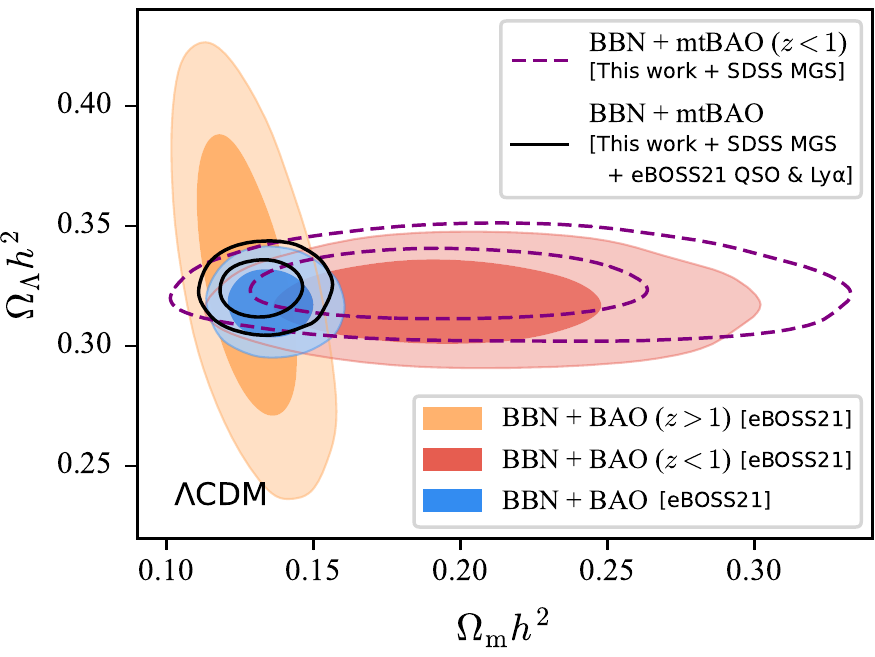}
\hspace{2pc}
\includegraphics[width=235pt]{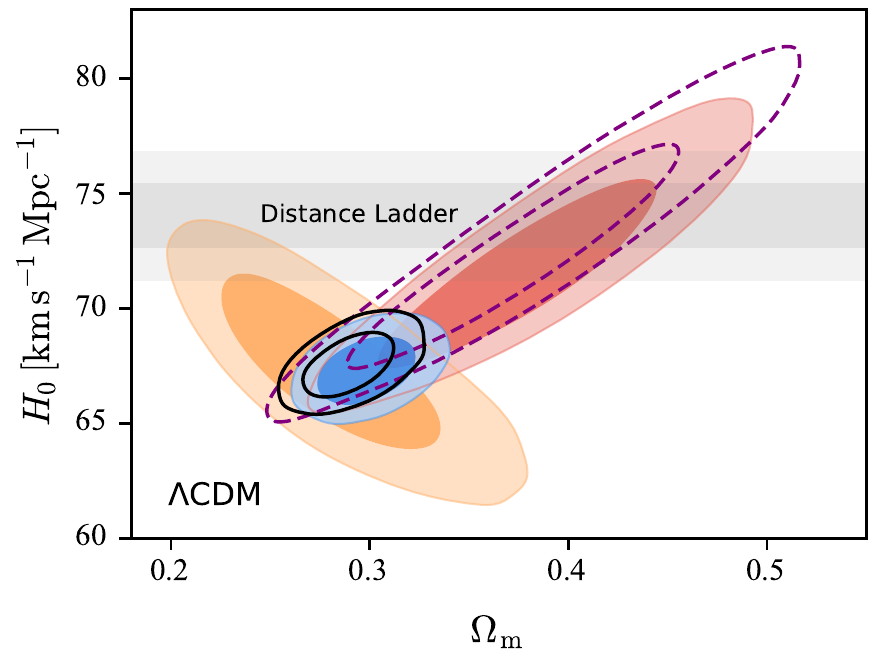}
\begin{picture}(0,0)
\setlength{\fboxsep}{0.3pt}
%\put(-108.8,162.8){\scalebox{0.8}{\colorbox{white}{\textsf{\citetalias[][]{eBOSS2021}}}}}
%\put(-108.8,151.5){\scalebox{0.8}{\colorbox{white}{\textsf{\citetalias[][]{eBOSS2021}}}}}
%\put(-131.4,140.6){\scalebox{0.8}{\colorbox{white}{\textsf{\citetalias[][]{eBOSS2021}}}}}
%\put(-44.1,66.2){\scalebox{0.8}{\colorbox{white}{\textsf{\citetalias[][]{Ross2015}}}}}
%\put(-44.1,47.7){\scalebox{0.8}{\colorbox{white}{\textsf{\citetalias[][]{Ross2015}}}}}
%\put(-70.8,40){\scalebox{0.8}{\colorbox{white}{\textsf{\citetalias[][]{eBOSS2021}}}}}
%\put(-61,103){\scalebox{1}{\colorbox{gray!23.5}{\textsf{\citetalias[][]{Riess2019}}}}}
\put(-299.2,61.8){\scalebox{0.8}{\colorbox{white}{\textsf{\citetalias[][]{eBOSS2021}}}}}
\put(-299.2,50.8){\scalebox{0.8}{\colorbox{white}{\textsf{\citetalias[][]{eBOSS2021}}}}}
\put(-321,39.8){\scalebox{0.8}{\colorbox{white}{\textsf{\citetalias[][]{eBOSS2021}}}}}
\put(-305.6,155){\scalebox{0.8}{\colorbox{white}{\textsf{\citetalias[][]{Ross2015}}}}}
\put(-305.6,135.4){\scalebox{0.8}{\colorbox{white}{\textsf{\citetalias[][]{Ross2015}}}}}
\put(-331,127.4){\scalebox{0.8}{\colorbox{white}{\textsf{\citetalias[][]{eBOSS2021}}}}}
\put(-166.8,116.3){\scalebox{1}{\colorbox{gray!23.5}{\textsf{\citetalias[][]{Riess2019}}}}}
\end{picture}
\caption{Constraints on $\Omega_{\Lambda} h^2$, $\Omega_{\rm m} h^2$, $H_0$ and $\Omega_{\rm m}$ in a flat-$\Lambda$CDM model, obtained from the combination of BBN and different BAO measurements. Orange, red, and blue contours show results taken from \citetalias[][]{eBOSS2021}, for SDSS 2D BAO measurements in different redshift bins. Dashed and solid lines indicate the multi-tracer BAO measurements of galaxies and voids in this work, together with measurements from the other SDSS tracers, including MGS, QSOs, and Ly$\alpha$ forests. Grey bands refer to the ($68\,\%, 95\,\%$) $H_0$ measurements from the distance ladder method \citep[][]{Riess2019}.}
\label{fig:H0_Om}
\end{figure*}

The joint BBN and BAO constraints in a flat-$\Lambda$CDM cosmology are shown in Figure~\ref{fig:H0_Om}, for both `mtBAO' and \citetalias[][]{eBOSS2021} BAO in different redshift ranges.
In general, the results from `mtBAO' and \citetalias[][]{eBOSS2021} BAO are consistent.
The left panel shows that the $\Omega_\Lambda h^2$ constraint is dominated by the low redshift ($z < 1$) BAO measurements, while BAOs of tracers with effective redshifts above one contribute mainly to the $\Omega_{\rm m} h^2$ measurement.
As the result, with `mtBAO' the 1\,$\sigma$ error of $\Omega_\Lambda h^2$ is reduced by $\sim 17$ per cent, compared to the results with \citetalias[][]{eBOSS2021} BAO, while the uncertainties of the $\Omega_{\rm m} h^2$ parameter are similar for the two cases.
The same posterior distributions projected on the $H_0$--$\Omega_{\rm m}$ plane are shown in the right panel of Figure~\ref{fig:H0_Om}, with the marginalized results listed in Table~\ref{tab:cosmo_param}. Though the 1D `mtBAO' at $z < 1$ yields worse constraints for both parameters than the ones from the 2D \citetalias[][]{eBOSS2021} BAO, the overall 1\,$\sigma$ uncertainties of $H_0$ and $\Omega_{\rm m}$, with the QSOs and Ly$\alpha$ forests taken into account, are both reduced by around 6 per cent.

\begin{figure*}
\centering
\includegraphics[width=1.9\columnwidth]{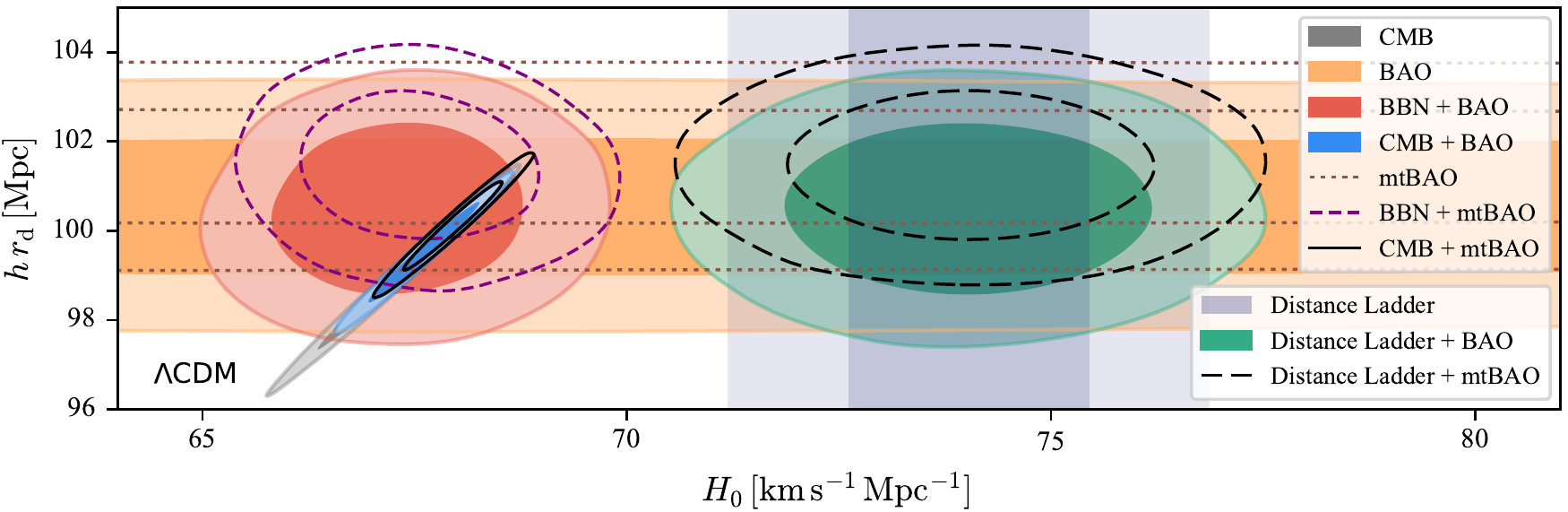}
\caption{Constraints on $(h\,r_{\rm d})$ and $H_0$ in a flat-$\Lambda$CDM model, for SDSS BAO measurements in \citetalias[][]{eBOSS2021} (`BAO'), and the ones with multi-tracer 1D BAO from LRGs and ELGs measured in this work (`mtBAO', see also Figure~\ref{fig:H0_Om}). The results are also combined with BBN, the Planck CMB temperature and polarization data, or the distance ladder measurements.}
\label{fig:rd_H0}
\end{figure*}

The $H_0$ value measured from BBN and BAO is consistent with the prediction of Planck CMB temperature and polarization data \citep[][]{Planck2020}, but significantly smaller than the local distance ladder measurement \citep[][]{Riess2019}. This is known as the `$H_0$ tension'.
To understand better the contributions of multi-tracer BAOs to the $H_0$ measurements, we further examine the constraints on $H_0$ and the sound horizon $r_{\rm d}$ from different probes, as well as their combinations. The 2D confidence intervals and marginalized statistics are shown in Figure~\ref{fig:rd_H0} and Table~\ref{tab:cosmo_param} respectively.
It can be seen that BAO only constrains $(h\,r_{\rm d})$, or $r_{\rm d}$ expressed in the unit of $\mpc$.
Thus, it does not improve the precision of $H_0$ measurement when combined with distance ladder results.
Nevertheless, the constraint on $(h\,r_{\rm d})$ can be translated to that of $H_0$, when $r_{\rm d}$ measurements are available.
This can be achieved with the CMB data, or through the $\omega_{\rm b}$ value constrained by BBN. Actually, the $\omega_{\rm b}$ measured from BBN is consistent with that from the Planck CMB data \citep[][]{Cooke2018}.
Therefore, they produce consistent $H_0$ constraints when combined with BAO.
Compared to the results with \citetalias[][]{eBOSS2021} BAO, the 1\,$\sigma$ uncertainty of $(h\,r_{\rm d})$ measured from `mtBAO' is around 12 pre cent smaller. This corresponds to 6 or 10 per cent reduction of the $H_0$ uncertainty, when combined with BBN or CMB data, respectively.

\begin{table*}
\centering
\begin{threeparttable}
\setlength{\tabcolsep}{\tabcolsep}
\caption{Marginalized best-fitting values and 68\,\% confidence limits of cosmological parameters in the flat-$\Lambda$CDM model and its two one-parameter extensions, obtained using different probes, including Planck CMB temperature and polarization data \citep[][]{Planck2020}, Pantheon SNe Ia sample \citep[][]{Scolnic2018}, and distance ladder measurements \citep[][]{Riess2019}. `BAO' denotes the SDSS BAO results used in \citetalias[][]{eBOSS2021}, where the BOSS/eBOSS LRG BAOs are measured in 2D. In contrast, `mtBAO` indicates BAO measurements from the same SDSS data, but with the BOSS/eBOSS LRG and ELG results replaced by the multi-tracer 1D BAO measurements in this work.
}
\begin{tabular}{clScScScScSc}
\toprule
Model & \multicolumn{1}{c}{Probe} & $H_0$\,(${\rm km}\,{\rm s}^{-1}\,{\rm Mpc}^{-1}$) & $\Omega_{\rm m}$ & $\Omega_\Lambda h^2$ & $\Omega_k$ & $w$ \\
\midrule
\multirowcell{7}[-2\aboverulesep]{$\Lambda$CDM} & BBN + BAO & $67.35 \pm 0.98$ & $0.299 \pm 0.016$ & $0.3179 \pm 0.0094$ & -- & -- \\
& BBN + mtBAO & $67.58 \pm 0.91$ & $0.290 \pm 0.015$ & $0.3241 \pm 0.0079$ & -- & -- \\
& CMB & $67.29 \pm 0.61$ & $0.3164 \pm 0.0084$ & $0.3096 \pm 0.0094$ & -- & -- \\
& CMB + BAO & $67.60 \pm 0.43$ & $0.3119 \pm 0.0058$ & $0.3145 \pm 0.0066$ & -- & -- \\
& CMB + mtBAO & $67.96 \pm 0.39$ & $0.3070 \pm 0.0051$ & $0.3201 \pm 0.0060$ & -- & -- \\
& Distance Ladder & $74.0 \pm 1.4$ & -- & -- & -- & -- \\
& Distance Ladder + BAO & $74.0 \pm 1.4$ & $0.299 \pm 0.016$ & $0.384 \pm 0.017$ & -- & -- \\
& Distance Ladder + mtBAO & $74.0 \pm 1.4$ & $0.290 \pm 0.015$ & $0.390 \pm 0.017$ & -- & -- \\
\midrule
\multirowcell{6}[-2\aboverulesep]{o$\Lambda$CDM} & BAO & -- & $0.285 \pm 0.023$ & $0.35_{-0.20}^{+0.11}$ & $0.079_{-0.10}^{+0.083}$ & -- \\
& mtBAO & -- & $0.284 \pm 0.022$ & $0.38_{-0.21}^{+0.12}$ & $0.04_{-0.12}^{+0.095}$ & -- \\
& CMB & $54.5 \pm 3.6$ & $0.483_{-0.069}^{+0.055}$ & $0.169_{-0.041}^{+0.030}$ & $-0.044_{-0.014}^{+0.019}$ & -- \\
& SN & -- & $0.317 \pm 0.071$ & -- & $-0.05 \pm 0.18$ & -- \\
& SN + BAO & -- & $0.289 \pm 0.021$ & $0.38_{-0.21}^{+0.11}$ & $0.037 \pm 0.069$ & -- \\
& SN + mtBAO & -- & $0.288 \pm 0.020$ & $0.39_{-0.21}^{+0.12}$ & $0.020 \pm 0.070$ & -- \\
\midrule
\multirowcell{6}[-2\aboverulesep]{$w$CDM} & BAO & -- & $0.271_{-0.017}^{+0.038}$ & $0.38_{-0.24}^{+0.11}$ & -- & $-0.69 \pm 0.15$ \\
& mtBAO & -- & $0.256_{-0.016}^{+0.048}$ & $0.39_{-0.25}^{+0.12}$ & -- & $-0.65 \pm 0.16$ \\
& CMB & -- & $0.199_{-0.057}^{+0.022}$ & $0.61_{-0.11}^{+0.23}$ & -- & $-1.58_{-0.35}^{+0.16}$ \\
& SN & -- & $0.309_{-0.056}^{+0.088}$ & -- & -- & $-1.07_{-0.18}^{+0.24}$ \\
& SN + BAO & -- & $0.292 \pm 0.015$ & $0.39_{-0.23}^{+0.11}$ & -- & $-0.950 \pm 0.061$ \\
& SN + mtBAO & -- & $0.286 \pm 0.014$ & $0.40_{-0.23}^{+0.12}$ & -- & $-0.943 \pm 0.060$ \\
\bottomrule
\end{tabular}
\label{tab:cosmo_param}
\end{threeparttable}
\end{table*}

\subsection{Curvature and dark energy}

\begin{figure}
\centering
\includegraphics[width=.98\columnwidth]{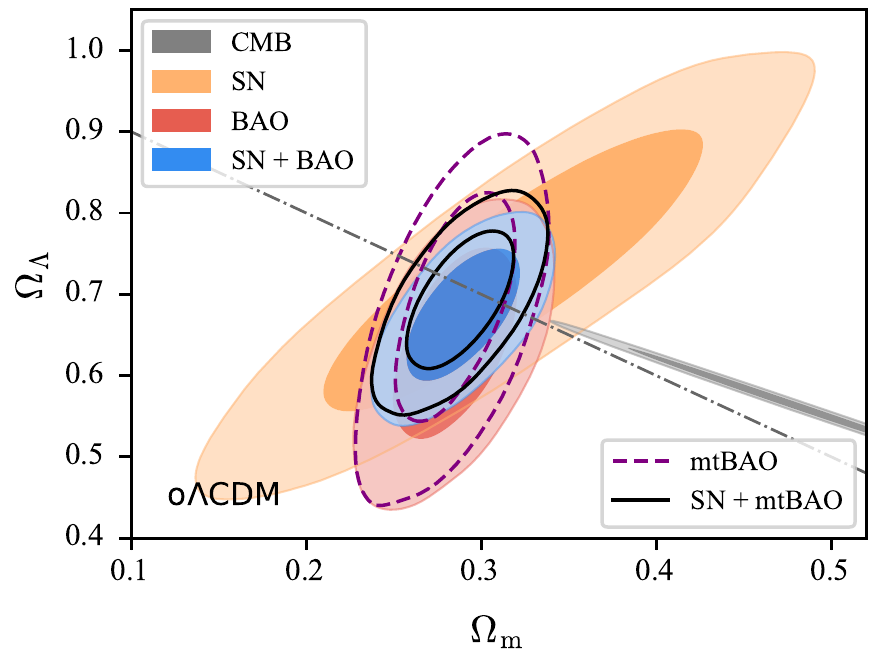}
\caption{Constraints on $\Omega_\Lambda$ and $\Omega_{\rm m}$ in a $\Lambda$CDM model with free curvature (o$\Lambda$CDM), for Planck CMB data, Pantheon SNe Ia sample, and the same SDSS BAO measurements as in Figure~\ref{fig:rd_H0}. The dash-dotted line indicates a flat-$\Lambda$CDM model.}
\label{fig:Ol_Om}
\end{figure}

\begin{figure}
\centering
\includegraphics[width=.98\columnwidth]{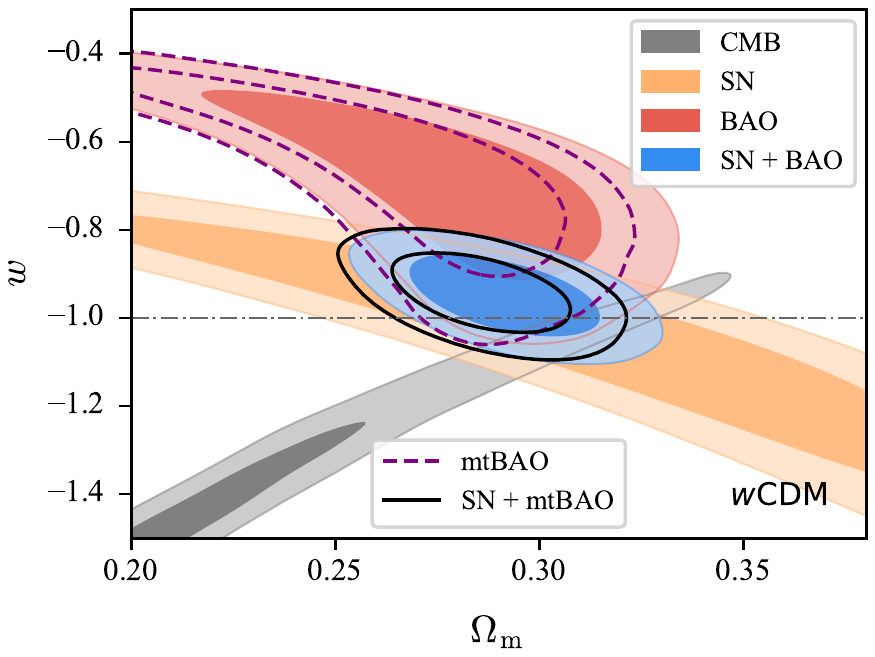}
\caption{Same as Figure~\ref{fig:Ol_Om}, but for $w$ and $\Omega_{\rm m}$ constraints in a flat-$w$CDM cosmology model.}
\label{fig:w_Om}
\end{figure}

We further examine the role of multi-tracer BAO measurements on two single-parameter extensions to the flat-$\Lambda$CDM model, i.e. o$\Lambda$CDM and $w$CDM, which are the cosmologies with free curvature and dark energy equation of state, respectively (see Section~\ref{sec:cosmo_par}).
Following \citetalias[][]{eBOSS2021}, we include also the Planck CMB temperature and polarization data \citep[][]{Planck2020} and the Pantheon SNe Ia sample \citep[][]{Scolnic2018}.
The results from single probes, as well as the combination of BAO and SNe Ia data, are shown in Figures~\ref{fig:Ol_Om} and \ref{fig:w_Om}, with the marginalized statistics listed in Table~\ref{tab:cosmo_param}.
Among these measurements, BAO provides the tightest single-probe constraints on the extension parameters $\Omega_k$ and $w$, which are also less coupled with $\Omega_{\rm m}$ compared to the results from CMB and SNe Ia data.
For both cosmological models, the posterior contours from BAOs are in around 2\,$\sigma$ tensions with those from CMB alone.
Therefore, we combine only the BAO and SNe Ia measurements, which are in better agreements.
With the combined constraints, a flat-$\Lambda$CDM model is favoured for both extended models.

For the BAO-only constraints on the $\Omega_k$ and $w$ parameters, the 1\,$\sigma$ errors from the multi-tracer 1D BAOs are both over 10 per cent larger than those from the \citetalias[][]{eBOSS2021} 2D BAOs.
It indicates that the anisotropic BAO measurements are crucial for the extended cosmological parameters.
Nevertheless, after combining with the SNe Ia data, the precisions of these parameters measured from `mtBAO' and \citetalias[][]{eBOSS2021} BAO are comparable.
In terms of the $\Omega_{\rm m}$ parameter, for both models the 1\,$\sigma$ uncertainties from `mtBAO' are $\sim 6$ per cent smaller the ones from \citetalias[][]{eBOSS2021} BAO, when the SNe Ia data are involved.

%%%%%%%%%%%%%%%%% NEW SECTION %%%%%%%%%%%%%%%%%%

\section{Conclusions}
\label{sec:conclusion}

We have performed a multi-tracer BAO analysis with different types of galaxies, as well as the corresponding cosmic voids identified by the \textsc{dive} algorithm \citep[][]{Zhao2016}, based on the SDSS BOSS DR12 and eBOSS DR16 data.
The datasets include over 1.6 million luminous red galaxies (LRGs) spanning in the redshift range of $0.2 < z < 1.0$ \citep[][]{Reid2016,Ross2020}, and 0.17 million star-forming emission line galaxies (ELGs) at $0.6 < z < 1.1$ \citep[][]{Raichoor2021}.
In addition, one thousand realizations of multi-tracer approximate mock catalogues are used for each galaxy sample \citep[][]{Kitaura2016,Zhao2021}, for method validations and covariance matrix estimations.
Since the footprints of the two types of galaxies overlap, we take into account also their cross correlation functions.

The void catalogues are constructed upon galaxy samples with the BAO reconstruction technique applied, which improves the BAO peak significance by suppressing nonlinear damping effects through the reverse of bulk flow motions.
To avoid contamination from voids-in-clouds which reside in overdensities \citep[][]{Liang2016,Zhao2016}, we select voids by the radii and use only large spheres for clustering analysis.
The radius thresholds are chosen to maximize the BAO signal-to-noise ratio measured from mocks.
As the result, the numbers of voids used in our studies are about twice the numbers of the corresponding galaxies.

To measure the BAO peak position with the contributions of both galaxies and voids, we have explored two different BAO fitting methods, including joint fits to galaxy--galaxy, galaxy--void, and void--void correlations with the same BAO dilation parameter $\alpha$, as well as combining individual correlation functions for the fit, with a negative weight applied to voids \citep[see][]{Zhao2020}. The optimal void weight is chosen to minimize the BAO fitting error for the mean combined 2PCFs of all mocks.
We find no significant differences on the BAO measurements from both methods for most cases, and the systematic errors estimated using mocks are generally small compared to the corresponding statistical uncertainties.
We then use the BAO fitting method with the combined 2PCFs for the SDSS data, due to the smaller length of the data vector, with which the covariance estimations are more accurate with a fixed number of mock realizations.

With the combined correlation functions of galaxies and voids, we observe on average around 10 per cent improvements on the precision of BAO measurements compared to the results from galaxies alone, for the mocks of all datasets.
Besides, for the majority of the individual mock realizations, improvements from the clustering of voids are detected.
It confirms that the clustering of underdensities encode additional information compared to that of galaxies alone.
Since the DT voids are defined by galaxies in a model-independent manner, the extra information is possibly exacted from higher-point clustering of the galaxy sample.

Interestingly, there are more mocks with improvements when the BAO uncertainty of the galaxy sample is smaller. This is consistent with findings in \citet[][]{Zhao2020}, and suggests that the contribution from voids may be more promising for future surveys with larger datasets, such as DESI.
In addition, cosmological fits in a flat-$\Lambda$CDM model recover well the $H_0$ and $\Omega_{\rm m}$ parameters of the mocks.
Both the cross correlation between different types of galaxies, and the clustering of voids, are found to improve the precision of the constraints. This confirms that a multi-tracer BAO analysis reduces cosmic variances.

For the SDSS data, the statistical uncertainties of BAO measurements are reduced by around 5 to 15 per cent for all the samples, when including the clustering statistics with voids.
The corresponding distance measurements are in good agreements with the BAO analyses of previous studies using the same galaxy samples \citep[e.g.][]{Alam2017,Bautista2021,Raichoor2021}.
Although only isotropic BAO measurements are considered in this work, the cosmological constraints can be better than those from the anisotropic BAO analysis with galaxies alone for certain parameters, especially when the results are combined with different datasets, such as the Planck CMB data and the Pantheon SNe Ia sample.
For instance, with the $D_{_{\rm V}} / r_{\rm d}$ values measured in this work, the uncertainties of the $H_0$, $\Omega_{\rm m}$, and $\Omega_\Lambda h^2$ parameters from a joint BBN and BAO analysis in the flat-$\Lambda$CDM cosmology, are reduced by around 6, 6, and 17 per cent respectively, compared to the results in \citetalias[][]{eBOSS2021}.
We foresee tighter cosmological constraints in the multi-tracer approach with anisotropic BAO measurements from cosmic voids, which may help in distinguishing different cosmological models. We leave relevant studies to a future work.

%%%%%%%%%%%%%%%%% NEW SECTION %%%%%%%%%%%%%%%%%%

\section*{Acknowledgements}

CZ, AV, DFS, AT, and JY acknowledge support from the Swiss National Science Foundation (SNF) grant 200020\_175751. FSK acknowledges the grants SEV-2015-0548, RYC2015-18693, and AYA2017-89891-P.
CT is supported by Tsinghua University and sino french CNRS-CAS international laboratories LIA Origins and FCPPL.
GR acknowledges support from the National Research Foundation of Korea (NRF) through Grants No. 2017R1E1A1A01077508 and No. 2020R1A2C1005655 funded by the Korean Ministry of Education, Science
and Technology (MoEST).

Clustering measurements and BAO fits for this work were performed at the National Energy Research Scientific Computing Center (NERSC)\footnote{\url{https://ror.org/05v3mvq14}}, a U.S. Department of Energy Office of Science User Facility operated under Contract No. DE-AC02-05CH11231.

Funding for the Sloan Digital Sky Survey IV has been provided by the Alfred P. Sloan Foundation, the U.S. Department of Energy Office of Science, and the Participating Institutions. SDSS-IV acknowledges support and resources from the Center for High-Performance Computing at the University of Utah. The SDSS web site is \url{www.sdss.org}.

SDSS-IV is managed by the Astrophysical Research Consortium for the Participating Institutions of the SDSS Collaboration including the 
Brazilian Participation Group, the Carnegie Institution for Science,
Carnegie Mellon University, the Chilean Participation Group,
the French Participation Group, Harvard-Smithsonian Center for Astrophysics, 
Instituto de Astrof\'isica de Canarias, The Johns Hopkins University,
Kavli Institute for the Physics and Mathematics of the Universe (IPMU) / University of Tokyo,
the Korean Participation Group, Lawrence Berkeley National Laboratory, 
Leibniz Institut f\"ur Astrophysik Potsdam (AIP),  
Max-Planck-Institut f\"ur Astronomie (MPIA Heidelberg), 
Max-Planck-Institut f\"ur Astrophysik (MPA Garching), 
Max-Planck-Institut f\"ur Extraterrestrische Physik (MPE), 
National Astronomical Observatories of China, New Mexico State University, 
New York University, University of Notre Dame, 
Observat\'ario Nacional / MCTI, The Ohio State University, 
Pennsylvania State University, Shanghai Astronomical Observatory, 
United Kingdom Participation Group,
Universidad Nacional Aut\'onoma de M\'exico, University of Arizona, 
University of Colorado Boulder, University of Oxford, University of Portsmouth, 
University of Utah, University of Virginia, University of Washington, University of Wisconsin, 
Vanderbilt University, and Yale University.

\section*{Data availability}

The SDSS galaxy data and the corresponding approximate mock catalogues used in this work can be downloaded from the SDSS Science Archive Server (see Section~\ref{sec:data} for details).
All software packages for the analysis, including the void finder, BAO fitter, and cosmological parameter sampler are all publicly available (see Section~\ref{sec:method}).
In addition, the void catalogues can be obtained directly upon request to CZ.

%%%%%%%%%%%%%%%%%%%%%%%%%%%%%%%%%%%%%%%%%%%%%%%%%%



\bibliographystyle{mnras}
\bibliography{SDSS_VoidBAO} % if your bibtex file is called example.bib

% \begin{thebibliography}{99}
% \bibitem[\protect\citeauthoryear{Author}{2012}]{Author2012}
% Author A.~N., 2013, Journal of Improbable Astronomy, 1, 1
% \bibitem[\protect\citeauthoryear{Others}{2013}]{Others2013}
% Others S., 2012, Journal of Interesting Stuff, 17, 198
% \end{thebibliography}

%%%%%%%%%%%%%%%%%%%%%%%%%%%%%%%%%%%%%%%%%%%%%%%%%%

%%%%%%%%%%%%%%%%% APPENDICES %%%%%%%%%%%%%%%%%%%%%

% If you want to present additional material which would interrupt the flow of the main paper,
% it can be placed in an Appendix which appears after the list of references.

\appendix

%%%%%%%%%%%%%%%%% NEW SECTION %%%%%%%%%%%%%%%%%%

\section{Covariances of clustering measurements}
\label{sec:cov_xi}

\begin{figure*}
\centering
\includegraphics[width=.9\textwidth]{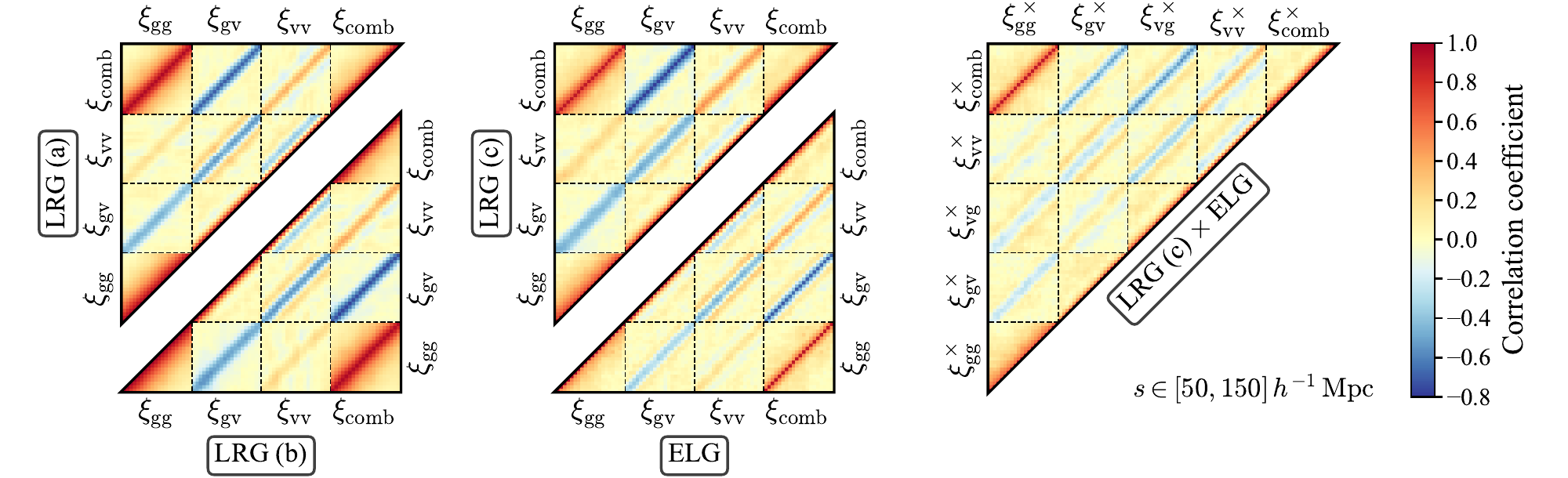}
\caption{Correlation coefficients of galaxy--galaxy, galaxy--void, void--void, and the combined correlation functions in $[50, 150]\,h^{-1}\,{\rm Mpc}$, measured from all mock realizations for each sample. Here, the combined correlation functions are computed with the optimal void weights found in Section~\ref{sec:optimal_weight}.}
\label{fig:xi_corr}
\end{figure*}

Since voids are indirect tracers resolved from galaxy samples, the distributions of voids and galaxies are highly correlated.
To quantify the correlations of their clustering measurements, we measure the correlation matrices of galaxy--galaxy ($\xi_{\rm gg}$), galaxy--void ($\xi_{\rm gv}$), and void-void ($\xi_{\rm vv}$) correlations from the 1000 mock realizations for each SDSS sample, and plot the results in Figure~\ref{fig:xi_corr}.
Moreover, since the pair counts for these 2PCFs are used for the evaluations of combined correlation functions of galaxies and voids ($\xi_{\rm comb}$; see Section~\ref{sec:cf_estimator}), the correlation coefficients involving $\xi_{\rm comb}$, which are computed with the optimal void weights found in Section~\ref{sec:optimal_weight}, are also shown.
In particular, we present results only in the separation range of $[50,150]\,h^{-1}\,{\rm Mpc}$, which is close to the fiducial BAO fitting range used in this work (see Section~\ref{sec:fit_range}).

As expected, the cross correlation coefficients between different clustering measurements are generally significant.
A typical diagonal correlation coefficient between $\xi_{\rm gg}$ and $\xi_{\rm gv}$ is $-0.4$. The negative value is due to the negative void biases \citep[][]{Zhao2016,Chuang2017}.
For the cross coefficients between $\xi_{\rm gg}$ and $\xi_{\rm vv}$, the diagonal values are usually around $0.2$.
In addition, $\xi_{\rm comb}$ is strongly correlated with the individual 2PCFs, as its diagonal coefficients of the cross covariances with $\xi_{\rm gg}$, $\xi_{\rm gv}$, and $\xi_{\rm vv}$ are roughly 0.9, 0.7, and 0.4 respectively.
These figures suggest that $\xi_{\rm gg}$ dominates $\xi_{\rm comb}$, which is consistent with the small amplitude of the optimal void weights.

\begin{figure}
\centering
\includegraphics[width=.98\columnwidth]{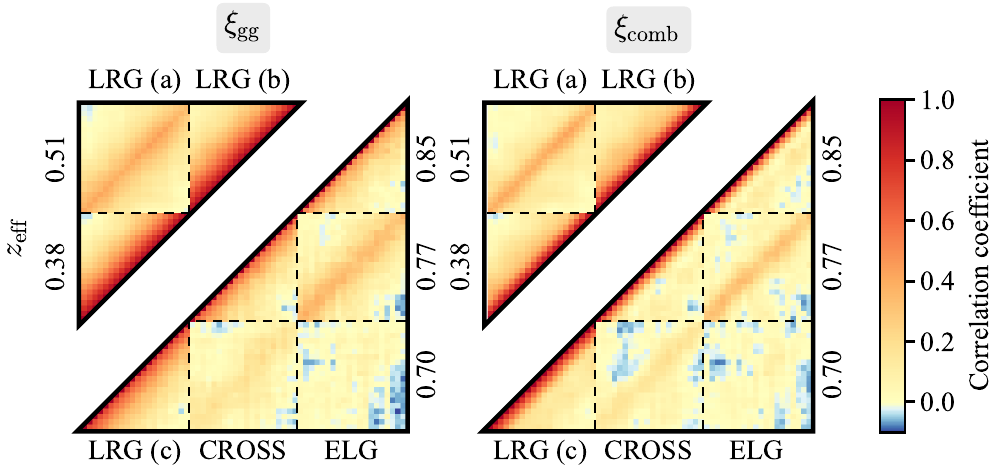}
\caption{Correlation coefficients of correlation function measurements in $[50, 150]\,h^{-1}\,{\rm Mpc}$, from different mock samples, for both galaxy 2PCFs ($\xi_{\rm gg}$) and combined correlations with the optimal void weights ($\xi_{\rm comb}$). `CROSS' denotes the cross correlations between `LRG (c)' and `ELG'.}
\label{fig:sample_xi_corr}
\end{figure}

We further examine the cross covariances between clustering measurements of different samples.
In particular, we consider the cross correlations between `LRG (a)' and `LRG (b)', as they share the same galaxies and voids due to the overlap of their redshift ranges.
Moreover, the cross covariances for `LRG (c)', `ELG', and their cross correlations are taken into account as well.
The results for $\xi_{\rm gg}$ and $\xi_{\rm comb}$ are shown in Figure~\ref{fig:sample_xi_corr}, where $\xi_{\rm comb}$ is again computed with the optimal weights found in Section~\ref{sec:optimal_weight}.
It can be seen that the correlation coefficients between `LRG (a)' and `LRG (b)' are relatively large.
Besides, clustering statistics of the `ELG' sample are strongly correlated with the cross correlations between `LRG (c)' and `ELG'.
This can be explained by the fact that a large fraction of the `ELG' sample is involved for the cross correlations, while many of the LRGs are too far away from the ELG footprints for correlations on BAO scale (see Figure~\ref{fig:footprint}).

%%%%%%%%%%%%%%%%% NEW SECTION %%%%%%%%%%%%%%%%%%

\section{Least squares fitting and chi-squared evaluation based on QR decomposition}
\label{sec:chi2_optimize}

Given a data vector $\boldsymbol{y}$, the best-fitting parameters of the the model
\begin{equation}
f(\boldsymbol{x}) = [ {\bf X} (\boldsymbol{x}) ]^{\sf T} \boldsymbol{a}
\end{equation}
can be solved linearly using the least squares method. Here, $\boldsymbol{a}$ indicates the free parameters to be fitted, and ${\bf X} (\boldsymbol{x})$ denotes a set of arbitrary fixed functions of $\boldsymbol{x}$.
This is because the parameters that minimize the chi-squared function for model $f$,
\begin{equation}
\chi_f^2 = ( \boldsymbol{y} - {\bf X} \boldsymbol{a} )^{\sf T} {\bf C}^{-1} ( \boldsymbol{y} - {\bf X} \boldsymbol{a} ) ,
\end{equation}
are the solution of the following linear equation \citep[e.g.][]{Press2007}
\begin{equation}
( {\bf X}^{\sf T} {\bf C}^{-1} {\bf X} ) \cdot \boldsymbol{a} = {\bf X}^{\sf T} {\bf C}^{-1} \boldsymbol{y} ,
\label{eq:least_square}
\end{equation}
where ${\bf C}$ denotes the covariance matrix.
Direct solution of this equation, as well as the inverse of the covariance matrix, can be numerically unstable. Moreover, since the chi-squared function (Eq.~\eqref{eq:baofit_chi2}) is typically evaluated again and again, with different parameters during a fitting process, solving Eq.~\eqref{eq:least_square} directly can be computational expensive in practice, especially when the dimension of $\boldsymbol{a}$ is high. 

If the covariance matrix is obtained in the form of ${\bf C} = {\bf M}^{\sf T} {\bf M}$, however, the problem can be largely simplified.
This is actually a common occurrence, as covariance matrices estimated using mocks and jackknife resampling can both be expressed in this way. The constant factors, such as the Hartlap correction (Eq.~\eqref{eq:hartlap}), can be seen as being absorbed by the ${\bf M}$ term here. This covariance generation matrix ${\bf M}$, can be QR-decomposed as
\begin{equation}
{\bf M} = {\bf Q R} ,
\end{equation}
where ${\bf Q}$ is an orthogonal matrix (${\bf Q}^{\sf T} {\bf Q} = {\bf I}$, with ${\bf I}$ being the identity matrix), and ${\bf R}$ is an upper triangular matrix. Thus,
\begin{equation}
{\bf C}^{-1} = ({\bf R}^{\sf T} {\bf Q}^{\sf T} {\bf Q R})^{-1} = {\bf R}^{-1} {\bf R}^{- {\sf T}} .
\label{eq:inverse_cov_qr}
\end{equation}
Eq.~\eqref{eq:least_square} can then be written as
\begin{equation}
( {\bf R}^{- {\sf T}} {\bf X} )^{\sf T} ( {\bf R}^{- {\sf T}} {\bf X} ) \cdot \boldsymbol{a} = ( {\bf R}^{- {\sf T}} {\bf X} )^{\sf T} {\bf R}^{- {\sf T}} \boldsymbol{y}.
\label{eq:least_square2}
\end{equation}
Let ${\bf Y} = {\bf R}^{- {\sf T}} {\bf X}$, it can be further QR-decomposed as ${\bf Y} = {\bf Q}' {\bf U}$, 
where ${\bf U}$ is another upper triangular matrix.
Then, Eq.~\eqref{eq:least_square2} can be simplified as
\begin{equation}
( {\bf U}^{\sf T} {\bf U} ) \cdot \boldsymbol{a} = {\bf Y}^{\sf T} {\bf R}^{- {\sf T}} \boldsymbol{y}.
\end{equation}
Let ${\bf Z} = {\bf U}^{-{\sf T}} {\bf Y}^{\sf T} {\bf R}^{- {\sf T}}$, we have finally
\begin{equation}
{\bf U} \cdot \boldsymbol{a} = {\bf Z} \boldsymbol{y} .
\end{equation}

In practice, since ${\bf M}$ and ${\bf X}$ are known, the matrices ${\bf U}$ and ${\bf Z}$ can be pre-computed. Typically, $\boldsymbol{y}$ is the fitting residual with the non-nuisance parameters. In this case, during the fitting procedure, one merely needs a matrix-vector production $\boldsymbol{b} = {\bf Z} \boldsymbol{y}$, and solving the equation
\begin{equation}
{\bf U} \cdot \boldsymbol{a} = \boldsymbol{b} .
\end{equation}
Since ${\bf U}$ is an upper triangular matrix, this equation can be solved efficiently with a backward substitution.
It is worth noting that ${\bf Y}$ and ${\bf Z}$ can be evaluated using forward and backward substitutions as well, in a column-by-column manner, to avoid the inverse of upper and lower triangular matrices.
Therefore, the whole least squares fitting process involves only QR decomposition, matrix production, as well as forward and backward substitutions. In particular, the two QR decompositions can both be computed in-place, as we do not use the orthogonal matrices explicitly. Thus, the algorithm here is highly numerically stable and efficient.

Furthermore, given the inverse covariance matrix expressed by Eq.~\eqref{eq:inverse_cov_qr}, the chi-squared function (Eq.~\eqref{eq:baofit_chi2}) becomes
\begin{equation}
\chi^2 = \boldsymbol{d}^{\sf T} {\bf R}^{-1} {\bf R}^{- {\sf T}} \boldsymbol{d}
= \lVert {\bf R}^{- {\sf T}} \boldsymbol{d} \rVert^2 ,
\end{equation}
which can be computed stably as well, by performing a forward substitution and taking the norm. Therefore, we benefit from QR decomposition even if the least squares method is not involved.

%%%%%%%%%%%%%%%%% NEW SECTION %%%%%%%%%%%%%%%%%%

\section{BAO measurements of mocks with different fitting ranges}
\label{sec:result_fit_range}

We perform BAO fits to the mean clustering measurements of the approximate mocks to check how fitting ranges affect BAO measurements.
The corresponding covariance matrices are all rescaled by $1/1000$, for better examinations on the performance of the models, as the parameters are generally well constrained with the rescaling.
Thus, the priors of the parameters for this study are all flat in ranges that are sufficiently large.
The bias of the $\alpha$ parameter, is then defined as
\begin{equation}
\delta \alpha = \frac{\alpha_{\rm fit, med} - \alpha_{\rm exp}}{ \sqrt{N_{\rm m}}\, \sigma_\alpha } ,
\label{eq:alpha_bias}
\end{equation}
where $N_{\rm m} = 1000$, $\alpha_{\rm fit, med}$ indicates the median value drawn from the fitted $\alpha$ posterior distribution, and $\alpha_{\rm exp}$ denotes the expected $\alpha$ value measured from the mocks, as is shown in Table~\ref{tab:fit_range}.

\begin{figure*}
\centering
\includegraphics[width=.9\textwidth]{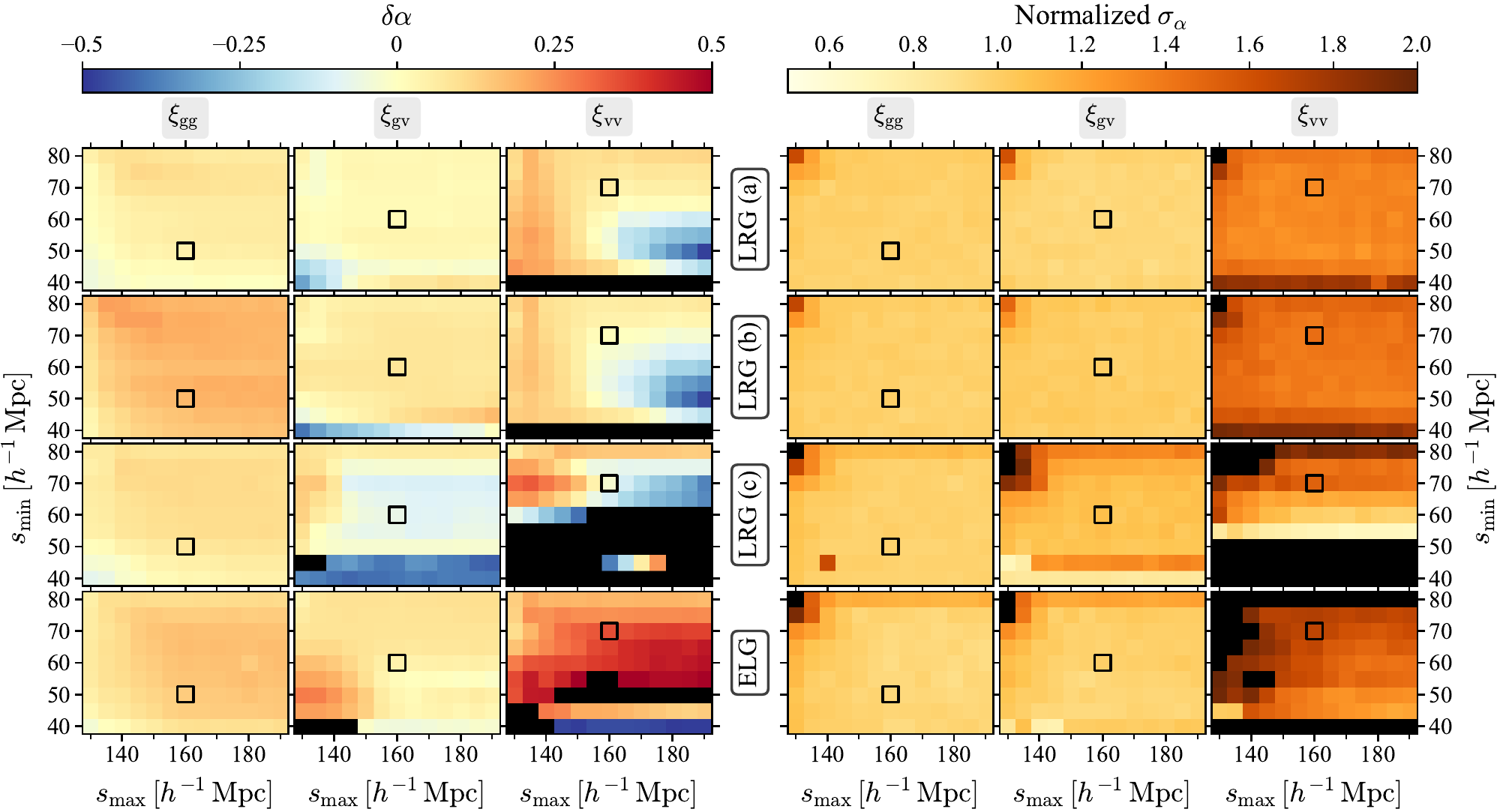}
\caption{The biases (Eq.~\eqref{eq:alpha_bias}) and fitted errors of $\alpha$, from fits to the mean clustering statistics of 1000 mock realizations. In particular, for illustration purposes, the errors on $\alpha$ in each row are normalized with the median $\sigma_{\alpha}$ from the corresponding galaxy auto correlation function ($\xi_{\rm gg}$) over all fitting ranges. Black regions indicate results outside the ranges of the colour maps, and open squares denote the fiducial fitting ranges used in this work.}
\label{fig:fit_range_main}
\end{figure*}

\begin{figure}
\centering
\includegraphics[width=.98\columnwidth]{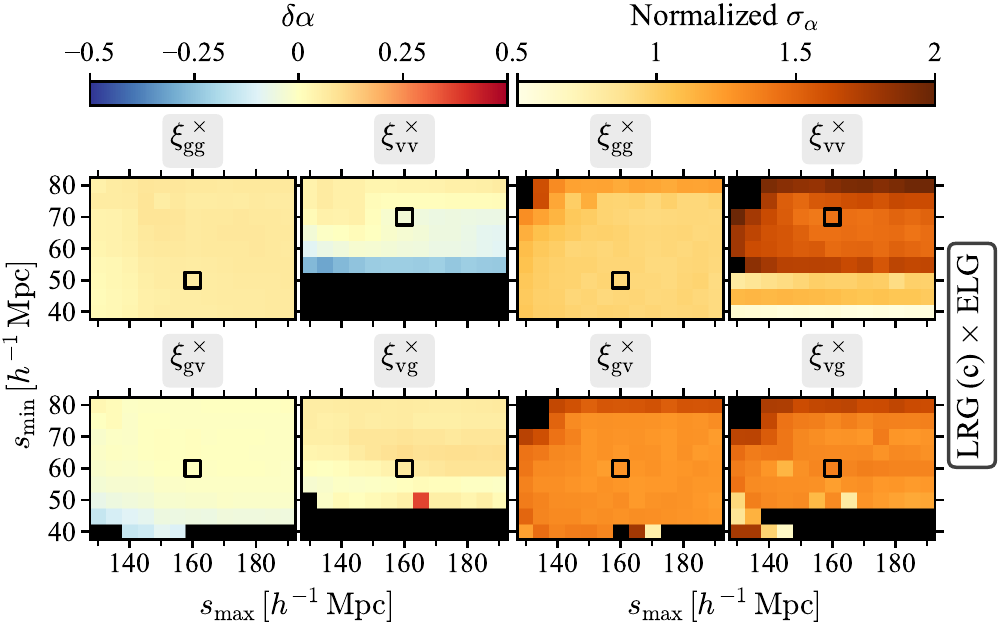}
\caption{Similar to Figure~\ref{fig:fit_range_main}, but the results are for the cross correlations between `LRG (c)' and `ELG' samples. Here, $\xi_{\rm gg}^\times$, $\xi_{\rm gv}^\times$, $\xi_{\rm vg}^\times$, and $\xi_{\rm vv}^\times$, are for the cross correlations between LRGs and ELGs, LRGs and ELG voids, LRG voids and ELGs, as well as LRG voids and ELG voids, respectively. The errors on $\alpha$ are all normalized by the median $\sigma_\alpha$ of $\xi_{\rm gg}^\times$, over all fitting ranges. The fiducial fitting ranges are indicated by open squares.}
\label{fig:fit_range_cross}
\end{figure}

We vary the lower and upper bounds of the fitting range in $[40, 80]$ and $[130, 190]\,\mpc$ respectively, both with the step size of $5\,\mpc$. Using models listed in Table~\ref{tab:fit_setting}, the resulting biases and fitted errors of $\alpha$, from galaxy--galaxy ($\xi_{\rm gg}$), galaxy--void ($\xi_{\rm gv}$), and void--void ($\xi_{\rm vv}$) correlations of different samples, are shown in Figure~\ref{fig:fit_range_main}.
In general, results for $\xi_{\rm gg}$ and $\xi_{\rm gv}$ are stable for a large set of fitting ranges, and the measured $\alpha$ are not significantly biased.
However, BAO measurements from $\xi_{\rm vv}$ are more sensitive to the fitting ranges, especially for the lower bound ($s_{\rm min}$). This is possibly because of the strong void exclusion patterns on small scales (see e.g. Figure~\ref{fig:xi_main}).
Similar trends are found for the cross correlations between the `LRG (c)' and `ELG' samples, as is shown in Figure~\ref{fig:fit_range_cross}.
Based on these results, the fiducial fitting ranges we choose are $[50, 160]$, $[60, 160]$, and $[70, 160]\,\mpc$, for $\xi_{\rm gg}$, $\xi_{\rm gv}$, and $\xi_{\rm vv}$ respectively, for all the samples.
In order to investigate whether the sensitivity of $\xi_{\rm vv}$ results to the fitting range bias the multi-tracer BAO measurements, we analyse two sets of stacked data vectors in Section~\ref{sec:mock_test}, with or without $\xi_{\rm vv}$ included.

\begin{figure}
\centering
\includegraphics[width=.98\columnwidth]{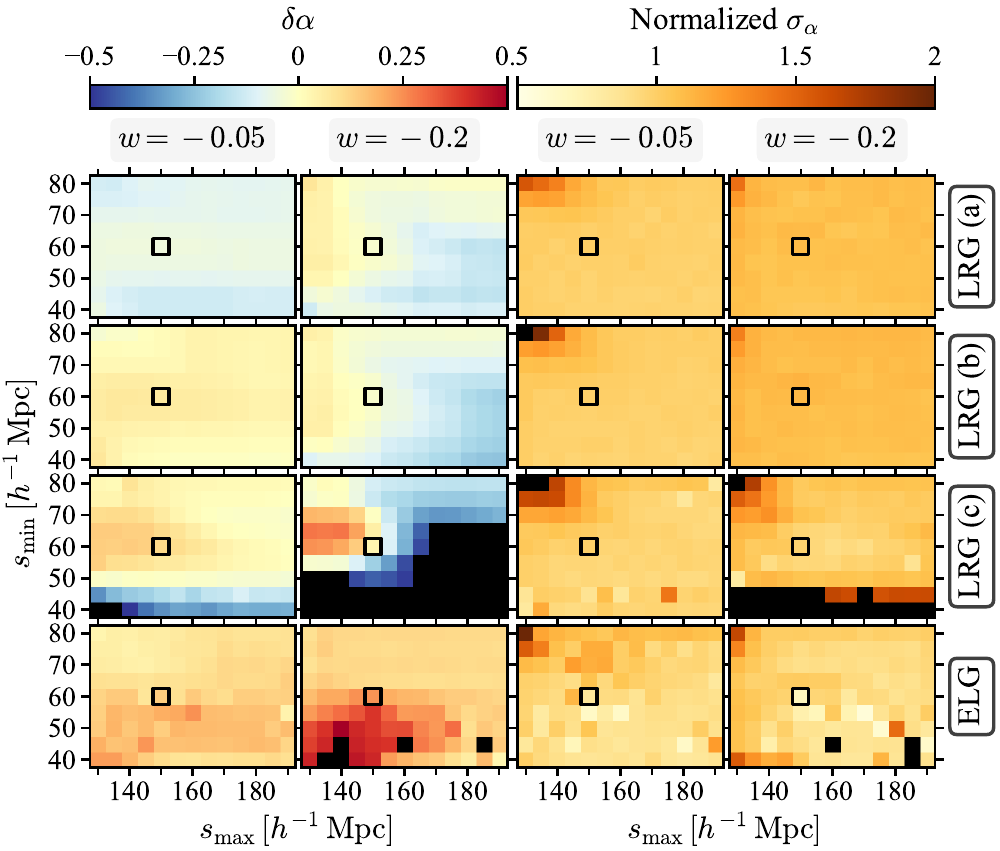}
\caption{Similar to Figure~\ref{fig:fit_range_main}, but the results are for the combined correlation functions evaluated using galaxy--galaxy, galaxy--void, and void--void pair counts, with two different weights applied to voids, i.e. $-0.05$ and $-0.2$. The fitted errors are normalized by the median $\sigma_\alpha$ with $w=-0.05$, for each sample. Open squares show our fiducial fitting ranges.}
\label{fig:fit_range_weight}
\end{figure}

For the multi-tracer approach with the combined correlation functions, we present the results with different fitting ranges and void weights in Figure~\ref{fig:fit_range_weight}.
The two weights are chosen to be close to the boundaries of the weight range studied in this work, i.e. $[-0.2, 0]$.
We find that the fitting results are more sensitive to the fitting ranges when the absolute weight value gets larger, especially for the `LRG (c)' sample with small $s_{\rm min}$.
This can be explained by the fact that the combined 2PCF becomes closer to the void auto correlation, and the BAO model expressed in Eq.~\eqref{eq:bao_model_para} is not able to describe the broad-band void exclusion pattern precisely.
Moreover, the void exclusion effect is more significant for `LRG (c)' on the scales of interest, due to the larger void radii than the other samples.
We then choose a fitting range of $[60,150]\,\mpc$ for all the samples, with which the fitted results are not significantly biased.

%%%%%%%%%%%%%%%%% NEW SECTION %%%%%%%%%%%%%%%%%%

\section{BAO measurements of mocks with different parameter priors}
\label{sec:result_fit_prior}

We explore in this section impacts of parameter priors on the fitted BAO scale $\alpha$, as well as its uncertainty inferred from the sampled posterior. This is done by performing BAO fits to all individual mock realizations with different parameter priors.
We examine the distribution of the best-fitting (median) values of $\alpha$ from individual fits ($\alpha_{\rm fit, med}$), and measure the median value $\langle \alpha \rangle$, as well as the $1\,\sigma$ dispersion $\alpha_{1\sigma}$, which is defined as half the width between the 16th and 84th percentiles of the cumulative distribution of $\alpha_{\rm fit, med}$.
This dispersion is then compared to the median value of the fitted error, $\langle\sigma_\alpha\rangle$, obtained from individual mocks.
The relative difference between these two quantities is defined as
\begin{equation}
\delta \sigma = \frac{\langle\sigma_{\alpha}\rangle - \alpha_{1\sigma}}{\alpha_{1\sigma}} .
\end{equation}
In addition, the pull statistic $g(\alpha)$, measured following Eq.~\eqref{eq:pull_alpha}, is also used for the prior selection.

For a given parameter $p$, we define its prior $p_{\rm prior}$ based on the best-fitting value ($\hat{p}$) and fitted error ($\hat{\sigma}_{p}$), obtained from fits to the mean 2PCFs of all mocks, with the covariance matrices rescaled by $1/1000$.
In particular, priors for fitting $\xi_{\rm gg}$, $\xi_{\rm gv}$, and $\xi_{\rm vv}$ of individual mocks, are all inferred from the joint fitting results with the data vector $\{ \boldsymbol{\xi}_{\rm gg}, \boldsymbol{\xi}_{\rm gv}, \boldsymbol{\xi}_{\rm vv} \}$, where the clustering statistics are the mean results from 1000 mock realizations.
Note also that $\hat{\sigma}_{p}$ is not rescaled here, so it is typically very small.
The following priors are then tested for different parameters:
\begin{enumerate}[label=(\roman*), leftmargin=*, align=left]
\item
flat prior with a range that is considered sufficiently large;
\item
flat prior $\mathcal{F} (\hat{p} \pm n \hat{\sigma}_{p} )$, where $\mathcal{F} (p_{\rm cen} \pm p_{\rm width})$ denotes a flat prior between $(p_{\rm cen} - p_{\rm width})$ and $(p_{\rm cen} + p_{\rm width})$, and $n$ is a fixed number;
\item
flat prior $\mathcal{U} (\hat{p} + n \hat{\sigma}_{p})$, where $\mathcal{U} (p_{\rm max})$ indicates a flat prior between 0 and $p_{\rm max}$;
\item
Gaussian prior $\mathcal{N} (\hat{p}, n \hat{\sigma}_{p})$, where $\mathcal{N} (p_{\rm cen}, p_{\rm sig})$ denotes a normal distribution centred at $p_{\rm cen}$, and with the standard deviation of $p_{\rm sig}$ ;
\item
fixed value of $\hat{p}$.
\end{enumerate}

In particular, we fix always the $c$ parameter in Eq.~\eqref{eq:bao_model_para} for fitting the combined 2PCFs.
The $(1 + c k^2)$ term is then equivalent to a template `non-wiggle' power spectrum. By determining the value of the $c$ parameter from the mean 2PCFs of the mocks, we essentially generate approximate templates for the combined 2PCFs with different void weights.
\citet[][]{Variu2021} show that this treatment enhances the robustness of the BAO measurement, without introducing significant systematic bias, compared to the result with a free $c$ parameter.
Nevertheless, we account for the potential systematic bias by varying the fixed $c$ value (see Section~\ref{sec:fit_prior}).

% as it is generally highly degenerate with the $B$ parameter \citep[see][]{Zhao2020}.
%\citet[][]{Variu2021} show that the fitted error on $\alpha$ from $\xi_{\rm vv}$ may be overestimated with a fixed $c$ parameter, while for $\xi_{\rm gv}$ the errors are reliable. This is not a problem for our study, as the contribution of $\xi_{\rm gv}$ to the combined correlation function is more significant than that of $\xi_{\rm vv}$ given the small void weights, and we check specifically the error estimates against the dispersion of best-fitting $\alpha$ from individual mocks.

\begin{table*}
\centering
\begin{threeparttable}
\newlength{\myaboverulesep}
\newlength{\mybelowrulesep}
\setlength\myaboverulesep{.7\aboverulesep}
\setlength\mybelowrulesep{.7\belowrulesep}
\setlength\aboverulesep{0.12ex}
\setlength\belowrulesep{0.12ex}
\caption{Fitting results of individual mocks with different priors. Numbers in the `clustering' column denote fitted errors of the mean 2PCFs of mocks, which are rescaled by $\sqrt{1000}$. `--' indicate sufficiently wide flat priors, and $\mathcal{F}$, $\mathcal{U}$ and $\mathcal{N}$ denotes different prior types detailed in Appendix~\ref{sec:result_fit_prior}.
$\langle\alpha\rangle$ and $\alpha_{1\sigma}$ are the median value and $1\,\sigma$ dispersion for the best-fitting $\alpha$ distribution of all mocks. $\langle \sigma_{\alpha}\rangle$ and $\langle \chi^2 \rangle$ denote the median value of $\sigma_{\alpha}$ and best-fitting $\chi^2$ of the fits.
$\delta \sigma$ is the relative difference between $\alpha_{1\,\sigma}$ and $\langle \sigma_\alpha \rangle$.
$\bar{g}(\alpha)$ and $\sigma_{g(\alpha)}$ indicate the mean value and standard deviation of the pull quantity.
Fiducial priors and the ones used for assessing systematic errors are indicated by `$\blacklozenge$' and `$\lozenge$' respectively.}
\begin{tabular}{ccccccccccc@{\hspace{3pt}}c@{\hspace{3pt}}}
\specialrule{\heavyrulewidth}{0pt}{\mybelowrulesep}
Sample & Clustering & $B_{\rm prior}$ & $\Sigma_{\rm nl, prior}$ & $\langle\alpha\rangle$ & $\alpha_{1\sigma}$ & $\langle\sigma_{\alpha}\rangle$ & $\delta \sigma$ & $\bar{g}(\alpha)$ & $\sigma_{g(\alpha)}$ & \multicolumn{1}{c}{$\langle\chi^2\rangle / {\rm d.o.f.}$} & \\
\specialrule{\lightrulewidth}{\myaboverulesep}{\mybelowrulesep}
%%%%% LRG (a) xi_gg %%%%%
\multirowcell{9}[-2\myaboverulesep]{MD-Patchy\\LRG (a)} & \multirowcell{3}{$\xi_{\rm gg}$\\(0.0125)} & -- & -- & 1.001 & 0.012 & 0.013 & $9.2\,\%$ & $-0.04$ & $0.97$ & $15.4/16=0.96$ \\
& & $\mathcal{F}(\hat{B}\pm10\hat{\sigma}_{_{\rm B}})$ & $\hat{\Sigma}_{\rm nl}$ & {\bf 1.000} & {\bf 0.012} & {\bf 0.012} & ${\bf 3.7\,\%}$ & ${\bf -0.02}$ & ${\bf 0.99}$ & ${\bf 16.6/17=0.98}$ & $\blacklozenge$ \\
& & $\mathcal{F}(\hat{B}\pm50\hat{\sigma}_{_{\rm B}})$ & $\mathcal{U}(\hat{\Sigma}_{\rm nl}+20\hat{\sigma}_{_{\Sigma_{\rm nl}}})$ & 0.999 & 0.012 & 0.012 & $0.7\,\%$ & $-0.03$ & $1.02$ & $15.5/16=0.97$ & $\lozenge$ \\
\cmidrule{2-12} %%%%% LRG (a) xi_gv %%%%%
& \multirowcell{3}{$\xi_{\rm gv}$\\(0.0120)} & -- & -- & 1.000 & 0.012 & 0.013 & $3.1\,\%$ & $-0.00$ & $0.97$ & $13.4/14=0.95$ & \\
& & $\mathcal{F}(\hat{B}\pm10\hat{\sigma}_{_{\rm B}})$ & $\hat{\Sigma}_{\rm nl}$ & {\bf 1.000} & {\bf 0.012} & {\bf 0.012} & ${\bf 0.8\,\%}$ & ${\bf -0.01}$ & ${\bf 0.99}$ & ${\bf 14.5/15=0.97}$ & $\blacklozenge$ \\
& & $\mathcal{F}(\hat{B}\pm50\hat{\sigma}_{_{\rm B}})$ & $\mathcal{U}(\hat{\Sigma}_{\rm nl}+20\hat{\sigma}_{_{\Sigma_{\rm nl}}})$ & 0.999 & 0.012 & 0.012 & $1.3\,\%$ & $-0.01$ & $1.01$ & $13.4/14=0.96$ & $\lozenge$ \\
\cmidrule{2-12} %%%%% LRG (a) xi_vv %%%%%
& \multirowcell{3}{$\xi_{\rm vv}$\\(0.0169)}  & -- & -- & 1.018 & 0.053 & 0.056 & $5.1\,\%$ & $-0.11$ & $4.42$ & $11.2/12=0.94$ & \\
& & $\mathcal{F}(\hat{B}\pm10\hat{\sigma}_{_{\rm B}})$ & $\hat{\Sigma}_{\rm nl}$ & {\bf 1.000} & {\bf 0.017} & {\bf 0.017} & ${\bf -1.1\,\%}$ & ${\bf -0.02}$ & ${\bf 1.01}$ & ${\bf 12.8/13=0.98}$ & $\blacklozenge$ \\
& & $\mathcal{F}(\hat{B}\pm50\hat{\sigma}_{_{\rm B}})$ & $\mathcal{U}(\hat{\Sigma}_{\rm nl}+20\hat{\sigma}_{_{\Sigma_{\rm nl}}})$ & 1.000 & 0.017 & 0.017 & $-2.2\,\%$ & $-0.02$ & $1.01$ & $11.6/12=0.97$ & $\lozenge$ \\
\specialrule{\lightrulewidth}{\myaboverulesep}{\mybelowrulesep}
%%%%% LRG (b) xi_gg %%%%%
\multirowcell{9}[-2\myaboverulesep]{MD-Patchy\\LRG (b)} & \multirowcell{3}{$\xi_{\rm gg}$\\(0.0115)} & -- & -- & 1.003 & 0.011 & 0.012 & $3.2\,\%$ & $-0.03$ & $0.98$ & $15.3/16=0.96$ & \\
& & $\mathcal{F}(\hat{B}\pm20\hat{\sigma}_{_{\rm B}})$ & $\hat{\Sigma}_{\rm nl}$ & {\bf 1.002} & {\bf 0.011} & {\bf 0.011} & ${\bf 1.0\,\%}$ & ${\bf -0.02}$ & ${\bf 0.99}$ & ${\bf 16.3/17=0.96}$ & $\blacklozenge$ \\
& & $\mathcal{F}(\hat{B}\pm20\hat{\sigma}_{_{\rm B}})$ & $\mathcal{U}(\hat{\Sigma}_{\rm nl}+20\hat{\sigma}_{_{\Sigma_{\rm nl}}})$ & 1.001 & 0.011 & 0.011 & $-0.9\,\%$ & $-0.02$ & $1.02$ & $15.6/16=0.98$ & $\lozenge$ \\
\cmidrule{2-12} %%%%% LRG (b) xi_gv %%%%%
& \multirowcell{3}{$\xi_{\rm gv}$\\(0.0115)} & -- & -- & 1.001 & 0.012 & 0.012 & $2.0\,\%$ & $-0.02$ & $0.99$ & $13.0/14=0.93$ & \\
& & $\mathcal{F}(\hat{B}\pm20\hat{\sigma}_{_{\rm B}})$ & $\hat{\Sigma}_{\rm nl}$ & {\bf 1.001} & {\bf 0.011} & {\bf 0.012} & ${\bf 1.2\,\%}$ & ${\bf -0.01}$ & ${\bf 0.99}$ & ${\bf 14.3/15=0.95}$ & $\blacklozenge$ \\
& & $\mathcal{F}(\hat{B}\pm50\hat{\sigma}_{_{\rm B}})$ & $\mathcal{U}(\hat{\Sigma}_{\rm nl}+20\hat{\sigma}_{_{\Sigma_{\rm nl}}})$ & 1.000 & 0.011 & 0.011 & $-2.2\,\%$ & $-0.02$ & $1.02$ & $13.1/14=0.94$ & $\lozenge$ \\
\cmidrule{2-12} %%%%% LRG (b) xi_vv %%%%%
& \multirowcell{3}{$\xi_{\rm vv}$\\(0.0167)} & -- & -- & 1.011 & 0.027 & 0.028 & $3.8\,\%$ & $-0.34$ & $1.83$ & $11.2/12=0.94$ & \\
& & $\mathcal{F}(\hat{B}\pm20\hat{\sigma}_{_{\rm B}})$ & $\hat{\Sigma}_{\rm nl}$ & {\bf 1.000} & {\bf 0.016} & {\bf 0.016} & ${\bf 3.8\,\%}$ & ${\bf -0.03}$ & ${\bf 0.99}$ & ${\bf 12.5/13=0.96}$ & $\blacklozenge$ \\
& & $\mathcal{F}(\hat{B}\pm50\hat{\sigma}_{_{\rm B}})$ & $\mathcal{U}(\hat{\Sigma}_{\rm nl}+20\hat{\sigma}_{_{\Sigma_{\rm nl}}})$ & 0.999 & 0.015 & 0.016 & $3.8\,\%$ & $-0.05$ & $1.02$ & $11.5/12=0.96$ & $\lozenge$ \\
\specialrule{\lightrulewidth}{\myaboverulesep}{\mybelowrulesep}
%%%%% LRG (c) xi_gg %%%%%
\multirowcell{10}[-2\myaboverulesep]{EZmock\\LRG (c)} & \multirowcell{3}{$\xi_{\rm gg}$\\(0.0152)} & -- & -- & 1.003 & 0.016 & 0.016 & $0.7\,\%$ & $-0.08$ & $0.94$ & $15.3/16=0.96$ & \\
& & $\mathcal{F}(\hat{B}\pm10\hat{\sigma}_{_{\rm B}})$ & $\hat{\Sigma}_{\rm nl}$ & {\bf 1.002} & {\bf 0.015} & {\bf 0.015} & ${\bf -0.4\,\%}$ & ${\bf -0.03}$ & ${\bf 0.99}$ & ${\bf 16.6/17=0.98}$ & $\blacklozenge$ \\
& & $\mathcal{F}(\hat{B}\pm50\hat{\sigma}_{_{\rm B}})$ & $\mathcal{U}(\hat{\Sigma}_{\rm nl}+20\hat{\sigma}_{_{\Sigma_{\rm nl}}})$ & 1.001 & 0.015 & 0.015 & $-1.0\,\%$ & $-0.03$ & $1.01$ & $15.5/16=0.97$ & $\lozenge$ \\
\cmidrule{2-12} %%%%% LRG (c) xi_gv %%%%%
& \multirowcell{3}{$\xi_{\rm gv}$\\(0.0169)} & -- & -- & 1.001 & 0.019 & 0.018 & $-2.1\,\%$ & $-0.00$ & $1.03$ & $13.1/14=0.94$ & \\
& & $\mathcal{F}(\hat{B}\pm10\hat{\sigma}_{_{\rm B}})$ & $\hat{\Sigma}_{\rm nl}$ & {\bf 1.000} & {\bf 0.017} & {\bf 0.017} & ${\bf 0.2\,\%}$ & ${\bf -0.01}$ & ${\bf 1.01}$ & ${\bf 14.8/15=0.99}$ & $\blacklozenge$ \\
& & $\mathcal{F}(\hat{B}\pm50\hat{\sigma}_{_{\rm B}})$ & $\mathcal{U}(\hat{\Sigma}_{\rm nl}+20\hat{\sigma}_{_{\Sigma_{\rm nl}}})$ & 0.999 & 0.017 & 0.017 & $-0.4\,\%$ & $0.01$ & $1.02$ & $13.4/14=0.96$ & $\lozenge$ \\
\cmidrule{2-12} %%%%% LRG (c) xi_vv %%%%%
& \multirowcell{4}{$\xi_{\rm vv}$\\(0.0241)} & -- & -- & 1.051 & 0.047 & 0.055 & $18.5\,\%$ & $-0.10$ & $1.70$ & $10.8/12=0.90$ & \\
& & $\mathcal{F}(\hat{B}\pm50\hat{\sigma}_{_{\rm B}})$ & $\mathcal{F}(\hat{\Sigma}_{\rm nl}\pm50\hat{\sigma}_{_{\Sigma_{\rm nl}}})$ & 0.996 & 0.016 & 0.018 & $14.3\,\%$ & $0.06$ & $0.87$ & $11.6/12=0.97$ & \\
& & $\mathcal{F}(\hat{B}\pm5\hat{\sigma}_{_{\rm B}})$ & $\hat{\Sigma}_{\rm nl}$ & {\bf 0.996} & {\bf 0.014} & {\bf 0.014} & ${\bf -0.1\,\%}$ & ${\bf -0.07}$ & ${\bf 1.00}$ & ${\bf 13.3/13=1.02}$ & $\blacklozenge$ \\
& & $\mathcal{F}(\hat{B}\pm20\hat{\sigma}_{_{\rm B}})$ & $\mathcal{U}(\hat{\Sigma}_{\rm nl}+10\hat{\sigma}_{_{\Sigma_{\rm nl}}})$ & 0.996 & 0.015 & 0.015 & $1.5\,\%$ & $-0.03$ & $0.99$ & $12.4/12=1.03$ & $\lozenge$ \\
\specialrule{\lightrulewidth}{\myaboverulesep}{\mybelowrulesep}
%%%%% ELG xi_gg %%%%%
\multirowcell{18}[-2.5\myaboverulesep]{EZmock\\ELG} & \multirowcell{6}{$\xi_{\rm gg}$\\(0.0402)} & -- & -- & 1.032 & 0.034 & 0.102 & $196.9\,\%$ & $-0.00$ & $0.58$ & $15.3/16=0.96$ & \\
& & -- & $\hat{\Sigma}_{\rm nl}$ & 1.008 & 0.035 & 0.055 & $56.8\,\%$ & $-0.04$ & $0.81$ & $16.2/17=0.95$ & \\
& & $\mathcal{F}(\hat{B}\pm5\hat{\sigma}_{_{\rm B}})$ & $\hat{\Sigma}_{\rm nl}$ & {\bf 1.009} & {\bf 0.040} & {\bf 0.041} & ${\bf 1.9\,\%}$ & ${\bf -0.10}$ & ${\bf 1.06}$ & ${\bf 16.7/17=0.98}$ & $\blacklozenge$ \\
& & $\mathcal{N}(\hat{B},3\hat{\sigma}_{_{\rm B}})$ & $\hat{\Sigma}_{\rm nl}$ & 1.009 & 0.040 & 0.041 & $1.8\,\%$ & $-0.10$ & $1.06$ & $17.0/17=1.00$ & \\
& & $\mathcal{F}(\hat{B}\pm5\hat{\sigma}_{_{\rm B}})$ & $\mathcal{N}(\hat{\Sigma}_{\rm nl},5\hat{\sigma}_{_{\Sigma_{\rm nl}}})$ & 1.010 & 0.040 & 0.041 & $3.8\,\%$ & $-0.10$ & $1.04$ & $16.7/16=1.04$ & \\
& & $\mathcal{F}(\hat{B}\pm10\hat{\sigma}_{_{\rm B}})$ & $\mathcal{U}(\hat{\Sigma}_{\rm nl}+10\hat{\sigma}_{_{\Sigma_{\rm nl}}})$ & 1.009 & 0.039 & 0.041 & $4.6\,\%$ & $-0.10$ & $1.03$ & $16.1/16=1.01$ & $\lozenge$ \\
\cmidrule{2-12} %%%%% ELG xi_gv %%%%%
& \multirowcell{6}{$\xi_{\rm gv}$\\(0.0424)} & -- & -- & 1.027 & 0.037 & 0.099 & $163.9\,\%$ & $-0.01$ & $0.56$ & $12.2/14=0.87$ & \\
& & -- & $\hat{\Sigma}_{\rm nl}$ & 1.007 & 0.036 & 0.064 & $77.6\,\%$ & $-0.04$ & $0.79$ & $14.1/15=0.94$ & \\
& & $\mathcal{F}(\hat{B}\pm10\hat{\sigma}_{_{\rm B}})$ & $\hat{\Sigma}_{\rm nl}$ & {\bf 1.008} & {\bf 0.044} & {\bf 0.044} & ${\bf -1.4\,\%}$ & ${\bf -0.10}$ & ${\bf 1.02}$ & ${\bf 14.4/15=0.96}$ & $\blacklozenge$ \\
& & $\mathcal{N}(\hat{B},10\hat{\sigma}_{_{\rm B}})$ & $\hat{\Sigma}_{\rm nl}$ & 1.008 & 0.043 & 0.044 & $2.2\,\%$ & $-0.09$ & $1.00$ & $14.5/15=0.97$ & \\
& & $\mathcal{F}(\hat{B}\pm10\hat{\sigma}_{_{\rm B}})$ & $\mathcal{N}(\hat{\Sigma}_{\rm nl},5\hat{\sigma}_{_{\Sigma_{\rm nl}}})$ & 1.007 & 0.043 & 0.043 & $-0.3\,\%$ & $-0.09$ & $1.01$ & $14.3/14=1.02$ & \\
& & $\mathcal{F}(\hat{B}\pm20\hat{\sigma}_{_{\rm B}})$ & $\mathcal{U}(\hat{\Sigma}_{\rm nl}+15\hat{\sigma}_{_{\Sigma_{\rm nl}}})$ & 1.008 & 0.043 & 0.044 & $4.4\,\%$ & $-0.09$ & $1.01$ & $13.6/14=0.97$ & $\lozenge$ \\
\cmidrule{2-12} %%%%% ELG xi_vv %%%%%
& \multirowcell{6}{$\xi_{\rm vv}$\\(0.0714)} & -- & -- & 1.073 & 0.034 & 0.098 & $188.1\,\%$ & $0.08$ & $0.63$ & $10.2/12=0.85$ & \\
& & -- & $\hat{\Sigma}_{\rm nl}$ & 1.027 & 0.045 & 0.089 & $98.7\,\%$ & $-0.01$ & $0.73$ & $12.3/13=0.94$ & \\
& & $\mathcal{F}(\hat{B}\pm3\hat{\sigma}_{_{\rm B}})$ & $\hat{\Sigma}_{\rm nl}$ & {\bf 1.035} & {\bf 0.057} & {\bf 0.065} & ${\bf 14.9\,\%}$ & ${\bf -0.04}$ & ${\bf 0.91}$ & ${\bf 12.7/13=0.98}$ & $\blacklozenge$ \\
& & $\mathcal{N}(\hat{B},\hat{\sigma}_{_{\rm B}})$ & $\hat{\Sigma}_{\rm nl}$ & 1.035 & 0.056 & 0.065 & $15.0\,\%$ & $-0.03$ & $0.91$ & $13.0/13=1.00$ & \\
& & $\mathcal{F}(\hat{B}\pm3\hat{\sigma}_{_{\rm B}})$ & $\mathcal{N}(\hat{\Sigma}_{\rm nl},5\hat{\sigma}_{_{\Sigma_{\rm nl}}})$ & 1.033 & 0.054 & 0.065 & $19.4\,\%$ & $-0.03$ & $0.89$ & $12.6/12=1.05$ & \\
& & $\mathcal{F}(\hat{B}\pm10\hat{\sigma}_{_{\rm B}})$ & $\mathcal{U}(\hat{\Sigma}_{\rm nl}+15\hat{\sigma}_{_{\Sigma_{\rm nl}}})$ & 1.032 & 0.055 & 0.067 & $23.6\,\%$ & $-0.06$ & $0.98$ & $11.6/12=0.96$ & $\lozenge$ \\
\specialrule{\lightrulewidth}{\myaboverulesep}{\mybelowrulesep}
%%%%% CROSS xi_gg %%%%%
\multirowcell{8}[-2.5\myaboverulesep]{EZmock\vspace{2pt}\\LRG (c)\\$\times$\\ELG} &  \multirowcell{2}{$\xi_{\rm gg}^\times$\\(0.0380)} & $\mathcal{F}(\hat{B}\pm10\hat{\sigma}_{_{\rm B}})$ & $\hat{\Sigma}_{\rm nl}$ & {\bf 1.008} & {\bf 0.039} & {\bf 0.039} & ${\bf -0.3\,\%}$ & ${\bf -0.08}$ & ${\bf 1.07}$ & ${\bf 16.2/17=0.96}$ & $\blacklozenge$ \\
& & $\mathcal{F}(\hat{B}\pm20\hat{\sigma}_{_{\rm B}})$ & $\mathcal{U}(\hat{\Sigma}_{\rm nl}+20\hat{\sigma}_{_{\Sigma_{\rm nl}}})$ & 1.008 & 0.037 & 0.039 & $5.2\,\%$ & $-0.07$ & $1.04$ & $15.5/16=0.97$ & $\lozenge$ \\
\cmidrule{2-12} %%%%% CROSS xi_gv %%%%%
& \multirowcell{2}{$\xi_{\rm gv}^\times$\\(0.0513)} & $\mathcal{F}(\hat{B}\pm10\hat{\sigma}_{_{\rm B}})$ & $\hat{\Sigma}_{\rm nl}$ & {\bf 1.008} & {\bf 0.051} & {\bf 0.055} & ${\bf 8.9\,\%}$ & ${\bf -0.06}$ & ${\bf 0.96}$ & ${\bf 14.3/15=0.95}$ & $\blacklozenge$ \\
& & $\mathcal{F}(\hat{B}\pm5\hat{\sigma}_{_{\rm B}})$ & $\mathcal{U}(\hat{\Sigma}_{\rm nl}+15\hat{\sigma}_{_{\Sigma_{\rm nl}}})$ & 1.007 & 0.051 & 0.054 & $6.1\,\%$ & $-0.07$ & $1.01$ & $13.8/14=0.98$ & $\lozenge$ \\
\cmidrule{2-12} %%%%% CROSS xi_vg %%%%%
& \multirowcell{2}{$\xi_{\rm vg}^\times$\\(0.0550)} & $\mathcal{F}(\hat{B}\pm3\hat{\sigma}_{_{\rm B}})$ & $\hat{\Sigma}_{\rm nl}$ & {\bf 1.011} & {\bf 0.050} & {\bf 0.054} & ${\bf 6.4\,\%}$ & ${\bf -0.05}$ & ${\bf 0.96}$ & ${\bf 14.8/15=0.99}$ & $\blacklozenge$ \\
& & $\mathcal{F}(\hat{B}\pm5\hat{\sigma}_{_{\rm B}})$ & $\mathcal{U}(\hat{\Sigma}_{\rm nl}+10\hat{\sigma}_{_{\Sigma_{\rm nl}}})$ & 1.010 & 0.049 & 0.051 & $5.0\,\%$ & $-0.07$ & $1.00$ & $13.8/14=0.99$ & $\lozenge$ \\
\cmidrule{2-12} %%%%% CROSS xi_vv %%%%%
& \multirowcell{2}{$\xi_{\rm vv}^\times$\\(0.0574)} & $\mathcal{F}(\hat{B}\pm3\hat{\sigma}_{_{\rm B}})$ & $\hat{\Sigma}_{\rm nl}$ & {\bf 1.008} & {\bf 0.053} & {\bf 0.060} & ${\bf 14.2\,\%}$ & ${\bf -0.07}$ & ${\bf 1.00}$ & ${\bf 12.7/13=0.98}$ & $\blacklozenge$ \\
& & $\mathcal{F}(\hat{B}\pm5\hat{\sigma}_{_{\rm B}})$ & $\mathcal{U}(\hat{\Sigma}_{\rm nl}+15\hat{\sigma}_{_{\Sigma_{\rm nl}}})$ & 1.009 & 0.052 & 0.062 & $18.0\,\%$ & $-0.07$ & $1.00$ & $11.8/12=0.99$ & $\lozenge$ \\
\specialrule{\heavyrulewidth}{\myaboverulesep}{0pt}
\end{tabular}
\label{tab:fit_prior}
\end{threeparttable}
\end{table*}

The fitting results with different priors are listed in Tables~\ref{tab:fit_prior} and \ref{tab:fit_prior_weight}, where comprehensive tests of the `ELG' sample are shown, as the results are sensitive to the choices of priors, due to the large 2PCF variances of the `ELG' sample.
In general, the best-fitting values can be biased when no prior is imposed, and the fitted errors are significantly larger than the dispersions of best-fitting $\alpha$, especially for measurements with large statistical errors (e.g. void--void clustering of the `ELG' sample).
This indicates that some restrictions of the parameters are necessary for improving the goodness of fits and ensuring robust error estimates.
Then, we choose ranges of the priors that make the standard deviation of $g(\alpha)$ as close to 1 as possible. When the $\sigma_{g(\alpha)}$ values are similar, we choose the one that minimize the differences between the rescaled $\alpha_{1\sigma}$ and $\langle\sigma_{\alpha}\rangle$.

To be more specific, fixing the $\Sigma_{\rm nl}$ parameter, as is done in most BAO studies \citep[e.g.][]{Xu2012, Alam2017, Bautista2021, deMattia2021}, mitigates the biases of $\alpha_{\rm med}$ in all cases.
Nevertheless, without a prior of the $B$ parameter, the median value of $\sigma_{\alpha}$ are still systematically larger than the dispersion of $\alpha_{\rm fit, med}$, and the distributions of $g(\alpha)$ do not agree with a standard normal distribution when the uncertainties are large. This is consistent with the findings in \citet[][]{Vargas2014}.
Both flat and Gaussian priors of the $B$ parameter can solve this problem. To be more conservative, we choose flat priors as our fiducial fitting scheme.
Besides, Gaussian priors of $\Sigma_{\rm nl}$ does not show improvements on the fitted results, but increases slightly the reduced chi-squared values. Thus, unvaried $\Sigma_{\rm nl}$ values are preferred in general.

When fitting to the observational data, fixed $\Sigma_{\rm nl}$ values require accurate estimates of the nonlinear BAO damping effect, as $\Sigma_{\rm nl}$ quantifies the broadening of the BAO peak. This is however only available for galaxy auto correlations $\xi_{\rm gg}$, for which the $\Sigma_{\rm nl}$ value can be measured from $N$-body simulations.
Since the BAO peak of approximate mocks is generally more damped, fitting results from the mean 2PCFs of approximate mocks can be used to estimate the upper limit of the $\Sigma_{\rm nl}$ value, for correlation functions involving voids.
Therefore, we check also flat $\Sigma_{\rm nl}$ priors in the form of $\mathcal{U} (\Sigma_{\rm nl, max})$, with only the upper limit estimated from mocks.
By adjusting a bit the priors on $B$, the results from this type of $\Sigma_{\rm nl}$ priors are roughly as good as traditional BAO fitting method, i.e., with fixed $\Sigma_{\rm nl}$ parameters, for both individual correlation functions and the combined 2PCF.

\begin{table*}
\centering
\begin{threeparttable}
\caption{Fitting results for the combined correlation functions of individual mocks, with different parameter priors. The $c$ parameter in Eq.~\eqref{eq:bao_model_para} is always fixed to the value for the corresponding sample in Table~\ref{tab:fit_setting}.
$\sigma_{\alpha, {\rm mean}}$ denotes the rescaled fitted errors with the mean 2PCFs of all mocks, and the rescaling factor is $\sqrt{1000}$. The other notations are the same as in Table~\ref{tab:fit_prior}.}
\begin{tabular}{ccccccccccc@{\hspace{3pt}}c@{\hspace{2pt}}}
\toprule
Sample & $\sigma_{\alpha, {\rm mean}}$ & $B_{\rm prior}$ & $\Sigma_{\rm nl, prior}$ & $\langle\alpha\rangle$ & $\alpha_{1\sigma}$ & $\langle\sigma_{\alpha}\rangle$ & $\delta \sigma$ & $\bar{g}(\alpha)$ & $\sigma_{g(\alpha)}$ & \multicolumn{1}{c}{$\langle\chi^2\rangle / {\rm d.o.f.}$} & \\
\midrule
%%%%% LRG (a) %%%%%
\multirowcell{2}[-0.5\aboverulesep]{MD-Patchy\\LRG (a)} & \multirowcell{2}[-0.5\aboverulesep]{0.0110}
& $\mathcal{F}(\hat{B}\pm20\hat{\sigma}_{_{\rm B}})$ & $\mathcal{U}(\hat{\Sigma}_{\rm nl}+20\hat{\sigma}_{_{\Sigma_{\rm nl}}})$ & {\bf 0.999} & {\bf 0.011} & {\bf 0.011} & ${\bf 0.7\,\%}$ & ${\bf -0.02}$ & ${\bf 1.01}$ & ${\bf 11.4/12=0.95}$  & $\blacklozenge$ \\
& & $\mathcal{F}(\hat{B}\pm20\hat{\sigma}_{_{\rm B}})$ & $\hat{\Sigma}_{\rm nl}$ & 1.000 & 0.011 & 0.011 & $1.3\,\%$ & $-0.01$ & $1.00$ & $12.2/13=0.94$  & $\lozenge$ \\
\midrule
%%%%% LRG (b) %%%%%
\multirowcell{2}[-0.5\aboverulesep]{MD-Patchy\\LRG (b)} & \multirowcell{2}[-0.5\aboverulesep]{0.0107}
& $\mathcal{F}(\hat{B}\pm20\hat{\sigma}_{_{\rm B}})$ & $\mathcal{U}(\hat{\Sigma}_{\rm nl}+10\hat{\sigma}_{_{\Sigma_{\rm nl}}})$ & {\bf 1.000} & {\bf 0.010} & {\bf 0.010} & ${\bf 0.6\,\%}$ & ${\bf -0.02}$ & ${\bf 1.04}$ & ${\bf 11.6/12=0.97}$  & $\blacklozenge$ \\
& & $\mathcal{F}(\hat{B}\pm15\hat{\sigma}_{_{\rm B}})$ & $\hat{\Sigma}_{\rm nl}$ & 1.001 & 0.010 & 0.010 & $2.9\,\%$ & $-0.02$ & $0.99$ & $12.5/13=0.96$  & $\lozenge$ \\
\midrule
%%%%% LRG (c) %%%%%
\multirowcell{2}[-0.5\aboverulesep]{EZmock\\LRG (c)} & \multirowcell{2}[-0.5\aboverulesep]{0.0136}
& $\mathcal{F}(\hat{B}\pm20\hat{\sigma}_{_{\rm B}})$ & $\mathcal{U}(\hat{\Sigma}_{\rm nl}+30\hat{\sigma}_{_{\Sigma_{\rm nl}}})$ & {\bf 1.002} & {\bf 0.014} & {\bf 0.014} & ${\bf -1.8\,\%}$ & ${\bf -0.02}$ & ${\bf 1.00}$ & ${\bf 11.8/12=0.98}$  & $\blacklozenge$ \\
& & $\mathcal{F}(\hat{B}\pm20\hat{\sigma}_{_{\rm B}})$ & $\hat{\Sigma}_{\rm nl}$ & 1.003 & 0.014 & 0.014 & $-3.0\,\%$ & $-0.01$ & $0.99$ & $12.5/13=0.96$  & $\lozenge$ \\
\midrule
%%%%% ELG %%%%%
\multirowcell{2}[-0.5\aboverulesep]{EZmock\\ELG} & \multirowcell{2}[-0.5\aboverulesep]{0.0368}
& $\mathcal{F}(\hat{B}\pm10\hat{\sigma}_{_{\rm B}})$ & $\mathcal{U}(\hat{\Sigma}_{\rm nl}+20\hat{\sigma}_{_{\Sigma_{\rm nl}}})$ & {\bf 1.010} & {\bf 0.039} & {\bf 0.039} & ${\bf 0.6\,\%}$ & ${\bf -0.07}$ & ${\bf 1.07}$ & ${\bf 11.9/12=0.99}$  & $\blacklozenge$ \\
& & $\mathcal{F}(\hat{B}\pm20\hat{\sigma}_{_{\rm B}})$ & $\hat{\Sigma}_{\rm nl}$ & 1.010 & 0.039 & 0.039 & $0.3\,\%$ & $-0.06$ & $0.98$ & $12.4/13=0.95$  & $\lozenge$ \\
\midrule
%%%%% CROSS %%%%%
\multirowcell{2}[-0.5\aboverulesep]{EZmock\\LRG (c) $\times$ ELG} & \multirowcell{2}[-0.5\aboverulesep]{0.0346}
& $\mathcal{F}(\hat{B}\pm10\hat{\sigma}_{_{\rm B}})$ & $\mathcal{U}(\hat{\Sigma}_{\rm nl}+20\hat{\sigma}_{_{\Sigma_{\rm nl}}})$ & {\bf 1.004} & {\bf 0.037} & {\bf 0.038} & ${\bf 0.6\,\%}$ & ${\bf -0.08}$ & ${\bf 1.06}$ & ${\bf 11.6/12=0.97}$ & $\blacklozenge$ \\
& & $\mathcal{F}(\hat{B}\pm20\hat{\sigma}_{_{\rm B}})$ & $\hat{\Sigma}_{\rm nl}$ & 1.004 & 0.037 & 0.038 & $1.3\,\%$ & $-0.06$ & $1.01$ & $12.0/13=0.92$  & $\lozenge$ \\
\bottomrule
\end{tabular}
\label{tab:fit_prior_weight}
\end{threeparttable}
\end{table*}

\begin{table*}
\centering
\begin{threeparttable}
\caption{Multi-tracer fitting results for combinations of correlation functions of individual mocks, with different parameter priors.
Rows marked with `$\blacklozenge$' and `$\lozenge$' are obtained with priors indicated by the same symbol in Table~\ref{tab:fit_prior}.
Meanings of columns with numbers are the same as in Tables~\ref{tab:fit_prior} and \ref{tab:fit_prior_weight}.}
\begin{tabular}{clcccccccccc@{\hspace{3pt}}c@{\hspace{2pt}}}
\toprule
Sample & Data vector & $\sigma_{\alpha, {\rm mean}}$ & $B_{\rm prior}$ & $\Sigma_{\rm nl, prior}$ & $\langle\alpha\rangle$ & $\alpha_{1\sigma}$ & $\langle\sigma_{\alpha}\rangle$ & $\delta \sigma$ & $\bar{g}(\alpha)$ & $\sigma_{g(\alpha)}$ & \multicolumn{1}{c}{$\langle\chi^2\rangle / {\rm d.o.f.}$} & \\
\midrule
%%%%% LRG (a) %%%%%
\multirowcell{4}[-\aboverulesep]{MD-Patchy\\LRG (a)} & \multirow{2}{*}{$\{\boldsymbol{\xi}_{\rm gg}, \boldsymbol{\xi}_{\rm gv}\}$} & \multirowcell{2}[-.5\aboverulesep]{0.0112}  & $\{\mathcal{F}\}$ & $\{ \hat{\Sigma}_{\rm nl} \}$ & {\bf 1.001} & {\bf 0.011} & {\bf 0.011} & ${\bf 3.9\,\%}$ & ${\bf -0.01}$ & ${\bf 0.98}$ & ${\bf 32.1/33=0.97}$ & $\blacklozenge$ \\
 & & & $\{\mathcal{F}\}$ & $\{\mathcal{U}\}$ & 1.000 & 0.011 & 0.010 & $-5.0\,\%$ & $-0.01$ & $1.05$ & $29.9/31=0.96$ & $\lozenge$ \\
\cmidrule{2-13}
& \multirow{2}{*}{$\{\boldsymbol{\xi}_{\rm gg}, \boldsymbol{\xi}_{\rm gv}, \boldsymbol{\xi}_{\rm vv}\}$} & \multirowcell{2}[-.5\aboverulesep]{0.0112}  & $\{\mathcal{F}\}$ & $\{ \hat{\Sigma}_{\rm nl} \}$ & {\bf 1.001} & {\bf 0.011} & {\bf 0.011} & ${\bf 0.5\,\%}$ & ${\bf -0.01}$ & ${\bf 1.00}$ & ${\bf 45.0/47=0.96}$ & $\blacklozenge$ \\
 & & & $\{\mathcal{F}\}$ & $\{\mathcal{U}\}$ & 1.000 & 0.011 & 0.010 & $-8.9\,\%$ & $-0.01$ & $1.08$ & $42.3/44=0.96$ & $\lozenge$ \\
\midrule
%%%%% LRG (b) %%%%%
\multirowcell{4}[-\aboverulesep]{MD-Patchy\\LRG (b)} & \multirow{2}{*}{$\{\boldsymbol{\xi}_{\rm gg}, \boldsymbol{\xi}_{\rm gv}\}$} & \multirowcell{2}[-.5\aboverulesep]{0.0104}  & $\{\mathcal{F}\}$ & $\{ \hat{\Sigma}_{\rm nl} \}$ & {\bf 1.002} & {\bf 0.010} & {\bf 0.010} & ${\bf 4.0\,\%}$ & ${\bf -0.02}$ & ${\bf 0.99}$ & ${\bf 31.9/33=0.97}$ & $\blacklozenge$ \\
 & & & $\{\mathcal{F}\}$ & $\{\mathcal{U}\}$ & 1.002 & 0.010 & 0.010 & $-2.9\,\%$ & $-0.02$ & $1.07$ & $30.3/31=0.98$ & $\lozenge$ \\
\cmidrule{2-13}
& \multirow{2}{*}{$\{\boldsymbol{\xi}_{\rm gg}, \boldsymbol{\xi}_{\rm gv}, \boldsymbol{\xi}_{\rm vv}\}$} & \multirowcell{2}[-.5\aboverulesep]{0.0106}  & $\{\mathcal{F}\}$ & $\{ \hat{\Sigma}_{\rm nl} \}$ & {\bf 1.002} & {\bf 0.010} & {\bf 0.010} & ${\bf 0.6\,\%}$ & ${\bf -0.02}$ & ${\bf 1.02}$ & ${\bf 45.0/47=0.96}$ & $\blacklozenge$ \\
 & & & $\{\mathcal{F}\}$ & $\{\mathcal{U}\}$ & 1.001 & 0.010 & 0.009 & $-6.9\,\%$ & $-0.02$ & $1.10$ & $43.0/44=0.98$ & $\lozenge$ \\
\midrule
%%%%% LRG (c) %%%%%
\multirowcell{4}[-\aboverulesep]{EZmock\\LRG (c)} & \multirow{2}{*}{$\{\boldsymbol{\xi}_{\rm gg}, \boldsymbol{\xi}_{\rm gv}\}$} & \multirowcell{2}[-.5\aboverulesep]{0.0144}  & $\{\mathcal{F}\}$ & $\{ \hat{\Sigma}_{\rm nl} \}$ & {\bf 1.002} & {\bf 0.014} & {\bf 0.014} & ${\bf -1.5\,\%}$ & ${\bf -0.02}$ & ${\bf 1.02}$ & ${\bf 32.6/33=0.99}$ & $\blacklozenge$ \\
 & & & $\{\mathcal{F}\}$ & $\{\mathcal{U}\}$ & 1.002 & 0.015 & 0.014 & $-7.2\,\%$ & $-0.01$ & $1.08$ & $30.2/31=0.97$ & $\lozenge$ \\
\cmidrule{2-13}
& \multirow{2}{*}{$\{\boldsymbol{\xi}_{\rm gg}, \boldsymbol{\xi}_{\rm gv}, \boldsymbol{\xi}_{\rm vv}\}$} & \multirowcell{2}[-.5\aboverulesep]{0.0136}  & $\{\mathcal{F}\}$ & $\{ \hat{\Sigma}_{\rm nl} \}$ & {\bf 1.001} & {\bf 0.011} & {\bf 0.011} & ${\bf -2.0\,\%}$ & ${\bf -0.04}$ & ${\bf 1.02}$ & ${\bf 46.3/47=0.99}$ & $\blacklozenge$ \\
 & & & $\{\mathcal{F}\}$ & $\{\mathcal{U}\}$ & 0.999 & 0.011 & 0.011 & $-4.9\,\%$ & $-0.02$ & $1.07$ & $42.9/44=0.97$ & $\lozenge$ \\
\midrule
%%%%% ELG %%%%%
\multirowcell{4}[-\aboverulesep]{EZmock\\ELG} & \multirow{2}{*}{$\{\boldsymbol{\xi}_{\rm gg}, \boldsymbol{\xi}_{\rm gv}\}$} & \multirowcell{2}[-.5\aboverulesep]{0.0376}  & $\{\mathcal{F}\}$ & $\{ \hat{\Sigma}_{\rm nl} \}$ & {\bf 1.012} & {\bf 0.039} & {\bf 0.038} & ${\bf -3.3\,\%}$ & ${\bf -0.09}$ & ${\bf 1.02}$ & ${\bf 31.8/33=0.96}$ & $\blacklozenge$ \\
 & & & $\{\mathcal{F}\}$ & $\{\mathcal{U}\}$ & 1.010 & 0.040 & 0.037 & $-7.0\,\%$ & $-0.09$ & $1.09$ & $30.7/31=0.99$ & $\lozenge$ \\
\cmidrule{2-13}
& \multirow{2}{*}{$\{\boldsymbol{\xi}_{\rm gg}, \boldsymbol{\xi}_{\rm gv}, \boldsymbol{\xi}_{\rm vv}\}$} & \multirowcell{2}[-.5\aboverulesep]{0.0379}  & $\{\mathcal{F}\}$ & $\{ \hat{\Sigma}_{\rm nl} \}$ & {\bf 1.014} & {\bf 0.040} & {\bf 0.037} & ${\bf -6.9\,\%}$ & ${\bf -0.11}$ & ${\bf 1.09}$ & ${\bf 45.3/47=0.96}$ & $\blacklozenge$ \\
 & & & $\{\mathcal{F}\}$ & $\{\mathcal{U}\}$ & 1.012 & 0.042 & 0.035 & $-17.9\,\%$ & $-0.12$ & $1.26$ & $42.8/44=0.97$ & $\lozenge$ \\
\midrule
%%%%% CROSS %%%%%
\multirowcell{4}[-\aboverulesep]{EZmock\\LRG (c)\\$\times$\\ELG} & \multirow{2}{*}{$\{\boldsymbol{\xi}_{\rm gg}^\times, \boldsymbol{\xi}_{\rm gv}^\times, \boldsymbol{\xi}_{\rm vg}^\times\}$} & \multirowcell{2}[-.5\aboverulesep]{0.0370}  & $\{\mathcal{F}\}$ & $\{ \hat{\Sigma}_{\rm nl} \}$ & {\bf 1.008} & {\bf 0.037} & {\bf 0.036} & ${\bf -2.1\,\%}$ & ${\bf -0.08}$ & ${\bf 1.06}$ & ${\bf 46.3/49=0.95}$ & $\blacklozenge$ \\
 & & & $\{\mathcal{F}\}$ & $\{\mathcal{U}\}$ & 1.007 & 0.039 & 0.031 & $-18.7\,\%$ & $-0.10$ & $1.20$ & $44.0/46=0.96$ & $\lozenge$ \\
\cmidrule{2-13}
& \multirow{2}{*}{$\{\boldsymbol{\xi}_{\rm gg}^\times, \boldsymbol{\xi}_{\rm gv}^\times, \boldsymbol{\xi}_{\rm vg}^\times, \boldsymbol{\xi}_{\rm vv}^\times\}$} & \multirowcell{2}[-.5\aboverulesep]{0.0370}  & $\{\mathcal{F}\}$ & $\{ \hat{\Sigma}_{\rm nl} \}$ & {\bf 1.009} & {\bf 0.036} & {\bf 0.035} & ${\bf -2.3\,\%}$ & ${\bf -0.10}$ & ${\bf 1.09}$ & ${\bf 59.0/63=0.94}$ & $\blacklozenge$ \\
 & & & $\{\mathcal{F}\}$ & $\{\mathcal{U}\}$ & 1.009 & 0.038 & 0.030 & $-21.6\,\%$ & $-0.15$ & $1.32$ & $56.1/59=0.95$ & $\lozenge$ \\
\bottomrule
\end{tabular}
\label{tab:fit_prior_multi}
\end{threeparttable}
\end{table*}

We then further check the BAO measurements with multiple clustering measurements fitted simultaneously, and the results are listed in Table~\ref{tab:fit_prior_multi}.
This time, the pull quantities with fixed $\Sigma_{\rm nl}$ prior are still consistent with those of the standard normal distribution.
However, for the $\Sigma_{\rm nl}$ prior with an upper limit, the fitted errors appear to be underestimated for the `ELG' sample, as well as the cross correlation measurements between the `LRG (c)' and `ELG' samples.
This is possibly because the clustering measurements in these cases are too noisy to have constraints on $\Sigma_{\rm nl}$.
%Moreover, since the $\Sigma_{\rm nl}$ value of $\xi_{\rm gg}$ can be estimated accurately using $N$-body simulations, the fitted errors with multiple 2PCFs are less affected by the overestimation of BAO damping with approximate mocks.
Thus, eventually we use fixed $\Sigma_{\rm nl}$ values for fits to multiple 2PCFs.
Nevertheless, accurate estimations of $\Sigma_{\rm nl}$ are crucial for the combined correlation functions, as there is no way so far to measure the values using $N$-body simulations.
Since the $\Sigma_{\rm nl}$ priors with upper limits work well for the combined 2PCFs (see Table~\ref{tab:fit_prior_weight}), we choose these priors as our fiducial fitting scheme for the combined correlation functions.
Moreover, we compare the results with the two types of $\Sigma_{\rm nl}$ priors, to assess the potential systematic biases (see Section~\ref{sec:fit_prior}).

With the fiducial fitting priors, the median values of $\alpha$, as well as their dispersions, are generally consistent with the results from fits to the mean 2PCFs of all mock realizations.
Obvious discrepancies are only observed for few cases, such as for the void auto correlations of the `LRG (c)' sample.
In this case, the pull distribution, as well as the comparison of the dispersion of $\alpha_{\rm fit, med}$ and the fitted $\sigma_\alpha$, both prefer the smaller BAO uncertainty from fits to individual mocks.
Even with prior widths of 100 times $\hat{\sigma}_{_B}$ and $\hat{\sigma}_{_{\Sigma_{\rm nl}}}$, the fitted $\sigma_\alpha$ from individual mocks are still significantly smaller than that of the mean 2PCFs of all mocks.
Besides, we have further performed a fit to the mean $\xi_{\rm vv}$ of all mocks, with our fiducial fitting priors, and the result is consistent with that of individual mocks.
It implies that a prior is always necessary for void clustering of the `LRG (c)' sample.
There is however no such problem for the combined 2PCFs. This suggests that a multi-tracer BAO fitting approach based on the combined 2PCFs is more robust, than that relying on the stacked data vectors of multiple 2PCFs.

%%%%%%%%%%%%%%%%% NEW SECTION %%%%%%%%%%%%%%%%%%

\section{BAO measurements for different Galactic caps}
\label{sec:bao_caps}

\begin{figure}
\centering
\includegraphics[width=.98\columnwidth]{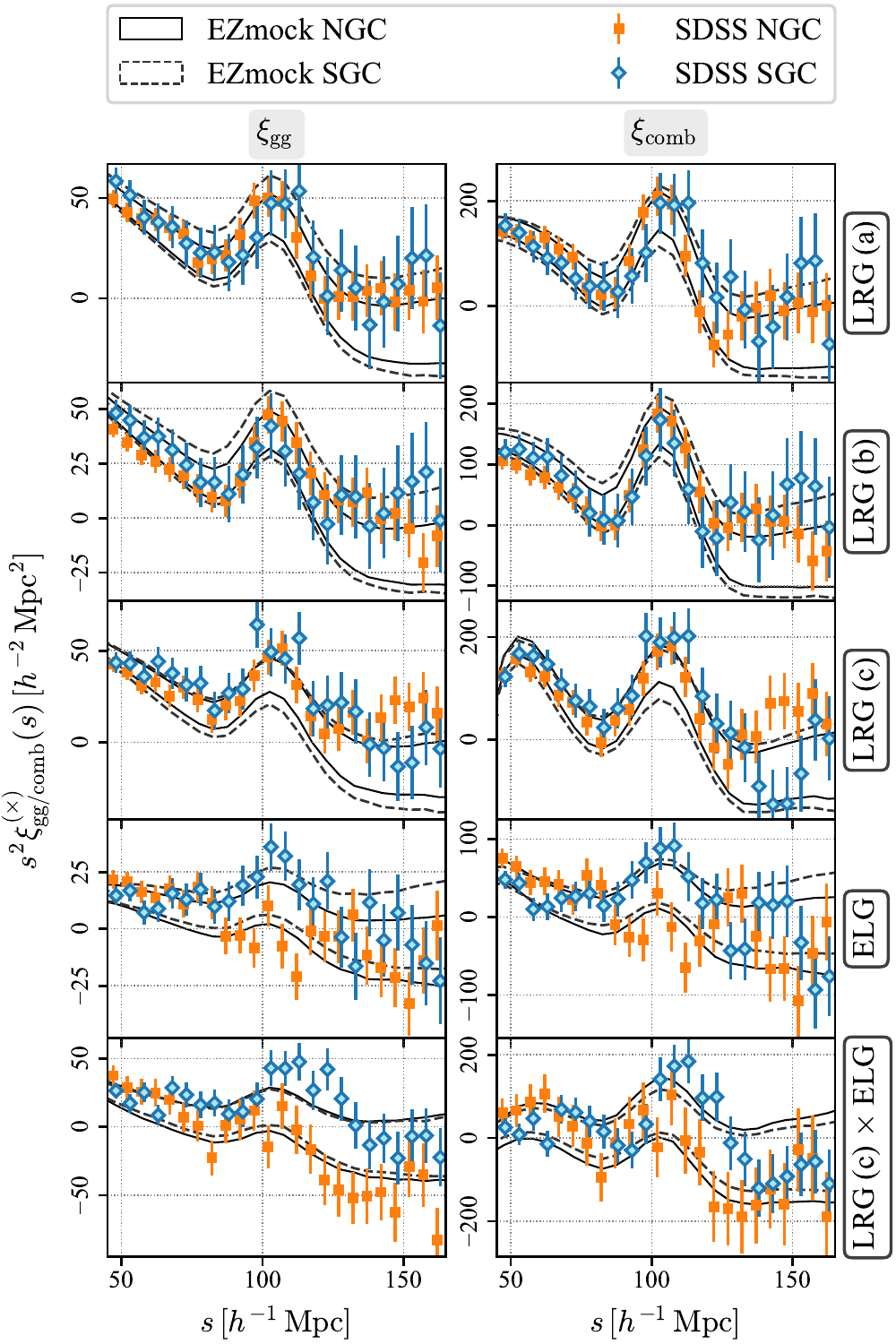}
\caption{Galaxy--galaxy and combined 2PCFs measured from samples in different Galactic caps. Envelopes show the 1\,$\sigma$ dispersions of 2PCFs from mocks. The abscissas of the SDSS NGC and SGC measurements are shifted by $-0.5$ and $0.5\,\mpc$ respectively for illustration purposes.}
\label{fig:xi_caps}
\end{figure}

\begin{figure}
\centering
\includegraphics[width=.98\columnwidth]{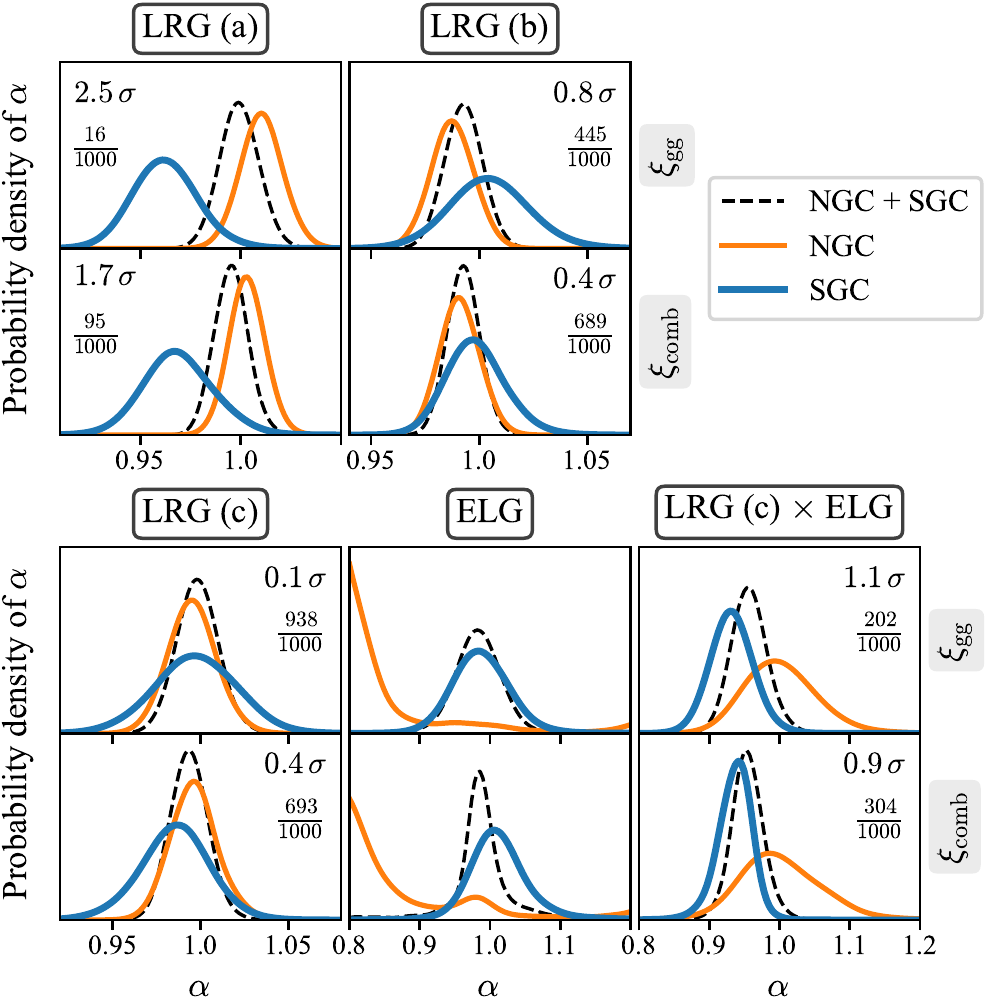}
\caption{Marginalized posterior distributions of $\alpha$ measured from the 2PCFs of SDSS data in different Galactic caps. The tensions between NGC and SGC measurements, as well as the fractions of mocks with larger tensions, are given in the subplots except for the `ELG' sample.}
\label{fig:alpha_caps}
\end{figure}

The 2PCFs of SDSS samples in different Galactic caps, as well as the 1\,$\sigma$ regions of the corresponding approximate mocks, are shown in Figure~\ref{fig:xi_caps}.
It can be seen that measurements from both Galactic caps are generally consistent, though the differences for those involving ELGs are relatively large \citep[see also][]{Raichoor2021}.
To examine the statistical consistencies of BAO measurements from different Galactic caps, we analyse the SDSS data in NGC and SGC separately with the fitting scheme shown in Table~\ref{tab:fit_setting}.
The resulting marginalized posterior distributions of $\alpha$ are shown in Figure~\ref{fig:alpha_caps}, in comparison with results from the combined NGC and SGC samples.
Here, the galaxy-only results for the eBOSS samples except ELGs in NGC are consistent with those from the joint BAO and RSD analysis in \citet[][]{Wang2020}.
However, the posterior distributions of $\alpha$ measured from the `ELG' sample in NGC seem to prefer values outside our prior range, for both galaxy auto correlation and the combined 2PCF of galaxies and voids.
This can be explained by the failure of BAO detection from this subsample, which is consistent with the relatively large cosmic variance of the `ELG' sample, based on the analyses of individual mocks \citep[][]{Raichoor2021}.

We define the tension between BAO measurements from NGC and SGC as
\begin{equation}
\delta \alpha_{_{\rm NS}} = \frac{| \alpha_{_{\rm NGC}} - \alpha_{_{\rm SGC}} |}{\sqrt{\sigma_{\alpha, \scriptscriptstyle{\rm NGC}}^2 + \sigma_{\alpha, \scriptscriptstyle{\rm SGC}}^2}} ,
\end{equation}
where $\alpha_{_{\rm NGC/SGC}}$ and $\sigma_{\alpha, \scriptscriptstyle{\rm NGC/SGC}}$ denote the median value and 1\,$\sigma$ errors of $\alpha$ respectively, which are obtained from the corresponding posterior distributions.
The tensions measured from different samples are shown in Figure~\ref{fig:alpha_caps} as well, together with the fractions of mock realizations that present larger differences between the two Galactic caps.
We observe a relatively large tension (2.5\,$\sigma$) for the galaxy auto correlation of the `LRG (a)' sample.
Nevertheless, the tension is significantly reduced when including the clustering of voids, in which case almost 10 per cent of the mocks show larger tensions.
For the rest of the samples, apart from `ELG', the BAO measurements from the combined correlation functions in NGC and SGC are well consistent, with the differences smaller than 1\,$\sigma$.
The results suggest that it is reasonable to combine the clustering statistics in NGC and SGC for the multi-tracer BAO measurements.

%%%%%%%%%%%%%%%%% NEW SECTION %%%%%%%%%%%%%%%%%%

\section{Cross covariances of distance measurements from different samples}
\label{sec:dist_cov}

Since some of our galaxy samples share the same volume, or even contain identical objects, it is crucial to take into account their cross correlations for cosmological constraints (see Figure~\ref{fig:sample_xi_corr}).
However, the cross covariances of distance measurements are not directly available from the BAO fits. Thus, we use the correlation matrices ${\bf R}$ of $D_{_{\rm V}} / r_{\rm d}$ obtained from individual mock realizations to approximate the covariances of measurements from different samples.
Denoting the statistical and systematic errors of distances measured from the SDSS data as $\boldsymbol{\sigma}$ and $\boldsymbol{s}$ respectively, where the vector components are results from individual samples, the covariance matrix for the measurements of all samples is computed as
\begin{equation}
{\bf C} = m_1 ( \boldsymbol{\sigma} \otimes \boldsymbol{\sigma} ) \circ {\bf R} +  \boldsymbol{s}^2 {\bf I} .
\end{equation}
Here, $\otimes$ and $\circ$ are the outer product and Hadamard (element-wise) product respectively, and ${\bf I}$ indicates the identity matrix.
The factor $m_1$ given by Eq.~\eqref{eq:m1} is for accounting the precision of the covariance matrices used for BAO fits. Since we use always the same number of mocks for covariance matrix estimates, as well as the same number of parameters and data points for the fits, $m_1$ is identical for all samples.
Furthermore, we assume no cross correlations between the systematic errors of different samples, so they are only added to the diagonal terms of the covariance matrix.

Using the correlation coefficients shown in Figure~\ref{fig:corr_coef}, the full covariance matrix of $D_{_{\rm V}} / r_{\rm d}$ measured from galaxy--galaxy correlations of the five SDSS samples is
\begin{equation}
\makeatletter\setlength\BA@colsep{1ex}\makeatother
\begin{aligned}
&{\bf C}_{\rm gg} (D_{_{\rm V}}/r_{\rm d}) =\\
&\begin{blockarray}{ccccccc}
{\scriptstyle z_{\rm eff}:} &
{\scriptstyle 0.38} &
{\scriptstyle 0.51} &
{\scriptstyle 0.70} &
{\scriptstyle 0.77} &
{\scriptstyle 0.85} & \\
\begin{block}{c(ccccc)l}
& 9.9313 & 4.6559 & & & & {\scriptstyle{\rm LRG\,(a)}} \\
& 4.6559 & 13.651 & & & & {\scriptstyle{\rm LRG\,(b)}} \\
10^{-3}
& & & 37.753 & 13.863 & 6.2144 & {\scriptstyle{\rm LRG\,(c)}} \\
& & & 13.863 & 195.24 & 86.093 & {\scriptstyle{\rm CROSS}} \\
& & & 6.2144 & 86.093 & 372.50 & {\scriptstyle{\rm ELG}}\\
\end{block}
\end{blockarray} ,
\end{aligned}
\end{equation}
where the effective redshifts of different samples are shown, and `CROSS' denotes the results from cross correlations between the `LRG (c)' and `ELG' samples. Similarly, the covariance matrix for distances measured from the combined correlations is
\begin{equation}
\makeatletter\setlength\BA@colsep{1ex}\makeatother
\begin{aligned}
&{\bf C}_{\rm comb} (D_{_{\rm V}}/r_{\rm d}) =\\
&\begin{blockarray}{ccccccc}
{\scriptstyle z_{\rm eff}:} &
{\scriptstyle 0.38} &
{\scriptstyle 0.51} &
{\scriptstyle 0.70} &
{\scriptstyle 0.77} &
{\scriptstyle 0.85} & \\
\begin{block}{c(ccccc)l}
& 7.2216 & 3.4095 & & & & {\scriptstyle{\rm LRG\,(a)}} \\
& 3.4095 & 9.8147 & & & & {\scriptstyle{\rm LRG\,(b)}} \\
10^{-3}
& & & 33.825 & 12.042 & 3.4234 & {\scriptstyle{\rm LRG\,(c)}} \\
& & & 12.042 & 149.02 & 61.581 & {\scriptstyle{\rm CROSS}} \\
& & & 3.4234 & 61.581 & 278.53 & {\scriptstyle{\rm ELG}}\\
\end{block}
\end{blockarray} .
\end{aligned}
\end{equation}

%%%%%%%%%%%%%%%%%%%%%%%%%%%%%%%%%%%%%%%%%%%%%%%%%%

% Don't change these lines
\bsp	% typesetting comment
\label{lastpage}
\end{document}